\documentclass[aps,prd,eqsecnum,floatfix,nofootinbib,preprint,tightenlines]{revtex4-1}

\usepackage{latexsym}
\usepackage{amsmath}
\usepackage{graphicx}
\usepackage{subcaption}
\usepackage{multirow}
\usepackage[dvipsnames]{xcolor}
\usepackage{esint}
\usepackage{hyperref}

\begin{document}

\begin{boldmath}
\begin{center}
{\large{\bf
Borel Representation of $\tau$ Hadronic Spectral Function Moments in Contour-Improved Perturbation Theory }
}\\[8mm]
Andr\'e H. Hoang,$^{a,b}$ Christoph Regner,$^{a,c}$  \\[8 mm]
$^a$University of Vienna, Faculty of Physics,\\Boltzmanngasse 5, 1090 Vienna, Austria\\
[5mm]
$^b$Erwin Schr\"odinger International Institute for Mathematics and Physics,\\
University of Vienna, Boltzmanngasse 9, 1090 Vienna, Austria\\
[5mm]
$^c$University of Vienna, Vienna Doctoral School in Physics,\\Boltzmanngasse 5, 1090 Vienna, Austria
\\[10mm]
\end{center}
\end{boldmath}
\begin{quotation}
We show that the Borel representations of $\tau$ hadronic spectral function moments based on contour-improved perturbation theory (CIPT) in general differ from those obtained within fixed-order perturbation theory (FOPT) in the presence of IR renormalons in the underlying Adler function. 
The Borel sums obtained from both types of Borel representations in general differ as well, and the apparently conflicting behavior of the FOPT and CIPT spectral function moment series at intermediate orders, which has been subject to many studies in the past literature, can be understood quantitatively using concrete Borel function models.
The difference between the CIPT and FOPT Borel sums, which we call the ``asymptotic separation", can be computed analytically for any Borel function model and is proportional to inverse exponential terms in the strong coupling. 
Even though moments can be designed where the asymptotic separation is strongly suppressed, it is as a matter of principle unavoidable. 
If the Borel function of the Euclidean Adler function has a sizeable gluon condensate renormalon cut, 
the asymptotic separation can explain the observed disparity of the CIPT and FOPT spectral function moments at the 5-loop level. The existence of the asymptotic separation implies that the power corrections in the operator product expansion for the spectral function moments in the CIPT expansion approach do not have the commonly assumed analytic standard form. 

\end{quotation}

\newpage

\tableofcontents

\section{Introduction}
\label{sec:intro}

Moments of the $\tau$ hadronic spectral functions obtained from the ALEPH~\cite{Barate:1998uf,Schael:2005am,Davier:2013sfa} and the OPAL~\cite{Ackerstaff:1998yj} collaborations have served as an important tool for precise determinations of the strong coupling $\alpha_s$. Theoretical predictions for the spectral function moments can be related to the QCD vacuum vector and axial-vector current correlator\footnote{For simplicity we assume massless quarks and neglect electroweak corrections throughout this article.}
\begin{equation}
\label{eq:correlator}
\left(p^\mu p^\nu - g^{\mu\nu} p^2 \right) \Pi(p^2) \,\equiv\,  i\!\int \! dx \, e^{ipx} \,
\langle\Omega|\,T\{ j_{v/av,jk}^{\mu}(x)\,j_{v/av,jk}^{\nu}(0)^\dagger\}
\Omega\rangle\,,
\end{equation}
where the vector and axial-vector currents are given by $j_{v/av,jk}^{\nu} = \bar q_j\gamma^\mu(\gamma_5)q_k$. Accounting only for first generation quarks and QCD corrections, the theoretical predictions for the moments in the massless quark limit are conventionally parametrized as~\cite{Braaten:1991qm,LeDiberder:1992jjr,Boito:2014sta,Pich:2016bdg}
\begin{equation}
\label{eq:momdef}
A_{W_i}(s_0) \, =\,  \frac{N_c}{2} |V_{ud}|^2 \biggl[\,
\delta^{\rm tree}_{W_i} + \delta^{(0)}_{W_i}(s_0) + 
\sum_{d\geq 2}\delta^{(d)}_{W_i}(s_0) +\delta^{\rm DV}_{W_i}(s_0)\,\biggr] \,,
\end{equation}
where $s_0$ is the upper bound of the spectral function integration and the index $W_i$ indicates the type of moment considered. Furthermore, $N_c=3$ and $V_{ud}$ is a CKM matrix element. The term $\delta^{\rm tree}_{W_i}$ is the tree-level contribution and $\delta^{(0)}_{W_i}(s_0)$ are higher order perturbative QCD corrections. The terms $\delta^{(d)}_{W_i}(s_0)$ represent (vacuum matrix element) condensate corrections in the framework of the operator product expansion (OPE)~\cite{Shifman:1978bx}, where the leading dimension $d=4$ term is related to the gluon condensate. The most common approach to define and compute the OPE corrections is to employ the $\overline{\mbox{MS}}$ scheme to separate the low-scale nonperturbative vacuum matrix elements and the high-scale perturbative Wilson coefficients. 
The term $\delta^{\rm DV}_{W_i}(s_0)$ stands for duality violation corrections related to nonperturbative contributions missed by the expansion in local condensate matrix elements. The validity of the ansatz of Eq.~(\ref{eq:momdef}) is based on an expansion in inverse powers of $s_0$, where both types of nonperturbative corrections are defined from the vacuum polarization function $\Pi(p^2)$. The perturbative QCD corrections can be written as a counterclockwise contour integral in the complex $p^2$-plane along a circle with radius $s_0$ around the origin~\cite{Braaten:1991qm,LeDiberder:1992jjr}:
\begin{equation}
\label{eq:deltadef}
\delta^{(0)}_{W_i}(s_0)\, =\,\frac{1}{2\pi i}\,\,\ointctrclockwise \limits_{|s|=s_0}\!\! \frac{{\rm d}s}{s}\,W_i({\textstyle \frac{s}{s_0}})\,\hat D(s)\,.
\end{equation}
Here $\hat D(s)$ is the (reduced) parton level Adler function defined as
\begin{equation}
\label{eq:redAdler}
\frac{1}{4\pi^2}\Big(1+\hat D(s)\Big) \, \equiv \, -\,s\,\frac{{\rm d}\hat\Pi(s)}{{\rm d} s}\,,
\end{equation}
where $\hat\Pi(s)$ stands for the parton level current correlator.
The OPE and duality violation corrections to the spectral function moments are computed in analogy to Eq.~(\ref{eq:deltadef}) by integration over the corresponding OPE and duality violation corrections of the Adler function.
At the level of the Adler function, the OPE corrections have the standard form 
\begin{equation}
\label{eq:DOPE}
D^{\rm OPE}(s) \, = \,
\frac{C_{4,0}(\alpha_s(-s))}{s^2} \langle  {\cal O}_{4,0} \rangle  +
\sum\limits_{d=6}^\infty \frac{1}{(-s)^{d/2}} \sum_i  C_{d,i}(\alpha_s(-s)) \langle  {\cal O}_{d,i}\rangle \,
\end{equation}
where $\langle {\cal O}_{d,i}\rangle$ are the nonperturbative condensate matrix elements and $C_{d,i}$ the Wilson coefficients. The terms $\delta^{(d)}_{W_i}(s_0)$ are obtained from integrals in analogy to Eq.~(\ref{eq:deltadef}). The index $i$ sums over operators of the same dimensions with different anomalous dimensions.
The integration path with distance $s_0$ from the origin in the complex $s$-plane may be deformed as long as the beginning and endpoints are at $s_0\pm i 0$, respectively, the path encloses the Landau pole of the strong coupling, does not cross cuts and stays within the perturbative regime.     
The weight function $W_i(x)$ is a polynomial that vanishes at $x=1$ and determines the moment considered.\footnote{The weight function $W_i(x)$ is connected to the weight function $w_i(x)$ used for integration over the measured $\tau$ spectral function by the relation $W_i(x)=2\int_x^1 {\rm d}\bar x\,w_i(\bar x)$ and arises when using integration by parts to relate the theoretical spectral function moments to the Adler function.} For $s_0=m_\tau^2$ and $W_\tau(x) =(1-x)^3(1+x)$ the moment directly applies to the normalized total hadronic $\tau$ decay width $R_\tau=\Gamma(\tau^-\to\mbox{hadrons}\,\nu_\tau(\gamma))/\Gamma(\tau^-\to e^-\bar{\nu}_e\nu_\tau(\gamma))$. The choices of the weight function $W_i(x)$ substantially affects the importance and size of the OPE and DV corrections. For example, if $W_i(x)$ does not contain a quadratic term, the effects of the dimension-4 gluon condensate is suppressed. Furthermore, DV effects are smaller, if $W_i(x)$ is ``pinched", i.e.\ if it vanishes  quadratically or with an even higher power at $x=1$~\cite{Boito:2014sta}. It should also be noted that the sizes of the OPE corrections $\delta^{(d)}_{W_i}(s_0)$, obtained from the terms in Eq.~(\ref{eq:DOPE}), are in one-to-one correspondence to contributions in $\delta^{(0)}_{W_i}(s_0)$, computed within dimensional regularization and the ${\overline {\rm MS}}$ renormalization scheme in QCD fixed-order perturbation theory, which diverge at large orders of perturbation theory~\cite{Gross:1974jv,tHooft:1977xjm,David:1983gz,Mueller:1984vh,Beneke:1998ui}. These diverging corrections make the series for $\delta^{(0)}_{W_i}(s_0)$ asymptotic, such that a concrete value for the entire series may only be assigned using the renormalon calculus and based on models for the Borel representation of the series $\delta^{(0)}_{W_i}(s_0)$~\cite{Ball:1995ni,Neubert:1995gd,Beneke:2008ad}. 
Such concrete values for $\delta^{(0)}_{W_i}(s_0)$ are usually called the ``Borel sum", and they 
involve making choices concerning the regularization of the nonanalytic infrared (IR) renormalon cuts of the Adler function's Borel function model.
The size and importance of the different renormalon contributions in the spectral function moments depend on the normalization of the various IR renormalons cuts and the choice of the weight function $W_i(x)$.

Using input from QCD multiloop calculations of the vacuum correlator at four (i.e.\ ${\cal O}(\alpha_s^3)$)~\cite{Gorishnii:1990vf,Surguladze:1990tg} and five loops (i.e.\ ${\cal O}(\alpha_s^4)$)~\cite{Baikov:2008jh} an impressive precision of around $1.5\%$ has been achieved for $\alpha_s(M_Z)$, see Ref.~\cite{Boito:2014sta,Pich:2016bdg,Boito:2020xli,ParticleDataGroup:2020ssz} for recent results and reviews. This current uncertainty of $\alpha_s$ determinations from $\tau$ hadronic spectral function moments is dominated by perturbative uncertainties associated to calculations of $\delta^{(0)}_{W_i}(s_0)$. At the present time one of the major limitations arises from the fact that two different perturbative prescriptions to evaluate $\delta^{(0)}_{W_i}(s_0)$ lead to systematic differences that do not seem to be covered by the conventional perturbative uncertainty estimates related to renormalization scale variations. 

Starting from the QCD perturbation series of the reduced Adler function\footnote{We adopt the notation from Ref.~\cite{Beneke:2008ad}, where the coefficients $c_{n,k}$ are defined from the perturbation series of the vacuum polarization function $\Pi(s)$.} 
\begin{eqnarray}
\label{eq:AdlerseriesCIPT}
\hat D(s) & \, =  \,& \, \sum\limits_{n=1}^\infty
   c_{n,1} \,\big({\textstyle \frac{\alpha_s(-s)}{\pi}}\big)^n\,, 
\\ \label{eq:AdlerseriesFOPT}
& \, = \, &
\, \sum\limits_{n=1}^\infty\,
\big({\textstyle \frac{\alpha_s(s_0)}{\pi}}\big)^n \, \sum\limits_{k=1}^{n} k\, c_{n,k}\,\ln^{k-1}({\textstyle \frac{-s}{s_0}}) \,,
\end{eqnarray}
the first expansion approach, called contour-improved perturbation theory (CIPT), is directly based on the evaluation of Eq.~(\ref{eq:deltadef}) using the Adler function series in Eq.~(\ref{eq:AdlerseriesCIPT}) in powers of $\alpha_s(-s)$, where the complex-valuedness of the Adler function is encoded  entirely in the strong coupling. The perturbation series for the CIPT moments adopt the form~\cite{Pivovarov:1991rh}
\begin{equation}
\label{eq:deltaCIPT}
\delta^{(0),{\rm CIPT}}_{W_i}(s_0)\, =\, 
\,\frac{1}{2\pi i} \,  \sum\limits_{n=1}^\infty c_{n,1} \,
\ointctrclockwise\limits_{|x|=1}\!\! \frac{{\rm d}x}{x}\,W_i(x)\,\big({\textstyle \frac{\alpha_s(-x s_0)}{\pi}}\big)^n\,.
\end{equation}
Here, the contour integral is written in terms of the dimensionless variable $x=s/s_0$, and the CIPT series is obtained by truncating the sum over $n$.
It has been argued in Ref.~\cite{LeDiberder:1992jjr} that the CIPT approach sums large perturbative corrections (involving factors of the phase of $s$) along the path in the complex plane and that this leads to a fast decrease of the size of the series coefficients with $n$. The second expansion approach, called fixed-order perturbation theory (FOPT), is based on the expansion in powers of $\alpha_s(s_0)$ shown in Eq.~(\ref{eq:AdlerseriesFOPT}), where the complex-valuedness of the Adler function is entirely encoded in explicit powers of the logarithm $\ln(-s/s_0)$, leading to 
\begin{equation}
\label{eq:deltaFOPT}
\delta^{(0),{\rm FOPT}}_{W_i}(s_0)\, =\, 
\,\frac{1}{2\pi i} \,  \sum\limits_{n=1}^\infty \,
\big({\textstyle \frac{\alpha_s(s_0)}{\pi}}\big)^n \,
\sum\limits_{k=1}^{n} k\, c_{n,k}\, \ointctrclockwise\limits_{|x|=1}\!\! \frac{{\rm d}x}{x}\,W_i(x)\, \ln^{k-1}(-x)
\,.
\end{equation}
The FOPT perturbation series is obtained by summing over $k$ as shown, but truncating the sum over $n$.
It has been argued in Ref.~\cite{Ball:1995ni} that the FOPT prescription leads to a more efficient realization of cancellations of asymptotic renormalon contributions at large orders in association to the suppression of OPE corrections (with respect to the vacuum polarization function) depending on the choice of the weight function.\footnote{In Refs.~\cite{Caprini:2009vf,Caprini:2011ya} `new/modified contour-improved' and `new/modified fixed-order' moment series were defined from the Borel sums over series of functional approximations to the Borel function of the Adler function. The actual CIPT and FOPT series expansions we consider here are fundamentally different from the `new/modified' expansions of Refs.~\cite{Caprini:2009vf,Caprini:2011ya}. The results of our work do not in any way apply to these `new/modified' expansions.}
In general, for physically well-motivated weight functions (including $W_\tau$) the CIPT and FOPT series each seems to approach definite values. It has also been observed that for many weight functions the CIPT series seems to ``converge'' somewhat more quickly and leads to smaller renormalization scale variations. However, with the advent of the five loop coefficient~\cite{Baikov:2008jh}, it became apparent for the total hadronic $\tau$ width that the values 
both series seem to approach are incompatible within their respective renormalization scale variation, where CIPT in general leads to the smaller result. As a result, and in absence of an argument that would imply differences for the OPE and DV corrections within both expansion approaches, which could reconcile the difference in the CIPT and FOPT perturbation series, strong coupling determinations based on FOPT lead to systematically smaller fitted values for $\alpha_s$. 

This FOPT-CIPT discrepancy problem has motivated a number of theoretical studies to explore the higher-order behavior of the CIPT and FOPT series beyond the concretely known five-loop level based on concrete models for the Borel (transformation) function of the Adler function~\cite{Jamin:2005ip,Beneke:2008ad,DescotesGenon:2010cr,Beneke:2012vb}. Part of the motivation was to learn more about the potential underlying systematics and to possibly identify which of the two expansion methods may be `better'. Making no assumption on whether the known (and apparently converging) five-loop series for the Adler function already contains sizeable contributions from the gluon condensate renormalon,\footnote{This assumption implies that the gluon condensate OPE correction of the Adler function may not be the dominant source of the Adler function's ambiguity in perturbation theory and that the IR renormalon cut associated to the gluon condensate OPE correction in $B[\hat D](u)$ is strongly suppressed. This view was argued to be implausible in Ref.~\cite{Beneke:2012vb}.} it was shown in Ref.~\cite{DescotesGenon:2010cr} that the construction of Borel models allows for so much freedom that the models' Borel sum can be made to agree with either the FOPT or the CIPT series' intermediate large order behavior. So no real insight could be gained. Adopting this view, the apparent disparity in the behavior of the FOPT and CIPT spectral function moment series at the 5-loop level may be considered as an artifact of a low-order trunction that
may be only reconciled (or better understood) through the computation of additional corrections beyond the 5-loops level.

On the other hand, if one argues that the gluon condensate renormalon already has a sizeable contribution to the known five-loop series, the Borel function can be shown to contain a gluon condensate renormalon cut with a sizeable normalization~\cite{Beneke:2012vb}. In this context more definite studies of the higher-order behavior could be carried out. It was found that FOPT in general approaches to the calculated model's Borel sum, while CIPT seems to approach a value as well, which can, however, be significantly different from the Borel sum~\cite{Beneke:2012vb}. In Ref.~\cite{Beneke:2012vb} several plausibility arguments were discussed to support their view, but no strict prove on the (in)validity of either one of these two views concerning the structure of the Borel function exists in the literature. We also stress that the purpose of this article is not to advocate or dismiss one of them. Rather, in this article we explore the question how a systematic disparity between the FOPT and CIPT series can arise in the presence of IR renormalons and if this could possibly explain the observed disparity in the behavior of the FOPT and CIPT spectral function moments at the 5-loop level as a systematic effect.

Interestingly, in all previous studies on the FOPT-CIPT discrepancy problem it has been commonly taken for granted that the Borel representations (and thus the corresponding Borel sums) for the perturbative moments in the FOPT as well as the CIPT expansion approach are identical and have the concrete form\footnote{For the QCD $\beta$-function in the $\overline{\rm MS}$ scheme we adopt the convention 
$\frac{d\alpha(\mu)}{d\ln\mu}=\beta(\alpha_s(\mu))= -2\alpha_s(\mu)\sum_{n=0}^\infty \beta_n(\frac{\alpha_s(\mu)}{4\pi})^{n+1}$ with $\beta_0=11-\frac{2}{3}n_f$ being the one-loop coefficient.}
\begin{equation}
\label{eq:BorelFOPT}
\delta_{W_i,{\rm Borel}}^{(0),{\rm FOPT}}(s_0) = {\rm PV} \int_0^\infty \!\! {\rm d} u \,\, 
\frac{1}{2\pi i}\,\ointctrclockwise\limits_{|x|=1} \frac{{\rm d}x}{x} \, W_i(x) \,
B[\hat D](u)\,e^{-\frac{4\pi u}{\beta_0\alpha_s(-x s_0)}}\,.
\end{equation}
Here and in the rest of this paper $B[\hat D](u)$ is the Borel function of the perturbation series of the real-valued Euclidean Adler function for negative real $s=-s_0$ with respect to the expansion in powers of $\alpha_s(s_0)$, where the two expansions in Eqs.~(\ref{eq:AdlerseriesCIPT}) and (\ref{eq:AdlerseriesFOPT}) are identical. Its Taylor series around the origin of the Borel plane has the form 
\begin{equation}
\label{eq:BorelD}
B[\hat D](u)=\sum_{n=1}^\infty \frac{4^n\,c_{n,1}}{\beta_0^n\,\Gamma(n)}\,u^{n-1}\,.
\end{equation}
The Taylor series converges absolutely for $|u|<1$, the distance of the ultraviolet (UV) renormalon cut along the negative real axis for $u\le -1$. If all orders $n$ would be available, the full form of $B[\hat D](u)$ including the IR renormalon cuts along the positive real axis for $u\ge 2$ and the ultraviolet (UV) renormalon cuts could be obtained by analytic continuation. In practice we have to rely on models for $B[\hat D](u)$ accounting for the known coefficients $c_{n,1}$.
The Euclidean Adler function series is recovered from the inverse Borel transform 
\begin{equation}
\label{eq:invBorelD}
\hat D(-s_0) = \int_0^\infty \!\! {\rm d} u \,\, 
B[\hat D](u)\,e^{-\frac{4\pi u}{\beta_0\alpha_s(s_0)}}\,
\end{equation}
adopting the Taylor series for $B[\hat D](u)$ given in Eq.~(\ref{eq:BorelD}).
In Eq.~(\ref{eq:BorelFOPT}), the canonical way to define the $u$-integration over the IR renormalon cuts contained in $B[\hat D](u)$ is by taking the average of deforming the path above and below the real axis, which corresponds to the principal value (PV) prescription in the case of poles (which appear in the large-$\beta_0$ approximation).
For simplicity we refer to this way of defining the value of the Borel integral for the rest of this article as the ``PV prescription", and we indicate it by the prefix ``PV".
It is the value obtained from Eq.~(\ref{eq:BorelFOPT}) defined with the PV prescription,
which has been used as the Borel sum in the theoretical studies~\cite{Jamin:2005ip,Beneke:2008ad,DescotesGenon:2010cr,Beneke:2012vb} mentioned above.\footnote{Using the PV prescription to define Eq.~(\ref{eq:BorelFOPT}) is a particular choice and not unique, so that Eq.~(\ref{eq:BorelFOPT}) has an ambiguity. This ambiguity is discussed in Sec.~\ref{sec:asymptoticseparation} and unrelated to the FOPT-CIPT discrepancy problem.}

The confusing aspect explored in these studies was 
that using the Taylor expansion for $B[\hat D](u)$ in powers of $u$ and carrying out the inverse Borel transformation integral for each term (prior to carrying out the contour integral) in Eq.~(\ref{eq:BorelFOPT}) leads to the CIPT series in Eq.~(\ref{eq:deltaCIPT}). On the other hand, with an additional expansion in powers  of $\alpha_s(s_0)$ (again prior to carrying out the contour integral) 
one can recover the FOPT series in Eq.~(\ref{eq:deltaFOPT}). So the apparent disparity in the asymptotic behavior of the FOPT and CIPT spectral function moments series
for models with a gluon condensate renormalon is bewildering, under the assumption that the remaining contour integration does not play an essential conceptual role and if one considers the difference between both expansions as a systematic effect and not as a quantification of the perturbative error. None of the previous theoretical studies offered any satisfying resolution or explanation of this matter.

This is the point from which we start the discussions of this article. We treat the contour integration as an essential aspect of the characterization of both expansion methods. In this context we show that Eq.~(\ref{eq:BorelFOPT}) with the PV prescription is the correct Borel representation for the FOPT series of the $\tau$ spectral function moments, hence employing the superscript `FOPT'. However, the correct Borel representation for the CIPT spectral function moments differs and has the form
\begin{equation}
\label{eq:BorelCIPT}
\delta_{W_i,{\rm Borel}}^{(0),{\rm CIPT}}(s_0) = \int_0^\infty \!\! {\rm d} \bar u \,\,  
\frac{1}{2\pi i}\,\ointctrclockwise\limits_{{\cal C}_x} \frac{{\rm d}x}{x} \, W_i(x) \,
\big({\textstyle \frac{\alpha_s(-x s_0)}{\alpha_s(s_0)}}\big)\,
B[\hat D]\Big({\textstyle \frac{\alpha_s(-x s_0)}{\alpha_s(s_0)}} \bar u\Big)
\,e^{-\frac{4\pi \bar u}{\beta_0\alpha_s(s_0)}}\,.
\end{equation}	
The two Borel representations in Eqs.~(\ref{eq:BorelFOPT}) and  (\ref{eq:BorelCIPT}) are equivalent perturbatively, i.e.\ when one considers only the Taylor expansion of the Borel function $B[\hat D](u)$.
However, beyond the perturbative expansion, if the Borel function $B[\hat D](u)$ contains IR renormalons, $\delta_{W_i,{\rm Borel}}^{(0),{\rm CIPT}}(s_0)$ differs from the FOPT Borel representation $\delta_{W_i,{\rm Borel}}^{(0),{\rm FOPT}}(s_0)$.
The path ${\cal C}_x$ in the complex $x$-plane has its beginning and endpoints at $x_\mp=1\pm i0$, respectively, 
but must be deformed away from the circular path $|x|=1$ to account for a modified singularity structure that arises in the integrand in Eq.~(\ref{eq:BorelCIPT}). Furthermore, since $\alpha_s(-x s_0)$ is complex-valued along the path, no regularization prescription is needed for the Borel integration over $\bar u$ along the real axis.

In this article we explore the properties of the CIPT Borel representation in Eq.~(\ref{eq:BorelCIPT}). In a study of Borel function models we show that the CIPT series indeed generally approach the Borel sum of Eq.~(\ref{eq:BorelCIPT}) in the same way as the FOPT series approach the Borel sum of Eq.~(\ref{eq:BorelFOPT}), regardless of the concrete form of the Borel model. The difference of the two Borel sums, which we call the ``asymptotic separation", can be computed analytically. It can  be traced back to the fact that the form of Eq.~(\ref{eq:BorelCIPT}) inherently implies a regularization of the nonanalytic IR renormalon cuts that differs from the PV prescription used in Eq.~(\ref{eq:BorelFOPT}). The asymptotic separation thus consists of
terms that scale as powers of $\Lambda_{\rm QCD}/s_0$
and involves exponentials of the inverse strong coupling that vanish to all orders in the fixed-order expansion. 
For Borel models with a sizeable gluon condensate renormalon norm, the size of the asymptotic separation can be significantly larger than the ambiguity that is commonly assigned to the Borel sum of the FOPT series and may explain the FOPT-CIPT discrepancy problem. The difference in size between the asymptotic separation and the FOPT Borel sum ambiguity is related to a number of peculiar analytic properties inherent to (\ref{eq:BorelCIPT}) in the presence of IR renormalon cuts contained in $B[\hat D](u)$.
We also show that the different IR regularizations involved in the FOPT and CIPT Borel representations in Eqs.~(\ref{eq:BorelFOPT}) and (\ref{eq:BorelCIPT}) are not simply related to differing schemes for the condensate matrix elements in Eq.~(\ref{eq:DOPE}), as one may expected for different IR OPE regulariation schemes.
Rather, it turns out that the OPE power corrections associated to the CIPT Borel representation and the CIPT expansion method cannot be computed at all from the OPE terms with the standard analytic form given in Eq.~(\ref{eq:DOPE}). In other words, 
the OPE corrections for the CIPT spectral function moments do not have standard form.

The findings in this article are model-independent in the sense that they are valid for any Borel function model including the Adler function's `true' Borel function. In the context of the large-$\beta_0$ approximation, where all higher order corrections and the Borel function are known explicitly, we can show that the asymptotic separation is indeed the reason for the FOPT-CIPT discrepancy problem.
In full QCD the phenomenological and practical implications for the 5-loop spectral function moments 
depend on whether the Borel function of the Adler function indeed contains a gluon condensate renormalon cut with a sizeable normalization.
For the case that the normalization is sizeable (i.e.\ not strongly suppressed) the asymptotic separation may
explain the observed disparity between the FOPT and CIPT spectral function moment series at the 5-loop level. 
This observation provides concrete prospects that the FOPT-CIPT discrepancy problem may eventually be reconciled. This is being explored in Ref.~\cite{Benitez-Rathgeb:2022yqb}.
On the other hand, if the normalization of the gluon condensate renormalon is strongly suppressed, the asymptotic separation still exists, but it is numerically small, so that the observed disparity between the 5-loop CIPT and FOPT spectral function series is of an unrelated origin.  
In any case, we believe that the results and implications of this article contribute toward a more refined understanding of the conceptual aspects of the CIPT and FOPT expansion methods for $\tau$ hadronic spectral function moments.

This article is organized as follows:
In Sec.~\ref{sec:Borelrepresentation} we prove that Eqs.~(\ref{eq:BorelFOPT}) and (\ref{eq:BorelCIPT}) are the correct Borel representations for the FOPT and CIPT spectral function moment series expansions, respectively. We also discuss why they are not equivalent in the presence of IR renormalons.
In Sec.~\ref{sec:anatomy} we examine in detail the structure and the analytic properties of the Borel representation and Borel function of the CIPT expansion in comparison to the FOPT expansion. We discuss the perturbative construction of the CIPT Borel function in powers of the Borel space variable $\bar u$ (which is manifestly different from that of the FOPT series), the deformation of the contour ${\cal C}_x$ needed to define the Borel integral and the unusual and peculiar analytic properties of the CIPT Borel function. These properties already indicate that the OPE corrections for the CIPT moment series do not have standard form.
In Sec.~\ref{sec:asymptoticseparation} we calculate the asymptotic separation analytically and present its form in comparison with the expression for the ambiguity of the FOPT Borel sum. That the asymptotic separation provides the correct description of the observable difference in the asymptotic behavior of the FOPT and CIPT series in the large-$\beta_0$ approximation is demonstrated in Sec.~\ref{sec:large-beta}. Here we consider $R_\tau$ and several other moments, as well as the Adler function's exact Borel function and other Borel function models. The results corroborate that the OPE corrections for the CIPT moment series do not have standard form. 
In Sec.~\ref{sec:realistic} a similar analysis is carried out in full QCD and using different models for the Borel function $B[\hat D](u)$. We show that, again, the asymptotic separation provides the correct description of the observable difference in the asymptotic behavior of the FOPT and CIPT series. Finally, in
Sec.~\ref{sec:implications} we explore the possibility that the different analytic form of the FOPT and CIPT Borel representations may be already relevant at the level of the Adler function for complex $s$.
In Sec.~\ref{sec:conclusions} we conclude.

\section{FOPT and CIPT Borel Representations}
\label{sec:Borelrepresentation}

Consider a real-valued quantity having the perturbation series $\sigma=\sum_{n=1}^{\infty}c_n (\alpha_s(\mu^2)/\pi)^n$
with some definite real-valued choice for the renormalization scale $\mu$ of the strong coupling.
The Borel function of the series with respect to the expansion in powers of $\alpha_s(\mu^2)$ is defined by the Taylor series expansion $B[\sigma](u)=\sum_{n=1}^{\infty}(4^n c_n)/(\beta_0^n \Gamma(n))u^{n-1}$. The series for $\sigma$ is recovered by the inverse Borel transform $\sigma=\int_0^\infty du \,B[\sigma](u)\,e^{-4\pi u/(\alpha_s(\mu^2)\beta_0)}$ term-by-term for the Taylor series for $B[\sigma](u)$. This is the Borel representation of the quantity $\sigma$. The renormalon calculus~\cite{Gross:1974jv,tHooft:1977xjm,David:1983gz,Mueller:1984vh,Beneke:1998ui} is based on the properties that the Taylor series of the Borel function $B[\sigma](u)$ in powers of $u$ is absolute convergent in a circle around the origin of the complex Borel $u$-plane with a radius that equals the distance of the renormalon cut or pole that is located closest to the origin. The resummed function $B[\sigma](u)$ in this circle is unique and can be analytically continued unambiguously into the entire Borel plane (at least as far as information accessible to perturbation theory is concerned). If the coefficients $c_n$ would be known to all orders, all renormalon cuts or poles accessible through perturbation theory could in principle be recovered unambiguously by the analytic continuation. Using a regularization prescription for the cuts or poles located on the positive real Borel axis in the Borel representation, a definite ``resummed" value can be obtained for the series. This value is called the Borel sum. Furthermore, imposing a scheme change in the strong coupling in the Borel representation, e.g.\ when expanding $\alpha_s(\mu^2)$ in powers of the strong coupling at a different definite real-valued scale, $\alpha(\mu^{\prime 2})$, the Borel function's form with respect to $\alpha_s(\mu^{\prime 2})$ is modified in a computable way, but the Borel sum is unchanged. 

In this section we use these principles to show that Eq.~(\ref{eq:BorelCIPT}) is the correct Borel representation for the perturbation series of the $\tau$ hadronic spectral function moments in the CIPT approach given in Eq.~(\ref{eq:deltaCIPT}), and that Eq.~(\ref{eq:BorelFOPT}) is the correct Borel representation for the perturbation series of the $\tau$ hadronic spectral function moments in the FOPT approach given in Eq.~(\ref{eq:deltaFOPT}). We consider the contour integrations as an essential ingredient in both series expansions. The first thing to note is that switching from the FOPT expansion to the CIPT expansion is not related to a change from definite renormalization scale $s_0$ to another one, because the CIPT series does simply not involve powers of the strong coupling at a definite renormalization scale.  
The difference between the two Borel representations at this point arises from the simple fact that the CIPT and FOPT perturbation series are of a different kind, which subsequently leads to a
difference in the Taylor series in the corresponding Borel functions. 
This by itself does not yet mean that the corresponding Borel sums necessarily differ as well. But it turns out, that 
it is not possible to analytically convert between the two Borel representations if the full resummed Borel function contains nonanalytic IR renormalon cuts. This leaves the possibility that the corresponding Borel sums do not agree. This difference and its implications are the subject of the discussions in the subsequent sections.    

We start with the proof that Eq.~(\ref{eq:BorelCIPT}) is the correct Borel representation for the perturbation series of the $\tau$ hadronic spectral function moments in the CIPT expansion approach of Eq.~(\ref{eq:deltaCIPT}).  
Let us start from the observation that the results for the contour integrals over the functions $\frac{1}{x} W_i(x)(\frac{\alpha_s(-x s_0)}{\pi})^n$ involve a nontrivial interplay of the $x$-dependences of the weight functions and the functions $\alpha_s^n(-x s_0)$. It is therefore natural that the numbers obtained from contour integrals at each order $n$ are considered as part of the series coefficients. In other words, we argue that the contour integration is an essential intrinsic ingredient of the CIPT series and that one cannot simply consider the results for the CIPT spectral function moment series as an expansion in power of the complex-valued $\alpha_s(-x s_0)$.  
Rather, as a definite expansion parameter one can adopt $\alpha_s(s_0)$, which one can conveniently pull out of the series coefficients.\footnote{One can in fact pull out any definite small constant which leads to equivalent Borel representations through a {\it real-valued} rescaling of the Borel variable $\bar u$, which represents a trivial way a Borel representation can be rewritten without changing the Borel sum value.} So when defining the Borel representation for the CIPT series one should, instead of Eq.~(\ref{eq:deltaCIPT}), consider the series 
\begin{equation}
\label{eq:deltaCIPT2}
\delta^{(0),{\rm CIPT}}_{W_i}(s_0)\, =\, 
\,\frac{1}{2\pi i} \,  \sum\limits_{n=1}^\infty \bar c_{n} \,\Big[\,
\ointctrclockwise\limits_{|x|=1}\!\! \frac{{\rm d}x}{x}\,W_i(x)\,\Big({\textstyle \frac{a(-x)}{a_0}}\Big)^n\,\Big]\,
a_0^n\,,
\end{equation}
which is a definite series in powers of $\alpha_s(s_0)$ with coefficients that arise from the contour integrals.
Here we have defined
\begin{align}
\label{eq:adef}
a(x)\, & \equiv\, \frac{\beta_0\,\alpha_s(s)}{4\pi}  \, = \, \frac{\beta_0\,\alpha_s(x s_0)}{4\pi} \,,
\nonumber \\
a_0\, & \equiv\, \frac{\beta_0\,\alpha_s(s_0)}{4\pi}  \, = \, a(1)\,,
\nonumber \\
\bar c_n\, & \equiv\, \frac{4^n\,c_{n,1}}{\beta_0^n}
\end{align} 
for convenience. We will use these definitions throughout this article. The Borel representation of the CIPT series in Eq.~(\ref{eq:deltaCIPT2}) can now be determined through the associated Taylor series in the Borel variable $\bar u$:
\begin{eqnarray}
\label{eq:BorelCIPTTaylor}
\delta_{W_i,{\rm Borel}}^{(0),{\rm CIPT}}(s_0) & = &
\int_0^\infty \!\! {\rm d} \bar u \,
B\Big[\delta_{W_i}^{(0),{\rm CIPT}}(s_0)\Big](\bar u) \,  e^{-\frac{\bar u}{a_0}} 
\nonumber \\
& = &
\int_0^\infty \!\! {\rm d} \bar u \,\,  
\frac{1}{2\pi i}\, \sum\limits_{n=1}^\infty  \,\Big[\, \ointctrclockwise\limits_{|x|=1} \frac{{\rm d}x}{x} \, W_i(x) \,
\frac{\bar c_{n}}{\Gamma(n)}\,\Big({\textstyle \frac{a(-x)}{a_0}}\Big)^n\,
 \bar u^{n-1} 
\,\Big]
\,e^{-\frac{\bar u}{a_0}}\,.
\end{eqnarray}
At this point we recall that the $u$-series for the Euclidean Borel function $B[\hat D](u)$ given in Eq.~(\ref{eq:BorelD}) is absolute convergent for $|u|<1$, with the radius of convergence being determined by the UV renormalon located closest to the origin. As a consequence the $\bar u$-series in Eq.~(\ref{eq:BorelCIPTTaylor}) is absolute convergent for $|\bar u\,a(-x)/a_0|<1$ and $|\bar u | <1$. Since $|a(-x)/a_0|=|\alpha_s(-x s_0)/\alpha_s(s_0)|\le 1$ we therefore have convergence for $|x|=1$ (and in fact along any contour with $|x|\ge 1$). Thus, adopting the principles of Borel representations mentioned above, we can swap the sum over $n$ and the contour integral, and the analytically continued Borel function beneath the contour integral associated to Eq.~(\ref{eq:BorelCIPTTaylor}) has the form $(\frac{a(-x)}{a_0})B[\hat D]((\frac{a(-x)}{a_0})\bar u)$ in the entire complex $\bar u$ plane. It inherits the nonanalytic structures already contained in $B[\hat D](u)$ which are, however, modified due to the additional dependence on $a(-x)/a_0$.  
We thus immediately obtain Eq.~(\ref{eq:BorelCIPT}) as the correct Borel representation for the CIPT series of the $\tau$ hadronic spectral function moments. We remind the reader that the existence of the contour integration is a crucial aspect of our argumentation. In contrast, the traditional view, that Eq.~(\ref{eq:BorelFOPT}) would be the valid Borel representation of the CIPT moment series, is based on the assumption that the contour integration is not a crucial aspect for the properties of the CIPT expansion method and that $\alpha_s(-x s_0)$ can be considered to be a valid expansion parameter for the formulation of the CIPT Borel function. While it is obviously true that one can reproduce the CIPT series terms from this traditional view (as this only relies on the Taylor series of $B[\hat D](u)$), we argue that this traditional view is not appropriate beyond perturbation theory (when we account for the nonanalytic renormalon singularities in $B[\hat D](u)$) and when considering the Borel sum of the CIPT series.

At this point we can immediately spot an important subtlety related to the contour integration along the path $|x|=1$, that arises due to the IR renormalon cuts (or poles) of the form $1/(p-u)^\gamma$ (for $p=2,3,\ldots$) that are contained in the analytically continued Borel function $B[\hat D](u)$. In the Borel representation of Eq.~(\ref{eq:BorelCIPT}) this leads to cuts along the real $x$-axis for ${\rm Re}[\alpha_s(-x s_0)] > p\alpha_s(s_0)/\bar u$. 
For $\bar u \ge p$ (when the contour integral is carried out before the Borel integral as indicated in Eq.~(\ref{eq:BorelCIPT})) these cuts enforce a deformation of the contour further into the negative real complex $x$-plane such that is crosses the real axis at some value
$\tilde x<-1$ so that ${\rm Re}[\alpha_s(-\tilde x s_0)] < p\alpha_s(s_0)/\bar u$. This deformation leaves the definition of the underlying CIPT series given in Eq.~(\ref{eq:deltaCIPT2}) unchanged and can also be applied to the FOPT series (and its Borel sum) without leading to any modification. The necessity for this deformation in the Borel representation of Eq.~(\ref{eq:BorelCIPT}) is a highly unusual feature and reflects that the CIPT expansion has unusual features. This is one of the central aspect of this work. More details on the contour deformation are discussed in Sec.~\ref{sec:path}.  

Let us now prove that Eq.~(\ref{eq:BorelFOPT}) is the correct Borel representation for the perturbation series of the $\tau$ hadronic spectral function moments in the FOPT approach given in Eq.~(\ref{eq:deltaFOPT}) which is an expansion in powers of $\alpha_s(s_0)$.
The argumentation is subtle, as it may appear natural that Eq.~(\ref{eq:BorelFOPT}) is associated to an expansion in powers of $\alpha_s(-x s_0)$. However, this view is inappropriate given that all $x$-dependences should be considered as part of the series coefficients which involve the contour integration in addition. 
To this end we rewrite Eq.~(\ref{eq:BorelFOPT}) to explicitly constitute a Borel representation with respect to the expansion parameter $\alpha_s(s_0)$, 
\begin{equation}
\label{eq:BorelFOPTrewr}
\delta_{W_i,{\rm Borel}}^{(0),{\rm FOPT}}(s_0) = {\rm PV}\int_0^\infty \!\! {\rm d} u \,\, 
\frac{1}{2\pi i}\,\ointctrclockwise\limits_{|x|=1} \frac{{\rm d}x}{x} \, W_i(x) \,
B[\hat D^{\rm FOPT}](u,x)\,e^{-\frac{u}{a_0}}\,,
\end{equation}
where we define
\begin{equation}
\label{eq:BorelfctFOPT}
B[\hat D^{\rm FOPT}](u,x) \equiv B[\hat D](u)\,e^{-\frac{u}{a(-x)}+\frac{u}{a_0}}\,
\end{equation}
and where $B[\hat D](u)$ is the full Euclidean Borel function in the entire complex Borel plane and not just its Taylor expansion.
We show that (i) the Taylor expansion of $B[\hat D^{\rm FOPT}](u,x)$ correctly reproduces the  FOPT series in powers of $\alpha_s(s_0)$ in Eq.~(\ref{eq:deltaFOPT}) and that (ii) $B[\hat D^{\rm FOPT}](u,x)$ (which at the level of Eq.~(\ref{eq:BorelfctFOPT}) still has a nontrivial dependence on $\alpha_s(s_0)$) can be rewritten in terms of a pure function of $u$ without changing the value of the Borel sum and without relying on the Taylor expansion. 
The first property ensures that the FOPT series for the spectral functions moments are correctly reproduced. The second property ensures that we have determined the correct Borel function with respect to the expansion in $\alpha_s(s_0)$ (which correctly contains all nonanalytic structures in the complex $u$-plane) given that the summation of the $u$-Taylor series uniquely recovers the full Borel function through analytic continuation. The subtle point is that we must carry out all the required analytic manipulations at the level of the full function  $B[\hat D^{\rm FOPT}](u,x)$ with all its nonanalytic cuts and poles, {\it without referring to its Taylor series in powers of $u$}. If we had to rely to the $u$-Taylor series in our manipulations, we were back to the statement that the CIPT and FOPT series are two different expansion approaches to the same underlying (asymptotic) series, which is of course undisputed. However, this is not the point of this discussion, since we are interested in the correct Borel representation and Borel sum which are relevant beyond the perturbation series.

In the large-$\beta_0$ approximation the manipulation involved for demonstrating property (ii) is actually quite trivial, since there is not much to do. We simply use the equality  $1/a(-x) = 1/a_0 +\ln(-x)$ to identically rewrite the function $B[\hat D^{\rm FOPT}](u,x)$ in the form
\begin{align}
\label{eq:BorelDFOPT2}
B[\hat D^{\rm FOPT}](u,x) \, =\, &
B[\hat D](u) \,  e^{-u\ln(-x)}\,.
\end{align}
There is no dependence on $\alpha_s(s_0)$, no assumption has been made on the form of $B[\hat D](u)$ and the PV prescription is not affected either. It is also straightforward to show property (i), namely that Taylor expanding $B[\hat D^{\rm FOPT}](u,x)$ reproduces the FOPT moment series in the large-$\beta_0$ approximation:
\begin{align}
B[\hat D^{\rm FOPT}](u,x) \, =\, &
\sum\limits_{n=1}^\infty\,  
\frac{u^{n-1}}{\Gamma(n)}\bar c_n \, \sum\limits_{i=0}^\infty  \frac{u^i (-\ln(-x))^{i}}{i!} \\
\, =\, &
\sum\limits_{m=1}^\infty\, u^{m-1}\,
\sum\limits_{i=0}^{m-1} \, \frac{\bar c_{m-i}}{i!(m-i-1)!}\,(-\ln(-x))^{i}
\,.
\end{align}
Determining the inverse Borel transform of $B[\hat D^{\rm FOPT}](u,x)$ term by term gives
the Adler function $\hat D(s)$ series in powers of $\alpha_s(s_0)$,
\begin{equation}
\label{eq:AdlerseriesFOPTb0}
\hat D(s) \, =  \,
\, \sum\limits_{m=1}^\infty\,
a_0^m \, \sum\limits_{i=0}^{m-1} \frac{(m-1)!}{i!(m-i-1)!}\,\bar c_{m-i}\,(-\ln(-x))^{i} \,,
\end{equation}
which subsequently yields
\begin{equation}
\label{eq:deltaFOPTlargeb0}
\delta^{(0),{\rm FOPT}}_{W_i}(s_0)\, =\, 
\,\frac{1}{2\pi i} \,  \sum\limits_{m=1}^\infty \,
a_0^m \,
 \ointctrclockwise\limits_{|x|=1}\!\! \frac{{\rm d}x}{x}\,W_i(x)\, 
 \sum\limits_{i=0}^{m-1} \frac{(m-1)!}{i!(m-i-1)!}\,\bar c_{m-i}\,(-\ln(-x))^{i} \,.
\end{equation}
This is exactly the large-$\beta_0$ version of the FOPT series for the spectral function moments in Eq~(\ref{eq:deltaFOPT}). The essential point is that the manipulations we carried out do not result in the expression of Eq.~(\ref{eq:BorelCIPT}).

In full QCD, beyond the large-$\beta_0$ approximation, the corresponding manipulations to show properties (i) and (ii) are a bit more elaborate since the simple equality $1/a(-x) = 1/a_0 +\ln(-x)$ does not hold, but contains an infinite series on the right-hand side (rhs). To show that Eq.~(\ref{eq:BorelfctFOPT}) can be manipulated into a pure function of $u$ without changing the value of the Borel sum integral, we use integration by parts and the fact that $s_0$ can be chosen sufficiently large, so that $\alpha_s(-s)$ can be expanded in powers of $\alpha_s(s_0)$ in terms of an absolute convergent series.\footnote{We are aware of the possibility that the radius of convergence for $\alpha_s(m_\tau^2)$ for $s_0=m_\tau^2$ in full QCD could be slightly smaller than the experimental value, as was pointed out in Ref.~\cite{LeDiberder:1992jjr}. This possibility does, however, not invalidate our argumentation as we are free to adopt $s_0 > m_\tau^2$.
}
We start by rewriting the exponential on the rhs of Eq.~(\ref{eq:BorelfctFOPT}) in the form
\begin{equation}
\label{eq:exponential}
e^{-\frac{u}{a(-x)}+\frac{u}{a_0}} \, = \,  e^{-u\ln(-x)}\, \bigg[\,1 + \sum_{k=1}^\infty f_{k}(u,\ln(-x))\,a^k_{\rm LL}(-x)   \,\bigg]\,,
\end{equation}
where $a_{\rm LL}(-x)$ is the leading logarithmic coupling defined by $a_{\rm LL}(-x) \equiv a_0/(1 + a_0\ln(-x))$ and the functions $f_k(u,\ln(-x))$ are polynomials in $u$ of order $k$ with coefficients containing the QCD $\beta$-function coefficients beyond one-loop and powers of $\ln(-x)$. 
The expressions up order $k=4$ are shown in Appendix~\ref{app:exponential}. Using the absolute convergence of the exponential function and the expansion in $\alpha_s(s_0)$ ensures that the sum over $k$ converges as well, and that Eq.~(\ref{eq:exponential}) represents a true mathematical identity. 
Inserting Eq.~(\ref{eq:exponential}) into Eq.~(\ref{eq:BorelFOPTrewr}), we then obtain
\begin{align}
\label{eq:BorelFOPTrewr2}
\delta_{W_i,{\rm Borel}}^{(0),{\rm FOPT}}(s_0) \,= \, & \, {\rm PV} \int_0^\infty \!\! {\rm d} u \,\, 
\frac{1}{2\pi i}\,\ointctrclockwise\limits_{|x|=1} \frac{{\rm d}x}{x} \, W_i(x) \,
B[\hat D](u)\,\\\nonumber
& \times\,\bigg[\,1 + \sum_{k=1}^\infty f_k(u,\ln(-x))\,a^k_{\rm LL}(-x)   \,\bigg]\,e^{-\frac{u}{a_{\rm LL}(-x)}}\,,
\end{align}
where we point out the appearance of $a_{\rm LL}(-x)$ in the last factor $e^{-\frac{u}{a_{\rm LL}(-x)}}$.
Since the integral is properly regularized, we can swap the $u$ and $x$ integrations and use integration by parts to remove the powers of $a_{\rm LL}(-x)$ in the brackets using that the real part of the strong coupling is positive in the entire complex $s$-plane. Upon exchanging the $u$ and $x$ integrals back to the original order, the expression can thus be identically rewritten in the form of Eq.~(\ref{eq:BorelFOPTrewr}) with 
\begin{align}
\label{eq:BorelfctFOPT2}
B[\hat D^{\rm FOPT}](u,x) \, = \, &\, \bigg[\, B[\hat D](u) + \sum_{k=1}^\infty \bar B_k[\hat D](u,\ln(-x)) \,\bigg]\, e^{-u\ln(-x)}\,,\\
\bar B_k[\hat D](u,\ln(-x)) \, \equiv \, & B_k^{(k)}[\hat D](u,\ln(-x)) - 
\sum_{i=0}^{k-1} \frac{u^i}{i!}\,B_k^{(k-i)}[\hat D](0,\ln(-x))
\end{align}
where we have defined
\begin{align}
\label{eq:Bdefs}
B_k^{(0)}[\hat D](u,\ln(-x)) \, \equiv \, &  \,f_k(u,\ln(-x))\,B[\hat D](u)\,, \\ \nonumber
B_k^{(n+1)}[\hat D](u,\ln(-x)) \, \equiv \, & \int \!\! {\rm d}u\,  B_k^{(n)}[\hat D](u,\ln(-x)) \,.
\end{align}
Here, the subtractions contained in the definition of the functions $\bar B_k[\hat D](u,\ln(-x))$ systematically remove all (ambiguous) integration constants that arise for the indefinite integrals $B_k^{(n\ge 1)}[\hat D](u)$.
It is straightforward to check that the Taylor series of the  Borel function in Eq.~(\ref{eq:BorelfctFOPT2}) correctly reproduces the  Adler function series in powers of $\alpha_s(s_0)$ in Eq.~(\ref{eq:AdlerseriesFOPT}) via the Adler function Borel representation
\begin{align}
\label{eq:AdlerQCD2}
\hat D(s) \, = \, & \int_0^\infty \!\! {\rm d} u \,\, 
B[\hat D^{\rm FOPT}](u,x)\,e^{-\frac{u}{a_0}}\,,
\end{align}
and thus also the FOPT spectral function moments series in powers of $\alpha_s(s_0)$ in Eq.~(\ref{eq:deltaFOPT}). This shows the property (i).
The manipulations we have carried out to obtain Eq.~(\ref{eq:BorelfctFOPT2}) do not make any assumption concerning the form of $B[\hat D](u)$ (and are thus also valid in the presence of its renormalon cuts) and they show that the
Borel sum based on Eq.~(\ref{eq:BorelFOPTrewr}) with either using the expression in Eq.~(\ref{eq:BorelfctFOPT}) or in Eq.~(\ref{eq:BorelfctFOPT2}) for $B[\hat D^{\rm FOPT}](u,x)$ is identical. This shows the property (ii). We have thus proven that Eq.~(\ref{eq:BorelFOPT}) is the correct Borel representation for the $\tau$ hadronic spectral function moments in the FOPT approach.

Let us now come to the point why the different analytic forms of the Borel representations in Eqs.~(\ref{eq:BorelFOPT}) and (\ref{eq:BorelCIPT}) can lead to different Borel sums. As we have already pointed out, due to the contour integration
it is impossible to switch between the CIPT and FOPT expansion series through a renormalization scale change in the strong coupling. Rather, both Borel representations are related through the $x$-dependent change of variable $u=\frac{\alpha_s(-x s_0)}{\alpha_s(s_0)} \bar u$. If $\frac{\alpha_s(-x s_0)}{\alpha_s(s_0)}$ were a positive real number for all $x$, both representations would be equivalent since the Borel integral is unchanged upon a real-valued rescaling of the Borel variable. If the Borel function $B[\hat D](u)$ would be analytic in the entire positive real $u$-plane (or if one considers only its Taylor expansion) both representations would be equivalent even for complex-valued $\frac{\alpha_s(-x s_0)}{\alpha_s(s_0)}$, because  it has a positive real part and the contour deformation involved in switching between $u$ and $\bar u$ would not affect the value of the integration. However, $\frac{\alpha_s(-x s_0)}{\alpha_s(s_0)}$ is a complex-valued number with a positive real part along the contour integration {\it and} the Borel function $B[\hat D](u)$ has cuts along the positive real $u$-plane.  So the integration along the real $\bar u$ axis for the CIPT Borel representation in Eq.~(\ref{eq:BorelCIPT}) is already well-defined and unambiguous concerning the nonanalytic cuts (or poles) contained in the Borel function $B[\hat D](u)$ without imposing the PV prescription for ${\rm Im}[x]\neq 0$. This differs from the  FOPT Borel representation, where the $u$ integration requires a choice of prescription such as PV to yield a definite value for the Borel sum and where the Borel sum has an ambiguity.
This peculiar property of the CIPT Borel representation signifies that the Borel sums of  Eqs.~(\ref{eq:BorelFOPT}) and (\ref{eq:BorelCIPT}) can be different, and why the asymptotic separation exists. The analytic properties of the difference  
make the asymptotic separation behave completely different than the ambiguity that is commonly adopted for the FOPT Borel sum. These analytic properties are discussed in more detail in Secs.~\ref{sec:anatomy} and \ref{sec:asymptoticseparation}.

\section{Anatomy of the CIPT Borel Function}
\label{sec:anatomy}

In this section we discuss the anatomy of the CIPT Borel representation given in Eq.~(\ref{eq:BorelCIPT}) in comparison to the Borel representation of the FOPT series in Eq.~(\ref{eq:BorelFOPT}). The discussion sheds more light on the peculiar and unusual properties of the CIPT Borel representation. On the one hand, we discuss how the CIPT Borel sum can be calculated in the presence of these properties. On the other hand, we show that these properties suggest that the CIPT expansion is not compatible with the standard form of the OPE corrections shown in Eq.~(\ref{eq:DOPE}) and their association to IR renormalons.

To be definite, we consider generic terms in the Borel function $B[\hat D](u)$ of the reduced Euclidean Adler function, related to Eqs.~(\ref{eq:BorelD}) and (\ref{eq:invBorelD}),
of the form
\begin{equation}
\label{eq:BorelDir}
B^{\rm IR}_{\hat D,p,\gamma}(u) \, = \, \frac{1}{(p-u)^\gamma}
\end{equation}
for an IR renormalon (with $p$ being a positive integer and $\gamma$ being real) and 
\begin{equation}
\label{eq:BorelDuv}
B^{\rm UV}_{\hat D,-\tilde p,\gamma}(u) \, = \, \frac{1}{(\tilde p+u)^\gamma}
\end{equation}
for an UV renormalon  (with $\tilde p=-p$ being a positive integer and $\gamma$ being real). The Adler function's Borel function is known to be an infinite linear combination of such generic terms plus possible functions that are analytic everywhere in the $u$-plane  (see  e.g.\ Ref.~\cite{Beneke:1998ui}). The nonanalytic (or singular) structure  of an IR renormalon contribution (cut for $u>p$) is located on the positive real axis, while the nonanalytic (or singular) structure of an UV renormalon contribution  (cut for $u<-\tilde p= p$) is located on the negative real axis. 
The notation allows to formulate analytic expressions that apply both to IR and UV renormalons since we can write the generic Borel functions of Eqs.~(\ref{eq:BorelDir}) and (\ref{eq:BorelDuv}) collectively as $B^{\rm IR/UV}_{\hat D,p,\gamma}(u)=1/[{\rm sign}(p)(p-u)]^\gamma$. The nonanalytic structure of the generic IR Borel function term in Eq.~(\ref{eq:BorelDir}) is in one-to-one correspondence to an equal-sign factorially divergent behavior of the perturbation series and entails the arbitrariness in the Borel integral of Eq.~(\ref{eq:BorelFOPT}) for $u>p$, which is made well-defined through the PV prescription. The associated renormalon ambiguity of the Adler function is associated to a nonperturbative OPE term in Eq.~(\ref{eq:DOPE})  for $d=2p$. The nonanalytic structure of the generic UV Borel transform term in Eq.~(\ref{eq:BorelDuv}) is in one-to-one correspondence to a sign-alternating factorially divergent behavior of the perturbation series and does not affect the definition of the inverse Borel integral. Thus the corresponding sign-alternating factorially divergent perturbation series can be formally summed without an ambiguity by the Borel representation.

\subsection{Analytic Result for the CIPT Moment Series Coefficients}
\label{sec:radius}

In this section we provide explicit analytic expressions for the coefficients of the CIPT spectral function moment series (with the contour integrations being carried out) which to the best of our knowledge are not available in the literature. The results allow us to determine the Taylor series of CIPT moment Borel function $B[\delta_{W_i}^{(0),{\rm CIPT}}(s_0)](\bar u)$ defined in Eq.~(\ref{eq:BorelCIPTTaylor}) in powers of $\bar u$ directly from the coefficients of CIPT moment series. In Sec.~\ref{sec:form} we show that this series agrees with the Taylor expansion of the CIPT Borel function determined directly from Eq.~(\ref{eq:BorelCIPT}).
As a side result, we show that the radii of convergence of the CIPT and FOPT moment Borel functions with respect to the expansion in powers of $\alpha_s(s_0)$ differ. This difference demonstrates the different character of the CIPT and FOPT moment series at the level of the perturbative Borel functions and that the traditional view, that the equivalence of CIPT and FOPT Borel representations can be taken for granted, is not appropriate. We also introduce the $t$-variable notation~\cite{Hoang:2017suc} that allows us to account for the higher order corrections of the QCD $\beta$-function in a transparent analytic way. We use the $t$-variable notation extensively in later sections of this article.

\begin{figure} 
	\centering
	\begin{subfigure}[b]{0.37\textwidth}
		\includegraphics[width=\textwidth]{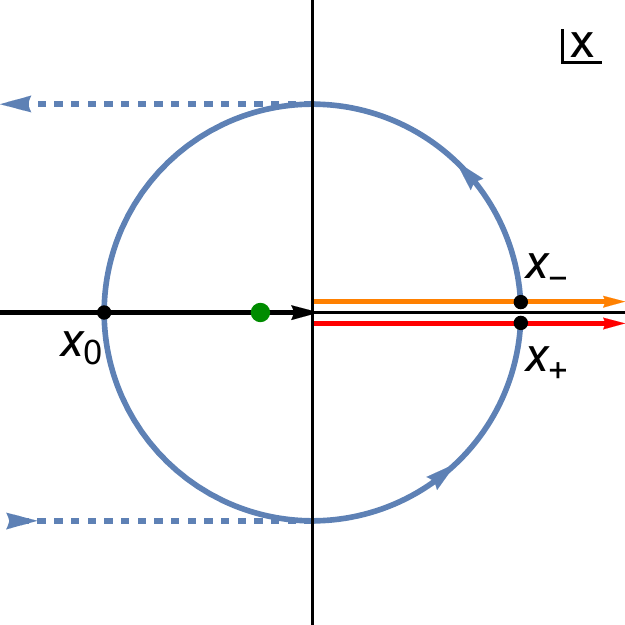}
		\caption{\label{fig:x-contour} complex $x$-plane  }
	\end{subfigure}
	 \qquad \qquad
	\begin{subfigure}[b]{0.37\textwidth}
		\includegraphics[width=\textwidth]{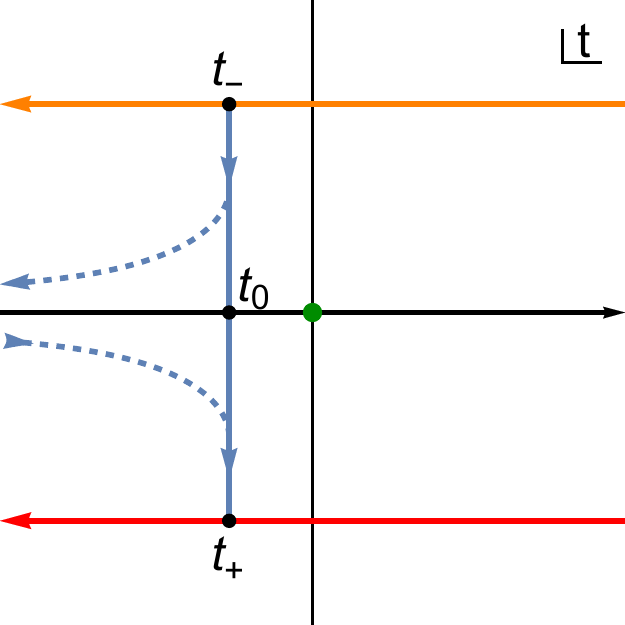}
		\caption{\label{fig:t-contour} complex $t$-plane }
	\end{subfigure}
\caption{\label{fig:contours} Panel (a): Path of the $|x|=1$ contour integration and characteristic points and lines in the complex $x$-plane. The upper and lower parts of the branch cut of the coupling $a(-x)$ are indicated by orange and red lines, respectively. Panel (b): Corresponding path and lines in the complex $t$-plane, where the arrows correspond to the directions shown in panel (a). }
\end{figure}

Let us consider the CIPT perturbation series for the spectral function moment with the monomial weight function $W(x)=(-x)^m$:
\begin{align}\label{eq:deltaCIPT3}
\delta^{(0),{\rm CIPT}}_{\{(-x)^m\}}(s_0)& \, =\,  
\sum\limits_{n=1}^\infty c_{n,1} \, J_{n,m}(s_0) \,\Big({\textstyle \frac{\alpha_s(s_0)}{\pi}}\Big)^n
\,,
\end{align}
where
\begin{align}
\label{eq:J1}
J_{n,m}(s_0) & \, = \, \frac{1}{2\pi i} \, \ointctrclockwise\limits_{|x|=1}\!\! \frac{{\rm d}x}{x}\,(-x)^m\,
\Big({\textstyle \frac{\alpha_s(-x s_0)}{\alpha_s(s_0)}}\Big)^n
\,.
\end{align}
We now change the integration variable $x$ to an integration over the $t$-variable
\begin{equation}
\label{eq:tdef}
t\,\equiv\,-\frac{2\pi}{\beta_0\alpha_s(-x s_0)} \, = \, -\frac{1}{2\, a(-x)}\,.
\end{equation}
We can then write ($t_0\equiv t(\alpha_s(\mu_0^2))$, $t_1\equiv t(\alpha_s(\mu_1^2))$)
\begin{equation}
\ln\Big(\frac{\mu_1^2}{\mu_0^2}\Big)
\, = \, 2 \int_{\alpha_s(\mu_0^2)}^{\alpha_s(\mu_1^2)}\,\frac{{\rm d}\alpha}{\beta(\alpha)}
\, = \, - 2 \int_{t_0}^{t_1}\,{\rm d}t\, \hat b(t) \, = \,2\,[ G(t_0)-G(t_1)]\,,
\end{equation}
where $\beta(\alpha_s)$ is the QCD $\beta$-function
and
\begin{align}
\hat b(t) & \equiv 1 +\sum_{k=1}^\infty \frac{\hat b_k}{t^k}
\end{align}
is a series in inverse powers of $t$ arising from the inverse of the QCD $\beta$-function.
The coefficients $\hat b_k$ are functions of the $\beta$-function coefficients and the function $G(t)$ is defined as the indefinite integral of $\hat b(t)$ of the form $G(t)=t+\hat b_1\ln(-t)-\sum_{k=2}^\infty \frac{\hat b_k}{(k-1)t^{k-1}}$. We refer to the appendix of Ref.~\cite{Hoang:2017suc} for the explicit analytic expressions for the coefficients $\hat b_k$. 
Using the $t$-variable notation, it is straightforward to derive explicit analytic expressions concerning the contour integral accounting for the evolution of the strong coupling according to the exact QCD $\beta$-function. Furthermore the change of variable provides a mapping of the complex $x$-plane onto a band around the real $t$-axis, where $t\to -\infty$ corresponds to $x\to -\infty$ and $t\to +\infty$ corresponds to $x\to 0$. The Landau pole at $s=\Lambda_{\rm QCD}^2$ corresponds to $t=0$. The cut along the positive real $x$ axis is split-mapped onto lines roughly parallel to the real $t$-axis, where the distance is related to the imaginary part of the strong coupling and $x\pm i0$ correspond to the line above/below the real axis. The mapping of these characteristics and also of the contour paths relevant for our discussions below are illustrated in Fig.~\ref{fig:contours}.

Using the definition of the scale $\Lambda_{\rm QCD}$ given by\footnote{Note that Eq.~(\ref{eq:LambdaQCD}) is related to the conventional $\overline{\rm MS}$ definition~\cite{ParticleDataGroup:2020ssz} by $\Lambda_{\rm QCD}^{\overline{\rm MS}}=2^{\hat b_1}\Lambda_{\rm QCD}$.} 
\begin{align}
\label{eq:LambdaQCD}
\Lambda_{\rm QCD} \, \equiv \, \mu \,e^{G(t(\alpha_s(\mu^2)))}\,,
\end{align}
we can rewrite Eq.~(\ref{eq:J1}) as
\begin{align}
\label{eq:J2}
J_{n,m}(s_0) & \, = \,
\frac{i}{\pi}\,\Big(\frac{\Lambda_{\rm QCD}^2}{s_0}\Big)^m 
\,\sum_{\ell=0}^\infty\,\tilde g_\ell^{(2m)}
(-t_0)^n\,\int_{t_{-}}^{t_{+}} {\rm d}t \,(-t)^{-2m\hat b_1-\ell-n} \,e^{-2mt}\,
\nonumber \\ & \,=  \,
\sum_{\ell=0}^\infty\,\tilde g_\ell^{(2m)}\,\tilde H(n,m,-2m\hat b_1-\ell-n,s_0)
\,,
\end{align}
with
\begin{align}
t_\pm \equiv & -\frac{2\pi}{\beta_0\alpha_s(-s_0\pm i0)} \, = \, -\frac{1}{2\, a(-1\pm i0)}\,,
\nonumber \\
t_0 \equiv &-\frac{2\pi}{\beta_0\alpha_s(s_0)} \, = \, -\frac{1}{2\, a(1)}\,.
\end{align}
The coefficients $\tilde g_\ell^{(2m)}$ are defined by the relation
\begin{align}
\hat b(t)\,e^{-2m G(t)}e^{2mt}(-t)^{2m\hat b_1} \, = \, \sum_{\ell=0}^\infty \tilde g_\ell^{(2m)} (-t)^{-\ell}\,.
\end{align}
We have $\tilde g_0^{(2m)}=1$, and the expressions for $\ell=1,2,\ldots$ can be obtained in a straightforward way from the functions $\hat b(t)$ and $G(t)$.
In the $\overline{\rm MS}$ scheme for the strong coupling the series in $\ell$ is infinite. However, the series terminates if a scheme is adopted, where the series for the function $\hat b(t)$ terminates, see e.g.\ Refs.~\cite{Brown:1992pk,Boito:2016pwf}.
In the large-$\beta_0$ approximation, where $\hat b(t)=1$ and $\hat b_k=0$ for all $k$, we also have $\tilde g_{\ell}^{(2m)}=\delta_{\ell 0}$.
The function $\tilde H$ can be readily evaluated and reads
\begin{align} 
\label{eq:Htilde1}
\tilde H(1,0,\eta,s_0)  & \,=\,  -\,\frac{i}{\pi}\,(-t_0)^{n}\,\ln\Big(\frac{t_+}{t_-}\Big)\,,\\
\label{eq:Htilde2}
\tilde H(n\ge 2,0,\eta,s_0)  & \,=\,  -\,\frac{i}{\pi}\,\frac{(-t_0)^{1+\eta+n}}{1+\eta} \bigg[\,\Big(\frac{t_+}{t_0}\Big)^{1+\eta}- \Big(\frac{t_-}{t_0}\Big)^{1+\eta}\,\bigg]\,,\\
\label{eq:Htilde3}
\tilde H(n,m\ge 1,\eta,s_0)  & \,=\,  \,\frac{i}{\pi}\,\Big(\frac{\Lambda_{\rm QCD}^2}{s_0}\Big)^m \,(2m)^{-1-\eta}\,
(-t_0)^n\,\bigg[\,
e^{i\pi(1+\eta)}\Gamma(1+\eta,2 m t_+) \nonumber\\ & \hspace{3cm}
-e^{-i\pi(1+\eta)} \Gamma(1+\eta,2 m t_-) 
-\frac{2\pi i}{\Gamma(-\eta)}
\,\bigg]\,.
\end{align} 
In the large-$\beta_0$ approximation ($\hat b_1=0$, $\ell=0$) the function $\tilde H$ can be concisely written in the form
\begin{align} 
\label{eq:Htildeb01}
\tilde H(n,0,-n,s_0)  & \,=\, {}_2{\rm F}_1\Big(\frac{n}{2}, \frac{n + 1}{2}, \frac{3}{2}, -a_0^2 \pi^2\Big)\,,\\
\label{eq:Htildeb02}
\tilde H(n,m\ge 1,-n,s_0)  & \,=\,  \,\frac{i}{2\pi}\, (-m)^{n-1}\,
(-2 t_0)^n\,e^{2 m t_0}\,\bigg[\, \Gamma(1-n,m (2 t_0 - i \pi))\nonumber\\ & \hspace{3cm} 
- \Gamma(1-n,m (2 t_0 + i \pi)) 
+(-1)^n\frac{2\pi i}{\Gamma(n)}
\,\bigg]\,.
\end{align} 

Together with Eq.~(\ref{eq:J2}) the expressions for the function $\tilde H$ provide explicit analytic results for the CIPT moment series coefficients, and we can now write down the $\bar u$ Taylor series of the CIPT moment series Borel function related to the Adler function contributions arising from $B^{\rm IR/UV}_{\hat D,p,\gamma}(u)$ given in Eqs.~(\ref{eq:BorelDir}) and (\ref{eq:BorelDuv}) for the monomial weight function $W(x)=(-x)^m$. Referring to the CIPT moment series as $\delta_{\{(-x)^m,p,\gamma\}}^{(0),{\rm CIPT}}(s_0)$, the Borel representation has the form
\begin{align}
\label{eq:BorelCIPT3}
& \delta_{\{(-x)^m,p,\gamma\},{\rm Borel}}^{(0),{\rm CIPT}}(s_0) \,= \, 
\int_0^\infty \!\! {\rm d} \bar u \,
B\Big[\delta_{\{(-x)^m,p,\gamma\}}^{(0),{\rm CIPT}}(s_0)\Big](\bar u) \,  e^{-\frac{\bar u}{a_0}} \,,
\end{align}
with
\begin{align}
\label{eq:BorelCIPT5}
B\Big[\delta_{\{(-x)^m,p,\gamma\}}^{(0),{\rm CIPT}}(s_0)\Big](\bar u) & = 
\sum\limits_{n=0}^\infty \frac{\Gamma(1-\gamma)}{\Gamma(1-\gamma-n)\Gamma(n+1)} \,
J_{n+1,m}(s_0)\, (\text{sign}(p)p)^{-\gamma} \,\Big(\frac{- \bar u}{p}\Big)^n\,,
\end{align}
This series agrees with the Taylor expansion of the CIPT Borel function determined directly from Eq.~(\ref{eq:BorelCIPT}), as we show in Sec.~\ref{sec:form}.
	
The corresponding FOPT Borel representation related to Eq.~(\ref{eq:BorelFOPT}) can be easily written down and reads
\begin{eqnarray}
\label{eq:BorelFOPT2}
\delta_{\{(-x)^m,p,\gamma\},{\rm Borel}}^{(0),{\rm FOPT}}(s_0) & =& 
 {\rm PV}\,
\int_0^\infty \!\! {\rm d}u \,  e^{-\frac{u}{a_0}}\,
B\Big[\delta_{\{(-x)^m,p,\gamma\}}^{(0),{\rm FOPT}}(s_0)\Big]( u)\nonumber \\
 & =&  {\rm PV}\, \int_0^\infty \!\! {\rm d}u \, 
B^{\rm IR/UV}_{\hat D,p,\gamma}(u)\,\, 
\frac{1}{2\pi i}\,\ointctrclockwise\limits_{|x|=1} \frac{{\rm d}x}{x} \,(-x)^m \,
e^{-\frac{u}{a(-x)}}\,.
\end{eqnarray} 
It is now straightforward to compare the radii of convergence for the Taylor series in $\bar u$ or $u$ of the Borel functions 
$B[\delta_{\{(-x)^m,p,\gamma\}}^{(0),{\rm CIPT}}(s_0)](\bar u)$ and $B[\delta_{\{(-x)^m,p,\gamma\}}^{(0),{\rm FOPT}}(s_0)]( u)$, respectively. 
For the FOPT moment series Borel function $B[\delta_{\{(-x)^m,p,\gamma\}}^{(0),{\rm FOPT}}(s_0)](u)$ we see that 
the contour integral modifies the norm of the cut contained in the generic Borel function $B^{\rm IR/UV}_{\hat D,p,\gamma}(u)$, but it does in general not affect the distance of the cut to the origin of the Borel plane.\footnote{In the large-$\beta_0$ approximation $\gamma$ can only adopt integer values so that the singular structures of the Adler function Borel transform only involve poles. This allows for the possibility of an elimination of a simple pole, which, however, cannot happen in the same way for the cuts that appear beyond the large-$\beta_0$ approximation.} 
Thus the radius of convergence of the $u$ Taylor expansion of $B[\delta_{\{(-x)^m,p,\gamma\}}^{(0),{\rm FOPT}}(s_0)]( u)$ agrees with the convergence radius of $B^{\rm IR/UV}_{\hat D,p,\gamma}(u)$ which is just $|p|$. Let us now have a look at the CIPT Borel function in Eq.~(\ref{eq:BorelCIPT5}).
Using the leading asymptotics for the incomplete $\Gamma$-function when its first argument adopts large negative values
$\Gamma(\alpha\to-\infty,z)\simeq |z|^\alpha e^{-x} e^{i\alpha \arg(z)}/\alpha$ and the fact that
the large-$n$ behavior of $J_{n,m}(s_0)$ is dominated by the $\ell=0$ term, one can see that 
$|J_{n,m}(s_0)|^{1/n}\simeq |\alpha_s(-s_0\pm i0)/\alpha_s(s_0)|$ for $n\to\infty$. 
Applying the root criterion for the $\bar u$ series in Eq.~(\ref{eq:BorelCIPT5})
we can see that the convergence radius is $|p\,\alpha_s(s_0)/\alpha_s(-s_0\pm i0)|=|p\,a_0/a(-1\pm i0)|>|p|$. In the large-$\beta_0$ approximation (or at the leading logarithmic approximation for $\alpha_s$) we have $\frac{1}{a(x)}=\frac{1}{a_0}+\ln(x)$, and the convergence radius reads $p|1\pm i a_0\pi|$.
The results show that the Borel function of the FOPT series can be obtained by summing its (convergent) series for $u<|p|$, and relies on an analytic continuation for $u>p$. In contrast, the Borel function of the CIPT series can be obtained from summing its Taylor series further out into the complex Borel plane.
This underlines the different character of the CIPT moments' Borel function. However, we note that the different convergence radii by themselves do not yet imply that the FOPT and CIPT Borel sums differ as well, since that depends on the existence of IR renormalons.

\subsection{Path of the Contour Integral in the Invariant Mass Plane}
\label{sec:path}

The invariant mass contour integration involved in the computation of the perturbative QCD corrections to the spectral function moments $A_{W_i}(s_0)$ conventionally involves a circular path in the complex $x$-plane with radius $|x|=1$ which begins/ends at the points located at $x_\mp=1\pm i 0$, see Fig.~\ref{fig:x-contour} for a graphical illustration.
The path applies both to the coefficients of the perturbation series for the CIPT and the FOPT approach, see Eqs.~(\ref{eq:deltaCIPT}) and (\ref{eq:deltaFOPT}), respectively, as well as for the Borel representation of the FOPT series given in Eq.~(\ref{eq:BorelFOPT}). There is the possibility to deform this path without changing the result as long as the path encloses the Landau pole of the strong coupling (illustrated by the green dot in Figs.~\ref{fig:contours}), does not cross the analyticity cuts of the Adler function and the strong coupling along the positive real $x$-axis, and stays within the perturbative regime. Which of such paths one actually picks is therefore a matter of practical choice. 

However, for the Borel representation of the CIPT series in Eq.~(\ref{eq:BorelCIPT}) additional restrictions arise on the contour of the $x$-integration for IR renormalons since the nonanalytic structures in the Adler function's Borel function affect the analytic properties of the integrand in the complex $x$-plane. Let us consider the CIPT Borel representation for the generic nonanalytic term in the Borel transform of the reduced Euclidean Adler function related to an IR or a UV renormalon (see Eqs.~(\ref{eq:BorelDir}) and (\ref{eq:BorelDuv})):
\begin{align}
\label{eq:BorelCIPT2}
\delta_{\{W_i,p,\gamma\},{\rm Borel}}^{(0),{\rm CIPT}}(s_0)  = &  
\int_0^\infty \!\! {\rm d} \bar u \,\,  
\frac{1}{2\pi i}\,\ointctrclockwise\limits_{{\cal C}_x} \frac{{\rm d}x}{x} \, W_i(x) \,
\Big({\textstyle \frac{a(-x)}{a_0}}\Big)\,
\frac{1}{\big[{\rm sign}(p)\big(p - \frac{a(-x)}{a_0} \bar u\big)\big]^\gamma}
\,e^{-\frac{\bar u}{a_0}}
\end{align}
For the case of a UV renormalon  ($p<0$) the pattern, where the nonanalytic structures appear in the complex $x$-plane, are the same as for the FOPT Borel representation. This is because the real part of the strong coupling is always positive as long as its scale remains in the perturbative regime, and thus the circular path with $|x|=1$ can also be adopted for a UV renormalon.
However, for an IR renormalon ($p>0$) we can see that, apart from the Landau pole and the cut along the positive real $x$ axis, which arise from the strong coupling function, there is an additional cut along the negative real $x$-axis for ${\rm Re}[\alpha_s(-x s_0)] > p\alpha_s(s_0)/\bar u$. As long as $\bar u< p$ this cut is still within the conventional circular path with radius $1$, but for $\bar u> p$ the path ${\cal C}_x$ must be deformed further away from the origin into the negative real complex plane to not cross the cut.  In the large-$\beta_0$ approximation the corresponding cuts reduce to poles located at $\tilde x(\bar u)=- e^{\frac{2(p-\bar u)t_0}{p}} = -(\Lambda_{\rm QCD}^2/s_0) e^{\frac{\bar u}{p a_0}}$, and the path must cross the negative real axis for $x<\tilde x(\bar u)$. Interestingly, for $\bar u\to\infty$ the allowed region where the path can cross the negative real $x$-axis is shifted toward negative infinity.  For the computation of the CIPT Borel sum of Eq.~(\ref{eq:BorelCIPT2}) this unusual property means that for IR renormalons the path of the contour integration must be deformed to minus negative real infinity for $\bar u\to\infty$. This will be an essential element for the explicit evaluation of the CIPT Borel sum and the asymptotic separation for IR renormalons that is discussed in Sec.~\ref{sec:asymptoticseparation}. From a physical perspective, the necessity of the contour deformation away from $|x|=1$ for the CIPT Borel representation is, again, quite peculiar and suggests an unphysical behavior given that the physical Adler function is analytic everywhere along the negative real $s$ axis.

\subsection{Form of the Borel Function}
\label{sec:form}

Let us now examine the full analytic expressions for the FOPT and CIPT spectral moment Borel representations of Eqs.~(\ref{eq:BorelFOPT}) and (\ref{eq:BorelCIPT}), respectively, arising from the generic IR and UV Borel function terms in Eqs.~(\ref{eq:BorelDir}) and (\ref{eq:BorelDuv}). Note that the implications of the form of the FOPT Borel representation are already know from previous literature (see e.g.~\cite{Ball:1995ni,Beneke:1998ui}). We review them for the purpose of comparison to the  CIPT Borel representation.
We again consider the generic monomial weight function $W(x)=(-x)^m$. 

For the FOPT approach the resulting generic expression is given in Eq.~(\ref{eq:BorelFOPT2}). Changing to the $t$-variable for the contour integration the result can be rewritten in the form
\begin{align}
\label{eq:BorelFOPT3}
&\delta_{\{(-x)^m,p,\gamma\},{\rm Borel}}^{(0),{\rm FOPT}}(s_0) \nonumber\\
  &\qquad  = \frac{i}{\pi}\, {\rm PV}\,\int_0^\infty \!\! {\rm d}u \, 
B^{\rm IR/UV}_{\hat D,p,\gamma}(u) 
\,\Big(\frac{\Lambda_{\rm QCD}^2}{s_0}\Big)^m 
\,\sum_{\ell=0}^\infty\,\tilde g_\ell^{(2m)}\int_{t_{-}}^{t_{+}} {\rm d}t \,(-t)^{-2m\hat b_1-\ell} \,e^{2(u-m)t}\, 
\end{align}
which gives
\begin{align}
\label{eq:BorelFOPT4}
B\Big[\delta_{\{(-x)^m,p,\gamma\}}^{(0),{\rm FOPT}}(s_0)\Big]( u) = 
B^{\rm IR/UV}_{\hat D,p,\gamma}(u)\,\sum_{\ell=0}^\infty\,\tilde g_\ell^{(2m)}\,\tilde F(m,-2m\hat b_1-\ell,s_0;u)\,.
\end{align}
In the complex $t$-plane, see Fig.~\ref{fig:t-contour}, the circular path of the $x$-contour integration corresponds to an essentially straight line connecting the points $t_-$ and $t_+$ (solid blue line) which are located in the negative real complex half plane on opposite sides of the real axis.
The function $\tilde F$ can then be readily evaluated giving
\begin{align} 
\label{eq:Ftilde}
\tilde F(m,\eta,& s_0;u)  \,=\,  \,\frac{i}{\pi}\,\Big(\frac{\Lambda_{\rm QCD}^2}{s_0}\Big)^m \,2^{-1-\eta}\,
e^{-2 u t_0}\,
\Big[
(u-m-i 0)^{-1-\eta} \, \Gamma(1+\eta,2(m-u)t_+)
\nonumber \\ &
-(u-m+i 0)^{-1-\eta} \,\Gamma(1+\eta,2(m-u)t_-) 
-\Theta(m-u)(m-u)^{-1-\eta}\,\frac{2\pi i}{\Gamma(-\eta)}
\Big]\,,
\end{align}
for real-valued $u$,
where the term involving the Heaviside step function $\Theta$ arises due to the cut of the incomplete $\Gamma$-function along the negative real axis in its second argument.
To the best of our knowledge the analytic result of Eq.~(\ref{eq:BorelFOPT4}) has not been given in the literature before. 
As can be seen from the leading asymptotics for the incomplete $\Gamma$-function when its second argument becomes large $\Gamma(\alpha,z\to\infty)\simeq z^{\alpha-1}e^{-z}$, we have $\tilde F\sim e^{2u(t_\pm-t_0)}/u$ for large values of $u$. Because ${\rm Re}[t_\pm-t_0]<0$, this provides an exponential suppression for large positive $u$. Furthermore the complex second arguments of the incomplete $\Gamma$-functions cause an additional oscillatory dependence with zeros at noninteger values for $u$. Thus the function $\tilde F$ modulates (and partially suppresses) the singular and nonanalytic structures contained in $B^{\rm IR}_{\hat D,p,\gamma}(u)$ on the real $u$ axis for $u>p$, but it does not eliminate them in general. This also visualizes the statement we have made for the convergence radius of the $u$ Taylor series of the FOPT Borel function discussed in Sec.~\ref{sec:radius}. To resolve the associated arbitrariness of the Borel integral of Eq.~(\ref{eq:BorelFOPT3}) in the case of IR renormalons for $u>p$ and to obtain a well-defined result, the PV prescription is therefore still needed in general.

In the large-$\beta_0$ approximation the exponential suppression does not arise because ${\rm Re}[t_\pm-t_0]=0$ and the $\Gamma$-functions acquire zeros at integer values for $u$. Here, only the $\ell=0$ term contributes and we have~\cite{Ball:1995ni}
\begin{align}
\label{eq:Ftildebeta0}
\tilde F(m,0,& s_0;u)  \,=\,  (-1)^m \,\frac{\sin(u\pi)}{\pi(u-m)}\,,
\end{align}
so $\tilde F$ has zeros at integer values for $u$, except for $u=m$.
Since in the  large-$\beta_0$ approximation only single or double poles arise in the Borel transform of the reduced Adler function, 
the function $\tilde F$ completely eliminates the single poles (and their associated renormalon ambiguity) and reduces the double poles to single poles if $p\neq m$. 
In the large-$\beta_0$ approximation the Adler function's Borel function, see Eq.~(\ref{eq:AdlerBorelb0}), contains a single (and no double) pole at $u=p=2$ which corresponds to the $d=4$ gluon condensate OPE correction. This $p=2$ IR renormalon has a sizeable impact on the behavior of the Adler function series already at intermediate orders and the corresponding gluon condensate OPE term represents the parametrically dominant OPE correction. Because the polynomial weight function for the $\tau$ hadronic decay rate $R_\tau$, $W_\tau(x) =(1-x)^3(1+x)=1 - 2 x + 2 x^3 - x^4$, does not contain a quadratic term with $m=2$, the effects of the $p=2$ IR renormalon are eliminated completely  in $\delta_{w_\tau}^{(0),{\rm FOPT}}(s_0)$. This renders the FOPT series related to the gluon condensate renormalon pole having a finite radius of convergence (i.e.\ being convergent for sufficiently small $\alpha_s(s_0)$). This is demonstrated explicitly in the numerical study of a generic simple pole $p=2$ IR renormalon series in Sec.~\ref{sec:large-beta-generic}. At the same time, the gluon condensate OPE correction (which only has a tree-level Wilson coefficient in the large-$\beta_0$ approximation) vanishes identically in the contour integration due to the residue theorem, see Eqs.~(\ref{eq:deltadef}) and (\ref{eq:DOPE}). In contrast, as we show below, the corresponding CIPT series is divergent and does not have a finite radius of convergence.

The complete removal of the $p=2$ renormalon divergence for the FOPT series is only possible within the large-$\beta_0$ approximation. In full QCD, when the higher loop corrections to the gluon condensate Wilson coefficient are accounted for, the effects of the $p=2$ IR renormalon are strongly suppressed but not completely eliminated~\cite{Beneke:1998ui}. 
As is also well known, the modulation/suppression of the IR renormalon structure in the FOPT series caused by the contour integration (and visualized in the form of the function $\tilde F$) is in one-to-one correspondence to analogous modulations/suppressions of the standard OPE corrections. This is because IR renormalon contributions in the FOPT series at high orders develop the same dependence on inverse powers of $s_0$ (and even logarithms of $s_0$) as the corresponding terms of the OPE corrections.  
Thus a suppression (or elimination) of an IR renormalon term in $\delta^{(0),{\rm FOPT}}(s_0)$ is accompanied by a corresponding suppression (or elimination) of the associated OPE correction term. 
So in the large-$\beta_0$ approximation for the FOPT series, the elimination of the $p=2$ IR renormalon in turn implies the absence of the gluon condensate OPE correction. This is in accordance with the standard analytic form of the OPE corrections to the Adler function shown in Eq.~(\ref{eq:DOPE}). 

For comparison, let us now have a close look at the spectral moment Borel representation in the CIPT approach for the generic IR  renormalon contribution ($p>0$) given in Eq.~(\ref{eq:BorelCIPT2}) and the monomial weight function $W(x)=(-x)^m$. 
As we will see, the CIPT Borel representation has again very unusual properties, that differ substantially from those of the FOPT Borel representation. Switching to the contour integration variable $t$, the CIPT moment Borel function (see Eq.~(\ref{eq:BorelCIPT3})) can be rewritten in the form
\begin{align}
\label{eq:BorelCIPT4}
B\Big[\delta_{\{(-x)^m,p,\gamma\}}^{(0),{\rm CIPT}}(s_0)\Big](\bar u)\, =\, &
-\frac{i\, |p|^{-\gamma}\,t_0 }{\pi}
\,\Big(\frac{\Lambda_{\rm QCD}^2}{s_0}\Big)^m \\ &
\times
\,\sum_{\ell=0}^\infty\,\tilde g_\ell^{(2m)}\int\limits_{{\cal C}_t} {\rm d}t \,(-t)^{-2m\hat b_1+\gamma-\ell-1} 
\,\frac{ e^{-2 m t}}{(\frac{t_0\bar u}{p}-t)^\gamma}\,.
\nonumber
\end{align}
We are not aware of a closed analytic solution of the generic integral, but we can still discuss its analytic properties.
The integration path ${\cal C}_t$, see Fig.~\ref{fig:t-contour}, starts and ends at $t_-$ and $t_+$, respectively, and furthermore crosses the real axis at $t<t_0\bar u/p$. Because $\mbox{Re}(t_0)<0$, this means that for $\bar u>p$ the path can in general not connect the points $t_\mp$ in a straight line and must be deformed further into the negative real complex 
half plane (dashed blue lines) -- as we have already mentioned in Sec.~\ref{sec:path} considering the path  ${\cal C}_x$  in the complex $x$-plane.
The expression shown in Eq.~(\ref{eq:BorelCIPT4}) also applies for a generic UV renormalon (where $p<0$). For a UV renormalon the path ${\cal C}_t$ crosses the real axis at $t<t_0\bar u/p$ too, but one can adopt 
a straight line between $t_-$ and $t_+$ because $\mbox{Re}(t_\pm)<0$ and $t_0\bar u/p>0$. 

Even though we can evaluate the ${\cal C}_t$ contour integrations only numerically, one can see that, due to the cut (or the pole for integer $\gamma$), the integral picks up a contribution $\simeq (-\frac{t_0 \bar u}{p})^{-2m\hat b_1+\gamma-\ell-1}\times e^{-2(\frac{m}{p})t_0\bar u}$. Interestingly, because $t_0$ is negative, we see that for IR renormalons and $m\ge p>0$ the Borel integral of Eq.~(\ref{eq:BorelCIPT3}), which involves the additional factor of $e^{-\bar u/a_0}=e^{2 t_0 \bar u}$,  does not have anymore the exponential suppression that is expected within the canonical renormalon calculus. The origin of the modified behavior is the monomial weight factor $W(x)=(-x)^m$ which causes an enhancement when the contour of the invariant mass integration is deformed further into the negative real complex half plane. Written in terms of the $t$ variable, this corresponds to the enhancement factor $e^{-2m t}$. 

Another interesting observation is that the expression in Eq.~(\ref{eq:BorelCIPT4}) involves nonanalytic cuts in the complex $\bar u$ plane that are generated at the start and endpoints $t_\mp$ of the integration path ${\cal C}_t$. These cuts arise for $t_0\bar u/p-t$ 
being negative and real, which corresponds to lines $\bar u = p \,\alpha_s(s_0)/\alpha(-s_0\pm i\pi)+z$, with $z>0$. They are responsible for the convergence radius determined in Sec.~\ref{sec:radius}. The fact that they are not located along the positive real axis is consistent with the observation we made already at the end of Sec.~\ref{sec:Borelrepresentation}, that the $\bar u$ Borel integration along the positive real axis can just be carried out without the need for the PV prescription. What is very peculiar as well is the fact that these cut arise in a completely unsuppressed way even for $p\neq m$, where an $u=p$ IR renormalon is strongly suppressed (or even eliminated) in the FOPT Borel representation. 

For the large-$\beta_0$ approximation ($\hat b_1=0$, $\tilde g_\ell^{(2m)}=\delta_{\ell 0}$, $\gamma=1,2$) these properties can be seen explicitly,  since the contour integration can be carried out analytically:
\begin{align}
\label{eq:BorelCIPTbeta0}
B\Big[\delta_{\{(-x)^m,p,\gamma\}}^{(0),{\rm CIPT},\beta_0}(s_0)\Big](\bar u)\,& =\, 
-\frac{i\, |p|^{-\gamma}\,t_0 }{\pi}
\,\Big(\frac{\Lambda_{\rm QCD}^2}{s_0}\Big)^m 
\int\limits_{{\cal C}_t} {\rm d}t \,(-t)^{\gamma-1} 
\,\frac{ e^{-2 m t}}{(\frac{t_0\bar u}{p}-t)^\gamma}\nonumber\\ & =
\,\tilde C(p,\gamma,m,s_0;\bar u)
\,,
\end{align}
where the expressions for the $\tilde C$-functions for single and double IR renormalon poles read
\begin{align}
\label{eq:Cfunc1}
\tilde C(p,1,m,s_0;\bar u) & = \frac{2\,t_0}{|p|}\, Q\Big(1,m,-2t_0(1-{\textstyle\frac{\bar u}{p}})\Big)  
\\ 
\label{eq:Cfunc2}
\tilde C(p,2,m,s_0;\bar u) & = \frac{2\,t_0}{p^2}\,\Big[ 
Q\Big(1,m,-2t_0(1-{\textstyle\frac{\bar u}{p}})\Big)  
-\frac{2 \bar{u} t_0}{p}\, Q\Big(2,m,-2t_0(1-{\textstyle\frac{\bar u}{p}})\Big)  
\Big]\,,
\end{align}
with ($n=1,2,3,\ldots$)
\begin{align}
\label{eq:Qfunc}
Q(1,0,\rho) & = \frac{i}{2\pi}\,\Big[ \ln(\rho+i\pi) -  \ln(\rho-i\pi)  \Big]\,,
\\ 
Q(n \ge 2,0,\rho) & = -\frac{i}{2\pi(n-1)}\,\Big[ (\rho+i\pi)^{1-n} - (\rho-i\pi)^{1-n}  \Big]\,,
\\
Q(n,m,\rho) & = m^{n-1}\,e^{-m\rho}\,\Big[\frac{(-1)^n\,i}{2\pi}\Big(
\Gamma(1-n,-m(\rho+i\pi)) \\
& \qquad \qquad \qquad -\Gamma(1-n,-m(\rho-i\pi))
\Big) -\frac{1}{\Gamma(n)} \Big]\,.  \nonumber
\end{align}
It is a straightforward exercise to check that for $\bar u<|p a_0/a(-1\pm i 0)|$ the $\bar u$ Taylor series of the CIPT Borel function in Eq.~(\ref{eq:BorelCIPT5}) correctly converges to the expression in Eq.~(\ref{eq:BorelCIPTbeta0}). In full QCD this is true as well, but the calculation is tedious. 
We clearly see that the cuts (and also poles) along the lines $\bar u = p \,\alpha_s(s_0)/\alpha(-s_0\pm i\pi)+z$, with $z>0$, arise from the analytic properties of the functions $Q$ irrespective of the values for $p$, $m$ and $\gamma$.

The fact that these cuts and poles arise for $p\neq m$, even when the Adler function's Borel function only has a single pole ($\gamma=1$), indicates that the corresponding CIPT moment series is asymptotic and does not have a finite radius of convergence. 
This is in contrast to the FOPT expansion series which has a finite radius of convergence in this case. Since the corresponding OPE correction based on the standard form of Eq.(\ref{eq:DOPE}) vanishes for this weight function, there is no OPE term that can ever compensate the divergent asymptotic character of the CIPT moment series. In full QCD, these OPE corrections do not vanish exactly (due to the QCD corrections to the Wilson coefficients), but they are still strongly suppressed, while the divergent asymptotic behavior of the CIPT moment series is contributing at full strength.  
This fact implies that the OPE corrections that have to be added to CIPT expansion method cannot be computed from the standard Adler function OPE corrections of Eq.(\ref{eq:DOPE}). In other words,
the CIPT expansion method is not compatible with the standard analytic form of the OPE corrections to the Adler function shown in Eq.~(\ref{eq:DOPE}) and their association with IR renormalons.

\subsection{Intermediate  Comments}
\label{sec:summary}

Before continuing, let us briefly summarize the findings we have made in this section and make some comments. We have shown that the $\bar u$ Taylor series for the Borel function of the CIPT moment series, when determined explicitly from the coefficient of the CIPT series terms, agrees with the $\bar u$ Taylor series calculated from the CIPT Borel representation of Eq.~(\ref{eq:BorelCIPT}). 
This agreement together with the proof given Sec.~\ref{sec:Borelrepresentation} show that the CIPT Borel representation of  Eq.~(\ref{eq:BorelCIPT}) cannot be simply dismissed and that its peculiar properties are a reflection of the properties of the CIPT expansion method itself. In the context of the canonical renormalon calculus, summarized e.g.\ in the standard reference~\cite{Beneke:1998ui}, these properties are quite unusual and arise in the presence of IR renormalons contained in the underlying Adler function: 
\begin{enumerate}
\item For moments where the FOPT expansion leads to a suppression of IR renormalons contained in the underlying Adler function, the
CIPT expansion does not exhibit an analogous suppression. This implies that the OPE corrections that need to be accounted for in the CIPT expansion cannot be parametrized using the standard OPE form for the Adler function given in Eq.~(\ref{eq:DOPE}).
\item The Borel function of the CIPT series has IR renormalon cuts (or poles) located away from the positive real Borel axis such that the Borel sum computed from the Borel integral along the positive real Borel axis provides an unambiguous value even though the CIPT series itself is asymptotic. The ambiguity of the CIPT series, which without doubt exists, can therefore not be computed from the procedures used in the canonical renormalon calculus based on infinitesimal path deformations away from the real axis.
\item In the computation of the full CIPT series Borel function (i.e.\ beyond its Taylor expansion) it is mandatory to deform the contour integral away from $|x|=1$ into the negative complex $x$-plane when the Borel variable $\bar u$ increases. The need for this deformation  suggests an unphysical behavior given that the physical Adler function is analytic everywhere along the negative real $x$ axis.   
\item For high power polynomial terms $(-x)^m$ in the weight function $W(x)$ the contour deformation entails that the CIPT moment series Borel function is exponentially enhanced for large $\bar u$, such that the inverse Borel integral along the real $\bar u$ axis may not converge.  
\end{enumerate} 
Property~1 is demonstrated explicitly in Sec.~\ref{sec:large-beta-generic} in the large-$\beta_0$ approximation, where the FOPT expansion leads to the elimination of the gluon condensate renormalon, while the CIPT expansion remains asymptotic.
Property~1 implies that the CIPT expansion method does not follow the canonical rules of the renormalon calculus, and it does also not follow the canonical association of IR renormalons with higher-dimensional OPE corrections. 
Concerning property~2, in this work we will not attempt to define a procedure how to properly define the ambiguity of the CIPT Borel sum, but hope to come back to this issue in an upcoming work. See also our comment prior to Eq.~(\ref{eq:IRBorelIntCIPT}). Properties~3 and 4 are relevant for the computation of the CIPT Borel sum and the asymptotic separation, which we discuss in the next section. Property~3 implies that the $x$ contour integration has to be deformed to minus real infinity when the Borel integral over $\bar u$ is carried out first. Property~4 implies that for high power polynomial terms in the weight function $W(x)$, the determination of the CIPT Borel sum involves an analytic continuation.

\section{The Borel Sum and the Asymptotic Separation}
\label{sec:asymptoticseparation}

In this section we focus on the computation of the Borel sums for the perturbative spectral function moments in FOPT and CIPT, associated to Eqs.~(\ref{eq:BorelFOPT}) and (\ref{eq:BorelCIPT}), respectively, and we present analytic formulas that allow to determine the
asymptotic separation for any Borel model (and even the exact Borel function of the Adler function, if it ever becomes known.)
To determine the FOPT and CIPT Borel sums we now take the approach to carry out the Borel integration prior to the contour integration.\footnote{The results for the CIPT Borel sums obtained by carrying out the Borel integration after the contour integration, i.e.\ by integrating over the Borel functions discussed in Sec.~\ref{sec:form}, lead to equivalent results. In this approach, the discussion concerning the analytic continuation for the case $m>p$ in the CIPT case is different.} 
We again consider the generic terms in the Borel transform of the reduced Adler function shown in Eqs.~(\ref{eq:BorelDir}) and (\ref{eq:BorelDuv}) for an IR and a UV renormalon, respectively. For the remaining contour integrations we provide analytic expressions, but the results can be readily obtained also by numerical evaluation.

\subsection{Borel Space Integrals}
\label{sec:borelspace}

We start considering the Borel space integral for the case of a UV renormalon. It is straightforward to show that the results for the CIPT and FOPT 
Borel representations give the same result, yielding the expression ($\tilde p>0$)
\begin{align}
\label{eq:UVBorelInt}
\int_0^\infty \!\! {\rm d} \bar u \, 
\Big({\textstyle\frac{a(-x)}{a_0}}\Big)\,
 \frac{e^{-\frac{\bar u}{a_0}}}{\big(\tilde p + \frac{a(-x)}{a_0} \bar u\big)^\gamma}
 \, = \,
\int_0^\infty \!\! {\rm d} u \, 
\frac{e^{-\frac{u}{a(-x)}}}{(\tilde p + u)^\gamma}
\, = \,
(a(-x))^{1-\gamma}\,e^{\frac{\tilde p}{a(-x)}}\,\Gamma\Big(1-\gamma,{\textstyle\frac{\tilde p}{a(-x)}}\Big)\,.
\end{align}
The result has a cut along the negative real $a(-x)$-axis which is outside the perturbative regime and a cut along the positive real $x$-axis from the strong coupling. The remaining contour integration can therefore be carried out along the path with $|x|=1$.

\begin{figure}[b] 
	 \centering
	\begin{subfigure}[b]{0.46\textwidth}
		\includegraphics[width=\textwidth]{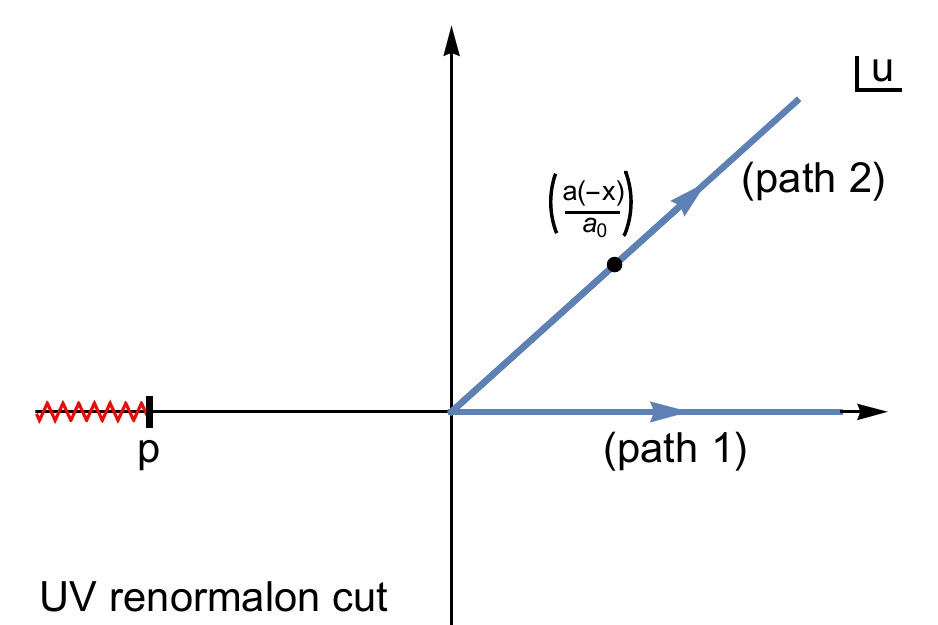}
		\caption{\label{fig:u-contourUV}}
	\end{subfigure}
	 \qquad 
	\begin{subfigure}[b]{0.46\textwidth}
		\includegraphics[width=\textwidth]{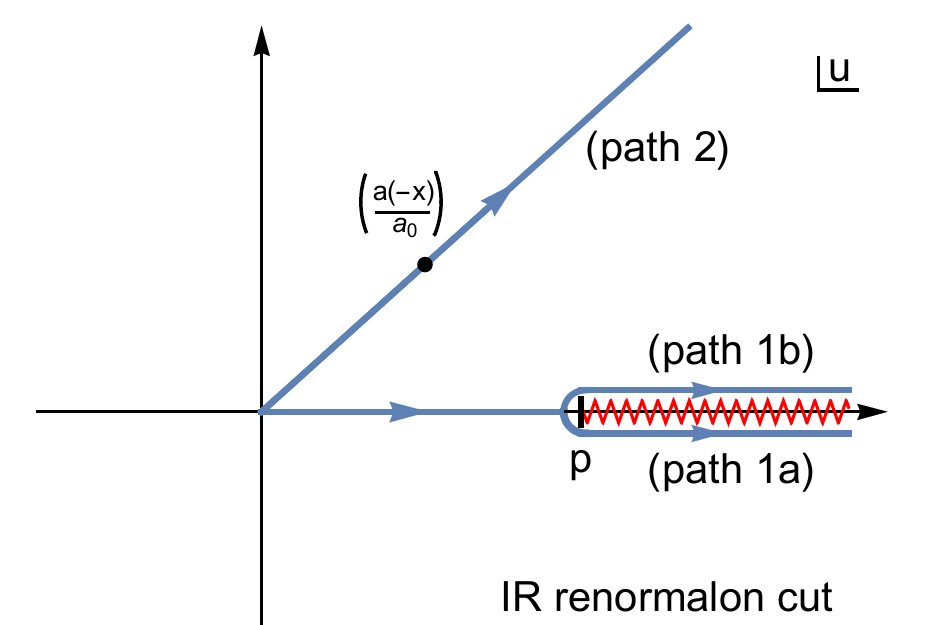}
		\caption{\label{fig:u-contourIR} }
	\end{subfigure}
\caption{\label{fig:ucontour} Graphical illustration of the Borel integration paths involved for the FOPT and CIPT Borel representations for ${\rm Im}[a(-x)]>0$  for the cases of a UV renormalon with $p<0$ (left panel) and an IR renormalon with $p>0$ (right panel).}
\end{figure}

It is tempting to assign the equality of the CIPT and FOPT expressions in Eq.~(\ref{eq:UVBorelInt}) to the fact that both integrals can be formally related through the change of variable $u =(\frac{a(-x)}{a_0}) \bar u$ as we have already mentioned in Sec.~\ref{sec:Borelrepresentation}. However, the argumentation is a bit more subtle, because $a(-x)$ is complex such that the integrals in $\bar u$ and $u$ are, in relation to each other, associated to two different paths in the complex Borel space $u$-plane: both run linearly from the origin to complex infinity, but with a relative angle that depends on the argument of the complex coupling $a(-x)$. This is visualized in Fig.~\ref{fig:u-contourUV} in the complex $u$-plane. Path 1 corresponds to the FOPT Borel integral and runs along the positive real axis. Path 2 is associated to $u =(\frac{a(-x)}{a_0}) \bar u$ with real positive $\bar u$ and corresponds to the CIPT Borel integral. It is a straight line starting at the origin and passing through the complex number $(\frac{a(-x)}{a_0})$. 
Both paths in the complex Borel space plane lead to the same result if the associated closed contour, which results from closing the two paths at positive real infinity, does not contain any poles or cuts. For a UV renormalon this is the case because its generic Borel function of Eq.~(\ref{eq:BorelDuv}) has a cut along the negative real Borel space axis, while both Borel integrals approach positive real infinity (because ${\rm Re}[a(-x)]>0$ and ${\rm Re}[a_0]>0$). This results in the equality shown in Eq.~(\ref{eq:UVBorelInt}). 
For an IR renormalon, however, the generic Borel function of Eq.~(\ref{eq:BorelDir}) has cuts (or poles) along the positive real Borel space axis as illustrated in Fig.~\ref{fig:u-contourIR}. So, path~2 is equivalent to path~1b if ${\rm Im}[a(-x)]>0$, and it is equivalent to path~1a if ${\rm Im}[a(-x)]<0$. It is this particular dependence of the result of the $\bar u$ Borel integral on the complex phase of $x$, which causes the difference of the CIPT Borel sum to the FOPT Borel sum, which is obtained from taking the average obtained from paths~1a and 1b. We stress that the way how the results of the integrations along paths~1a and 1b are handled by the CIPT Borel representation is a reflection of the properties of CIPT expansion method. However, the considerations related to Fig.~\ref{fig:ucontour} also show that, if one is willing to consider deformations of the Borel integration paths far away from either the real $u$-axis for FOPT or the $\bar u$-axis for CIPT, it is possible to make the FOPT Borel sum agree with CIPT Borel sum and vice versa. We do not discuss this possibility here since the integrations along the real Borel $u$- or $\bar u$-axes provide the correct description of the FOPT and CIPT moment series at intermediate orders as we see in Secs.~\ref{sec:large-beta} and \ref{sec:realistic}.

The CIPT Borel integral for an IR renormalon gives the expression ($p>0$)
\begin{align}
\label{eq:IRBorelIntCIPT}
\int_0^\infty \!\! {\rm d} \bar u \, 
({\textstyle\frac{a(-x)}{a_0}})\,
\frac{e^{-\frac{\bar u}{a_0}}}{\big(p - \frac{a(-x)}{a_0} \bar u\big)^\gamma}
\, = \, &
-(-a(-x))^{1-\gamma}\,e^{-\frac{p}{a(-x)}}\,\Gamma\Big(1-\gamma,{\textstyle - \frac{p}{a(-x)}}\Big)
\end{align}
in terms of simple incomplete $\Gamma$ and exponential functions when $a(-x)$ has a finite imaginary part. This imaginary part makes the integral over $\bar u$  well-defined without imposing any regularization prescription as we have already pointed out before.
Since for the CIPT approach the remaining contour integration is deformed such that it never crosses the negative real $x$-axis (see Sec.~\ref{sec:summary}), $a(-x)$ never becomes real-valued, and the result given in Eq.~(\ref{eq:IRBorelIntCIPT}) is sufficient. 

In contrast, for the FOPT Borel integration the cut coming from the Borel function of the reduced Adler function is located on the real $u$-axis and thus lies on the path of integration. The PV prescription corresponds to taking the average of using paths~1a and 1b.
These paths are	equivalent to the infinitesimal shifts $p\to p\pm i 0$ in the generic Borel function of Eq.~(\ref{eq:BorelDir}).
Together with the PV prescription the canonical definition of the ambiguity associated to the FOPT Borel integral used in the literature, is given by
half of the difference with respect to both paths multiplied by a factor of $i$ and a conventional factor $1/\pi$. 
The corresponding analytic expressions for the FOPT Borel integral and its renormalon ambiguity  
are given by
\begin{align}
\label{eq:IRBorelIntFOPTsum}
& {\rm PV} \int_0^\infty \!\! {\rm d} u \,  \frac{e^{-\frac{u}{a(-x)}}}{(p - u)^\gamma} \, = \, \frac{1}{2} \left[\, 
\int_0^\infty \!\! {\rm d} u \,  \frac{e^{-\frac{u}{a(-x)}}}{(p+i 0 - u)^\gamma}
\, + \,
\int_0^\infty \!\! {\rm d} u \,  \frac{e^{-\frac{u}{a(-x)}}}{(p-i 0 - u)^\gamma}
\,\right]  \\ 
 = & \,
 -(-a(-x))^{1-\gamma}\,e^{-\frac{p}{a(-x)}}\,\Gamma\Big(1-\gamma,{\textstyle - \frac{p}{a(-x)}}\Big)
 - {\rm sig}[{\rm Im}[a(-x)]]\,(i\pi)\,\frac{(a(-x))^{1-\gamma}}{\Gamma(\gamma)}\,e^{-\frac{p}{a(-x)}}
 \nonumber
\end{align}
and
\begin{align}
\label{eq:IRBorelIntFOPTambi}
& \frac{i}{2\pi} \left[\, 
\int_0^\infty \!\! {\rm d} u \,  \frac{e^{-\frac{u}{a(-x)}}}{(p+i 0 - u)^\gamma}
\, - \,
\int_0^\infty \!\! {\rm d} u \,  \frac{e^{-\frac{u}{a(-x)}}}{(p-i 0 - u)^\gamma}
\,\right] \, = \,
\frac{(a(-x))^{1-\gamma}}{\Gamma(\gamma)}\,e^{-\frac{p}{a(-x)}}\,,
\end{align}
where the function ${\rm sig[z]}$ gives the sign of $z$. Note that the rhs of Eq.~(\ref{eq:IRBorelIntFOPTambi}) is valid for any complex $a(-x)$ with a positive real part, while the rhs Eq.~(\ref{eq:IRBorelIntFOPTsum}) applies only if ${\rm Re}(a(-x))>0$ and  ${\rm Im}(a(-x))\neq 0$. 
Note that we also have ${\rm sig}[{\rm Im}[a(-x)]]={\rm sig}[{\rm Im}[x]]$, taking into account the analytic structure of the strong coupling. Further we point out that Eq.~(\ref{eq:IRBorelIntFOPTambi}) is proportional to $e^{-\frac{p}{a(-x)}}\sim \Lambda_{\rm QCD}^{2p}/s^p$, indicating that it has the same power-suppression 
as the nonperturbative OPE corrections in Eq.~(\ref{eq:DOPE}) that is associated to the IR renormalon.

Inspecting the analytic structure of the result for the CIPT Borel integral in Eq.~(\ref{eq:IRBorelIntCIPT}) we see that it exhibits a cut along the entire positive real axis in the complex $a(-x)$-plane. Together with the cut contained in the strong coupling itself, 
the expression has a cut along the entire real $x$-axis. 
The cut along the negative real $x$-axis originates from the cut already discussed in Sec.~\ref{sec:path} and for Eq.~(\ref{eq:BorelCIPT3}) and therefore stretches into the entire region accessible by the perturbative evolution of the strong coupling
(when the $\bar u$ Borel integral is carried out first). In the complex $t$-plane the cut covers the entire real axes as well.
Since the contour of the $x$ integration is not allowed to cross the real negative axis at any finite distance from the origin it must be deformed to infinity. Comparing to the result for the FOPT Borel integral in Eq.~(\ref{eq:IRBorelIntFOPTsum}) we see that
the first term agrees with the CIPT Borel integral result, and that the second term exhibits a cut along the entire real axis in the complex $a(-x)$-plane. Interestingly, the cut along the positive real $a(-x)$-axis precisely cancels in the sum of both terms in Eq.~(\ref{eq:IRBorelIntFOPTsum}), allowing to do the contour along the circular path $|x|=1$ when computing the FOPT Borel sum. The same is true for the expression for the FOPT Borel sum  ambiguity given in Eq.~(\ref{eq:IRBorelIntFOPTambi}).

It is the difference of Eqs.~(\ref{eq:IRBorelIntCIPT}) and (\ref{eq:IRBorelIntFOPTsum}), which is the second term on the rhs of Eq.~(\ref{eq:IRBorelIntFOPTsum})), that leads to the asymptotic separation. The result in Eq.~(\ref{eq:IRBorelIntFOPTambi}), which leads to the FOPT Borel sum ambiguity, has the same analytic form up to the additional factor ${\rm sig}[{\rm Im}[a(-x)]]={\rm sig}[{\rm Im}[x]]$. Both exhibit the same power suppression $\propto e^{-\frac{p}{a(-x)}}\sim \Lambda_{\rm QCD}^{2p}/s^p$, but it is the factor ${\rm sig}[{\rm Im}[x]]$ that causes the asymptotic separation to be much larger than the FOPT Borel sum ambiguity.

Note that the existence of the cut in Eq.~(\ref{eq:IRBorelIntCIPT}) along the negative real $x$-axis may be viewed as that the CIPT Borel representation suggests that the Borel sum of the Adler function would have a cut along the negative real $x$-axis. We stress, that the CIPT Borel representation does in principle not allow for this interpretation, because the contour integration over $x$ is an absolutely integral part of the CIPT Borel representation, and one should not interpret Eq.~(\ref{eq:IRBorelIntCIPT}) without it.
This means that the  CIPT Borel representation does not imply that the expression in Eq.~(\ref{eq:IRBorelIntCIPT}) has to be interpreted as a contribution to the Borel sum of the Adler function. This possibility can, however, not be excluded either. We discuss this topic in Sec.~\ref{sec:implications}, which is, however, not conclusive.

\subsection{Asymptotic Separation and FOPT Borel Sum Ambiguity}
\label{sec:asysepambiguity}

Let us now focus on the determination of the final analytic results for the asymptotic separation arising from the generic IR renormalon Borel function term shown in Eq.~(\ref{eq:BorelDir}). Considering again the monomial weight function $W(x)=(-x)^m$ the expression for the asymptotic separation reads
\begin{align}
\label{eq:Sepa1}
&  \Delta(m,p,\gamma,s_0) \, \equiv \,
\delta_{\{(-x)^m,p,\gamma\},{\rm Borel}}^{(0),{\rm CIPT}}(s_0) \, - \,
\delta_{\{(-x)^m,p,\gamma\},{\rm Borel}}^{(0),{\rm FOPT}}(s_0) \nonumber \\[3mm]
& =\,
\frac{1}{2 \Gamma(\gamma)} \,\ointctrclockwise\limits_{{\cal C}_x} \frac{{\rm d}x}{x} \, (-x)^m \,
{\rm sig}[{\rm Im}[a(-x)]]\,(a(-x))^{1-\gamma}\,e^{-\frac{p}{a(-x)}}
\,.
\end{align}
For comparison, the corresponding expression for the ambiguity of the FOPT Borel sum has the form\footnote{The numerical values obtained from the integral in Eq.~(\ref{eq:IRBorelIntFOPTambi2}) evaluate to real numbers with either sign. We define the FOPT Borel sum ambiguity for a given Adler function Borel function model as the size of the coherent sum of all individual terms $\delta^{\rm FOPT}(m,p,\gamma,s_0)$ that arise. In Tabs.~\ref{tab:largebeta0} and \ref{tab:5loopbeta} we have kept the resulting overall signs of the resulting values of the FOPT Borel sum ambiguity.}
\begin{align}
\label{eq:IRBorelIntFOPTambi2}
& \delta^{\rm FOPT}(m,p,\gamma,s_0) \, \equiv \,
\frac{1}{2 \pi i}\frac{1}{\Gamma(\gamma)} \,\ointctrclockwise\limits_{|x|=1} \frac{{\rm d}x}{x} \, (-x)^m \,
(a(-x))^{1-\gamma}\,e^{-\frac{p}{a(-x)}}\,.
\end{align}
As already mentioned, the integrand for the asymptotic separation differs from the FOPT Borel sum ambiguity due to the additional factor ${\rm sig}[{\rm Im}[a(-x)]]={\rm sig}[{\rm Im}[x]]$. The asymptotic separation $\Delta$ can therefore be sizeable even when the value for $\delta^{\rm FOPT}$ is strongly suppressed or even vanishes. As we show in the subsequent numerical analyses, for Borel function models with a sizeable gluon condensate renormalon cut this feature explains quantitatively why the discrepancy between the asymptotic behavior of the FOPT and CIPT spectral function moments series can exceed by far the size of the ambiguity assigned to the FOPT series. 

What remains to be discussed for the asymptotic separation is how to carry out the integration over the contour ${\cal C}_x$. 
The discussion is subtle because the convergence issues that we already discussed at the end of Sec.~\ref{sec:anatomy} for $m\ge p$
reemerge. The question of convergence can also be seen in the form of Eq.~(\ref{eq:Sepa1}) since the power-suppression coming from the
exponential term $e^{-p/a(-x)}$ competes with the power-enhancement from the monomial term $(-x)^m$ when the contour is deformed to negative real infinity. 

Let us first consider the asymptotic separation for the case $m<p$, where the exponential suppression wins and the contour integral in Eq.~(\ref{eq:Sepa1}) is convergent. Here the path ${\cal C}_x$ is split into two contributions. The first starts at $x_-=1+i 0$ and ends at negative real infinity in the positive imaginary half plane, i.e.\ at $x^\infty_-=-\infty+i\eta$ with $\eta$ being some positive real number. The second starts at $x^\infty_+=-\infty-i\eta$, runs in the negative imaginary half plane and ends at  $x_+=1-i 0$. Both paths are also visualized in Fig.~\ref{fig:x-contour} as the arrowed blue dotted lines.
Changing again to the $t$ variable defined in Eq.~(\ref{eq:tdef}) we can rewrite the expression for the asymptotic separation as
\begin{align}
\label{eq:Sepa2}
& \Delta(m,p,\gamma,s_0)\nonumber  \\ & \, = \,
-\,\Big(\frac{\Lambda_{\rm QCD}^2}{s_0}\Big)^m \,
\frac{2^{\gamma-1}}{\Gamma(\gamma)}\,
\,\sum_{\ell=0}^\infty\,\tilde g_\ell^{(2m)}\,
\left[\! \int_{t_-}^{t_-^\infty} + \int_{t_+}^{t_+^\infty}\,\right] {\rm d}t \,(-t)^{-2m\hat b_1+\gamma-\ell-1} \,e^{2(p-m)t}\,,
\end{align}
where the upper limits of the two integrals are $t_\mp^\infty=-\infty\pm i\eta$. The corresponding paths are visualized in Fig.~\ref{fig:t-contour}, again by the arrowed blue dotted lines.
The $t$-integrals can be readily evaluated giving 
\begin{align}
\label{eq:Sepa3}
 \Delta(m\neq p,p,\gamma,s_0) & \, = \,
\Big(\frac{\Lambda_{\rm QCD}^2}{s_0}\Big)^m \,
\,\sum_{\ell=0}^\infty\,\tilde g_\ell^{(2m)}\,
\frac{2^{2m\hat b_1+\ell}}{\Gamma(\gamma)}\,\\ &
\, \times\,{\rm Re}\Big[ (p-m+i 0)^{2m\hat b_1-\gamma+\ell} \,
\Gamma(-2m\hat b_1+\gamma-\ell,-2(p-m)t_-) \Big] \nonumber
\end{align}
For the case $m>p$ we cannot employ the integration path described above, because Eq.~(\ref{eq:Sepa2}) diverges. We therefore have to rely on an analytic continuation.
To define the result for $m>p$ we can add an infinitesimal imaginary contribution to $p$ of the form
$p\to p\pm i 0$ in the original integral in Eq.~(\ref{eq:Sepa2}) in the upper/lower complex $t$ plane. This leaves the value of the integral for $m<p$ unchanged. This modification now allows us to change the integration limits to $t_\mp^\infty=\pm i\infty$, again without modifying the integral value for $m<p$. With these modifications the integral can now be evaluated for $m>p$ and the resulting analytic expression is the one already given in Eq.~(\ref{eq:Sepa3}), where the previously mentioned infinitesimal imaginary part prescription is accounted for in the term  $(p-m+i 0)^{2m\hat b_1-\gamma+\ell}$. In other words, the analytic continuation for $m>p$ simply entails using the analytic result obtained for the case $m<p$ with a infinitesimal imaginary part prescription to render its dependence on the sign of $p-m$ unambiguous. 
Note that the results for the overall power-dependence of the asymptotic separation on $\Lambda_{\rm QCD}$ is $\sim (\Lambda_{\rm QCD}^{2}/s_0)^p$, which arises from the combination of the prefactor $\sim (\Lambda_{\rm QCD}^{2}/s_0)^m$ and the analytic form of the incomplete $\Gamma$-function.
We will show in our numerical analyses of Secs.~\ref{sec:large-beta} and \ref{sec:realistic} that the result for the asymptotic separation given in Eq.~(\ref{eq:Sepa3}) provides values for the CIPT Borel sums that are perfectly compatible with the CIPT series behavior for $m\neq p$ at intermediate orders where the series show a stable behavior.  

For the case $m=p$ we see that Eq.~(\ref{eq:Sepa2}) is convergent if $-2p\hat b_1+\gamma<0$. 
For a generic nonanalytic term $\sim 1/(p-u)^\gamma$ in the Borel function associated to pure $\Lambda_{\rm QCD}^{2p}$ ambiguity the term $\gamma=1+2p \hat b_1$ always arises, so the condition can in general not be satisfied and there is no obvious way to consistently define the asymptotic separation. However, for the case $m=p$ the IR renormalon is not suppressed in the FOPT expansion. Since this implies that the FOPT as well as the CIPT moment series behave quite badly, such that the issue of their discrepancy at some intermediate order where both series show a stable behavior is not arising in practice (where the case $p=2$ is numerically most relevant), the notion of the asymptotic separation does not arise for this case. 
We therefore take a practical approach and define the asymptotic separation to be zero for the case $m=p$:
\begin{align}
\label{eq:Sepa4}
\Delta(p,p,\gamma,s_0)  \, = \, 0\,.
\end{align}
We will show below that this practical definition for $\Delta(p,p,\gamma,s_0)$ provides results for the CIPT Borel sums that are perfectly compatible with the behavior of the FOPT and the CIPT series at intermediate orders when the case $m=p$ arises. 

In the large-$\beta_0$ approximation the corresponding results are quite compact. For $\gamma=1, 2$ they adopt the simple form: 
\begin{align}
\label{eq:Sepabeta0} 
\Delta_{\beta_0}(m\neq p,p,1,s_0) \, = \, &   \frac{ (-1)^{p-m}}{p-m}\,e^{-\frac{p}{a_0}} \,, 
\\
\nonumber
\Delta_{\beta_0}(m\neq p,p,2,s_0) \, = \, &  (-1)^{p-m}\,\bigg[ \frac{1}{(p-m)^2} +  \frac{1}{(p-m)a_0}\bigg]\,e^{-\frac{p}{a_0}}\,, 
\end{align}
where we remind the reader that the identity $e^{-\frac{p}{a_0}}=(\Lambda_{\rm QCD}^{2}/s_0)^p$ holds in the large-$\beta_0$ approximation.

It is instructive to compare the results for the analytic separation between the CIPT and the FOPT Borel sums to the corresponding ones for the ambiguity of the FOPT Borel sum given in Eq.~(\ref{eq:IRBorelIntFOPTambi2}). After changing variable from $x$ to $t$, the ambiguity of the FOPT Borel sum can be written as
\begin{align}
\label{eq:IRBorelIntFOPTambi3}
\delta^{\rm FOPT}(m,p,\gamma,s_0)  \, = \,
\frac{1}{\pi}\,\Big(\frac{\Lambda_{\rm QCD}^2}{s_0}\Big)^m \,
\frac{2^{\gamma-1}}{\Gamma(\gamma)}\,
\,\sum_{\ell=0}^\infty\,\tilde g_\ell^{(2m)}\,
\int_{t_-}^{t_+}  {\rm d}t \,(-t)^{-2m\hat b_1+\gamma-\ell-1} \,e^{2(p-m)t} \,.
\end{align}
Since ${\rm Re}[t_\pm]<0$ and the integrand is analytic in the negative real complex half plane, one can adopt a straight line for the integration path between $t_\pm$ for any values of $p$ and $m$. It is straightforward to do the integrals analytically giving
\begin{align}
\label{eq:IRBorelIntFOPTambi4}
\delta^{\rm FOPT}(m\neq p,p,\gamma,s_0) & \, = \, 
\frac{1}{\pi}\,
\Big(\frac{\Lambda_{\rm QCD}^2}{s_0}\Big)^m \,
\,\sum_{\ell=0}^\infty\,\tilde g_\ell^{(2m)}\,
\frac{2^{2m\hat b_1+\ell}}{\Gamma(\gamma)}\\ &
 \, \times\,\bigg\{\, {\rm Im}\Big[ (p-m+i 0)^{2m\hat b_1-\gamma+\ell} \,
\Gamma(-2m\hat b_1+\gamma-\ell,-2(p-m)t_-) \Big] \nonumber \\ &
\hspace{1cm}  
+\, \Theta(m-p)\, (m-p)^{2m\hat b_1-\gamma+\ell} \frac{\pi}{\Gamma(2m\hat b_1-\gamma+\ell+1)}
\,\bigg\}\, ,\nonumber
\end{align}
and
\begin{align}
\label{eq:IRBorelIntFOPTambi5}
\delta^{\rm FOPT}(p,p,\gamma,s_0)  \, = \,
\frac{1}{\pi}\,
\Big(\frac{\Lambda_{\rm QCD}^2}{s_0}\Big)^m \,
\frac{2^{\gamma}}{\Gamma(\gamma)}\,
\,\sum_{\ell=0}^\infty\,\tilde g_\ell^{(2m)}\,
\frac{1}{2m\hat b_1-\gamma+\ell}\,
{\rm Im}\Big[ (-t_-)^{-2m\hat b_1+\gamma-\ell} \Big]\,.
\end{align}
In the large-$\beta_0$ approximation the results are compact as well and read
\begin{align}
\label{eq:ambitbeta0}
\delta^{\rm FOPT}_{\beta_0}(m\neq p,p,1,s_0)  &=    0 \,, 
&
\delta^{\rm FOPT}_{\beta_0}(p,p,1,s_0) & =  e^{-\frac{p}{a_0}} \,, 
\\ \nonumber
\delta^{\rm FOPT}_{\beta_0}(m\neq p,p,2,s_0) & =   \frac{(-1)^{p-m}}{m-p}\,e^{-\frac{p}{a_0}}\,, 
&
\delta^{\rm FOPT}_{\beta_0}(p,p,2,s_0) & =  \frac{1}{a_0} \,e^{-\frac{p}{a_0}} \,. 
\end{align}
The results for the FOPT Borel sum ambiguity $\delta^{\rm FOPT}$ for a certain generic IR renormalon term in the Adler function's Borel function are sometimes taken as a proxy for the impact and the parametric size of the corresponding OPE correction (determined within the standard OPE method) in the spectral function moment. So the equality   
$\delta^{\rm FOPT}_{\beta_0}(m\neq p,p,1,s_0)=0$ expresses that the OPE correction associated to a simple pole renormalon $\sim 1/(p-u)$ vanishes for the weight function $W(x)=(-x)^{m\neq p}$.

It is instructive to compare the expressions for the FOPT Borel sum ambiguity $\delta^{\rm FOPT}$ with those for the asymptotic separation $\Delta$ for the same $p$ value. Both exhibit the same power suppression $\sim (\Lambda_{\rm QCD}^{2}/s_0)^p$ as a reminder of the fact that both stem from the same IR renormalon. It is straightforward to see that for integer values $m\neq p$ we always have $\Delta > \delta^{\rm FOPT}$. More importantly, it is even possible that parametrically $\Delta \gg \delta^{\rm FOPT}$, as can be easily seen in the large-$\beta_0$ approximation where a single pole renormalon ambiguity in the Adler function is always eliminated for $m\neq p$ (i.e.\ $\delta^{\rm FOPT}_{\beta_0}(m\neq p,p,1,s_0)=0$), while the corresponding asymptotic separation $\Delta_{\beta_0}$ is finite.  
We thus see that the asymptotic separation due to some IR renormalon in the Adler function can be parametrically larger than the size of the OPE corrections computed in the standard OPE approach. This is a reflection of the peculiar and unusual properties of the CIPT expansion we have already mentioned in Sec.~\ref{sec:anatomy}.

\section{Application in the Large-$\beta_0$ Approximation}
\label{sec:large-beta}

In this section we demonstrate that the result for the asymptotic separation derived in Sec.~\ref{sec:asymptoticseparation} correctly quantifies the discrepancy between the FOPT and CIPT series at intermediate orders in the large-$\beta_0$ approximation.
The behavior of the CIPT series also corroborates our conceptual considerations of Sec.~\ref{sec:anatomy} that the CIPT expansion method is not consistent with the standard form of the Adler function's OPE corrections shown in Eq.~(\ref{eq:DOPE}).

In the  large-$\beta_0$ approximation the all-order perturbative series and the corresponding exact Borel functions are known for the vacuum polarization, the reduced Adler function and many other observables, and furthermore many results can be given in terms of brief and simple analytic expressions. The large-$\beta_0$ approximation is useful as it shares important qualitative properties with the corresponding exact QCD results. Note that all our numerical analyses (here and in Secs.~\ref{sec:realistic} and \ref{sec:implications}) are carried out in the $n_f=3$ flavor scheme for the strong coupling with $\alpha_s(m_\tau^2)=0.34$ and for $s_0=m_\tau^2$.

\subsection{Generic $p=2$ Renormalon}
\label{sec:large-beta-generic}

\begin{figure} 
	\centering
	\begin{subfigure}[b]{0.48\textwidth}
		\includegraphics[width=\textwidth]{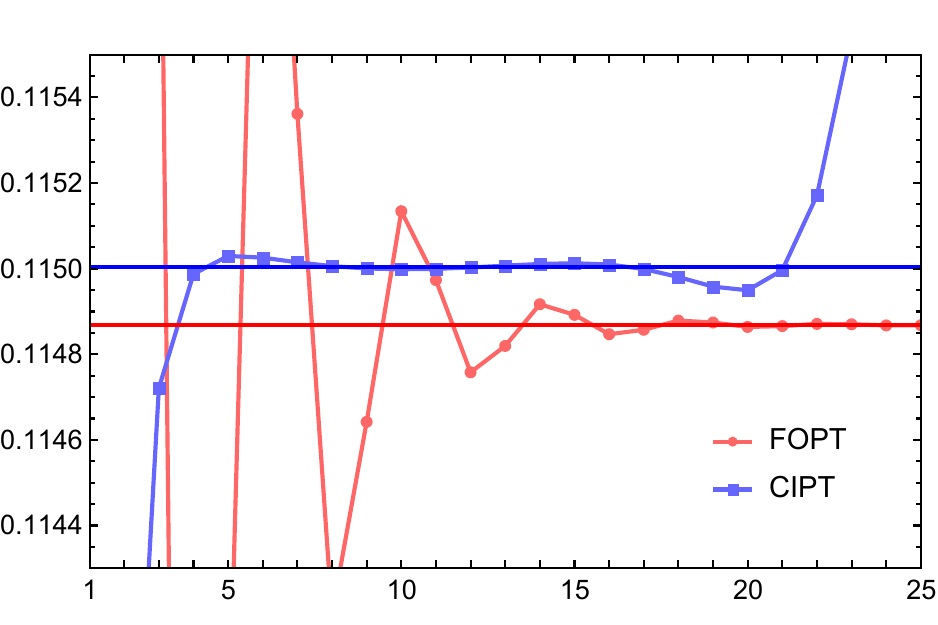}
		\caption{\label{fig:beta0simple0} Simple pole, p=2, $W(x)=1$, large-$\beta_0$}
	\end{subfigure}
	~ 
	\begin{subfigure}[b]{0.48\textwidth}
		\includegraphics[width=\textwidth]{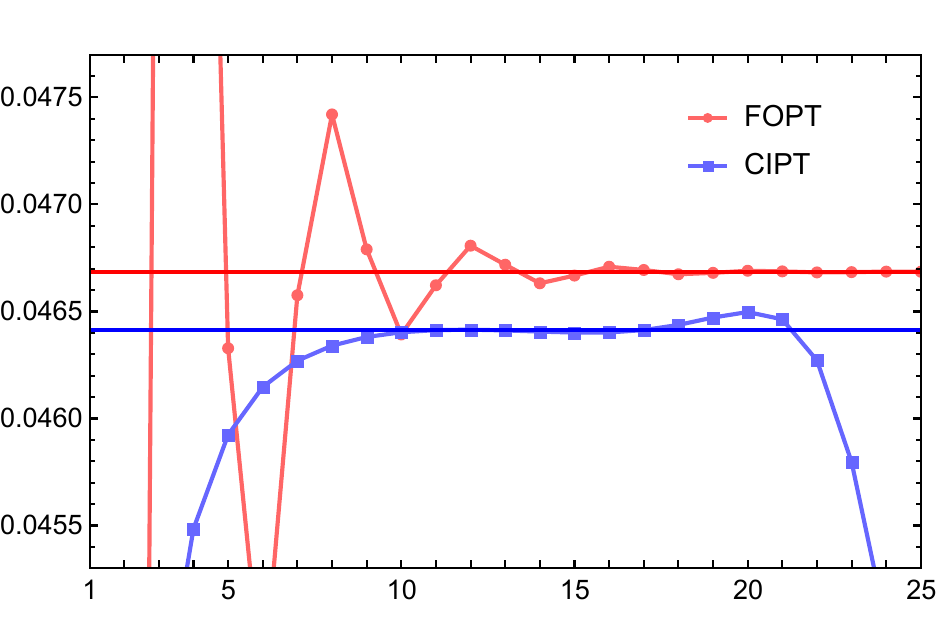}
		\caption{\label{fig:beta0simple1}  Simple pole, p=2, $W(x)=(-x)$, large-$\beta_0$}
	\end{subfigure}
	
	\begin{subfigure}[b]{0.48\textwidth}
		\includegraphics[width=\textwidth]{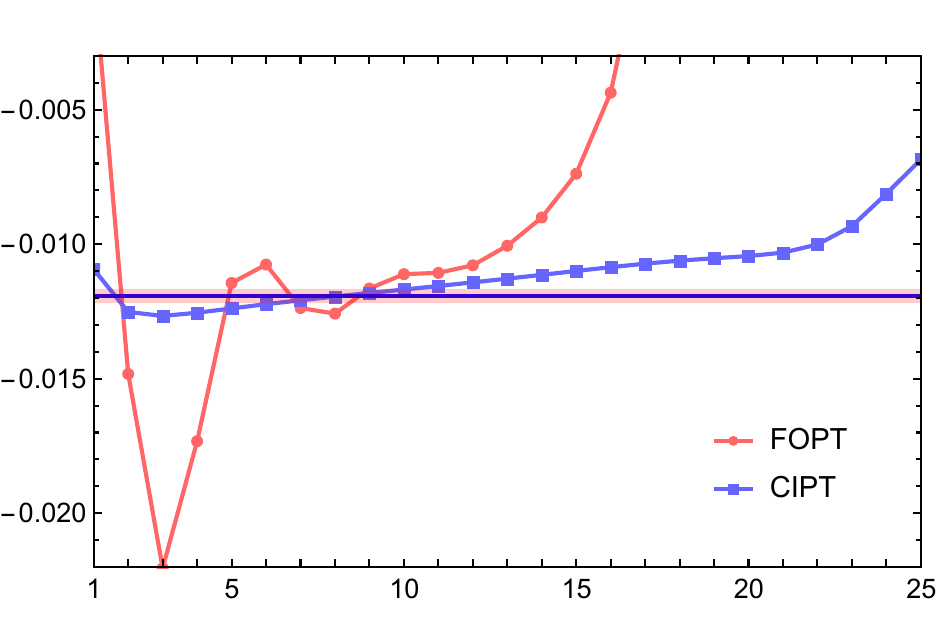}
		\caption{\label{fig:beta0simple2} Simple pole, p=2, $W(x)=(-x)^2$, large-$\beta_0$}
	\end{subfigure}
	~
	\begin{subfigure}[b]{0.48\textwidth}
		\includegraphics[width=\textwidth]{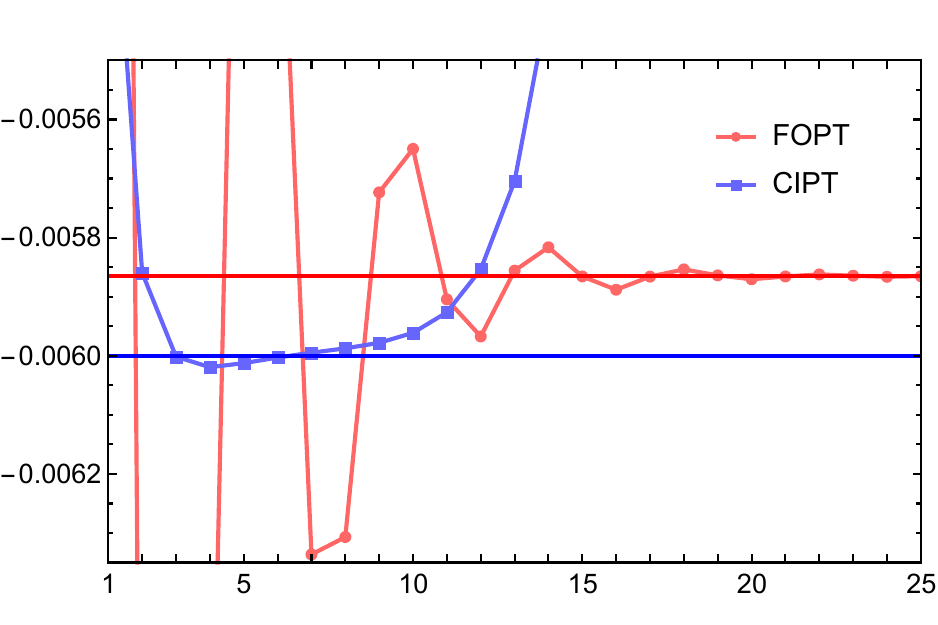}
		\caption{\label{fig:beta0simple4} Simple pole, p=2, $W(x)=(-x)^4$, large-$\beta_0$}
	\end{subfigure}
	\caption{\label{fig:beta0simplepole} 
	Moments $\delta_{\{(-x)^m,2,1\}}^{(0),{\rm FOPT}}(m_\tau^2)$ (red) and $\delta_{\{(-x)^m,2,1\}}^{(0),{\rm CIPT}}(m_\tau^2)$ (blue) in the large-$\beta_0$ approximation for a pure $p=2$ single renormalon pole and weight functions $W(x)=(-x)^m$ with $m=0,1,2,4$ as a function of the order up to which the series are summed. The red and blue horizonal lines represent the Borel sums of the FOPT and CIPT series, respectively, and the red band indicates the conventional Borel ambiguity of the FOPT series for $m=2$.}
\end{figure}

Let us start with the spectral function moment series arising from a generic single pole IR renormalon Borel term related to a $p=2$ gluon condensate term in the spectral function moment's OPE, $B^{\rm IR}_{\hat D,2,1}(u)=1/(2-u)$, see Eq.~(\ref{eq:BorelDir}).  It constitutes the dominant IR renormalon in the Borel function of the reduced Adler function and, due to its sizeable normalization, provides sizeable contributions to Adler function's perturbative coefficients already at low orders, see Sec.~\ref{sec:large-beta-TauWidth}.  
The FOPT series can be obtained directly from expanding the $\ell=0$ term in Eq.~(\ref{eq:BorelFOPT3}) with the expression of Eq.~(\ref{eq:Ftildebeta0}) in powers of $u$ and carrying out the Borel integral. The CIPT series can be obtained by determining the $\alpha_s$ series of Eq.~(\ref{eq:AdlerseriesCIPT}) associated to $B^{\rm IR}_{\hat D,2,1}(u)$ and then using Eq.~(\ref{eq:deltaCIPT3}). The resulting FOPT and CIPT moment series are shown in Figs.~\ref{fig:beta0simplepole} as the red and blue dots, respectively, as a function of the truncation order $n$ for the monomial weight functions $W(x)=(-x)^m$ for $m=0,1,2,4$. 
We note that physical weight functions vanish at $x=1$ and are always linear combinations of such monomials. The study of moments based on simple monomials thus allows to examine the interplay of their respective contributions. We use the same color assignments in all subsequent figures in this article. The respective numerical values for this analysis and all the others of Sec.~\ref{sec:large-beta} are collected for convenience in Tab.~\ref{tab:largebeta0}.
 
\begin{table}[t]
	\center
	\begin{tabular}{|c|c|c|c|r|r|r|r|}
		\hline
	    \multicolumn{8}{|c|}{large-$\beta_0$}\\
		\hline
		$B(u)$ & $W(x)$ & Figure & Scheme &  \multicolumn{1}{|c|}{$\delta^{(0),{\rm FOPT}}_{\rm Borel}$}  &  \multicolumn{1}{|c|}{$\delta^{(0),{\rm CIPT}}_{\rm Borel}$} &
		\multicolumn{1}{|c|}{$\delta^{\rm FOPT}$}  & \multicolumn{1}{|c|}{$\Delta$}  \\
		\hline
		$\frac{1}{(2-u)}$ & $1$  & \ref{fig:beta0simple0} & $\overline{\mbox{MS}}$ & $0.11487$  & $0.11500$ & $0$ & $0.00014$ \\
		$\frac{1}{(2-u)}$ & $(-x)$ & \ref{fig:beta0simple1} & $\overline{\mbox{MS}}$ & $0.04668$  & $0.04641$ & $0$ & $-0.00027$ \\
		$\frac{1}{(2-u)}$ & $(-x)^2$ & \ref{fig:beta0simple2} & $\overline{\mbox{MS}}$ & $-0.01193$  & $-0.01193$ & $0.00027$ & $0$ \\
		$\frac{1}{(2-u)}$ & $(-x)^4$ & \ref{fig:beta0simple4} & $\overline{\mbox{MS}}$ & $-0.00587$  & $-0.00600$ & $0$ & $-0.00014$ \\
		$\frac{1}{(2-u)^2}$ & $1$ & \ref{fig:beta0double0} & $\overline{\mbox{MS}}$ & $0.06340$  & $0.06402$ & $-0.00014$ & $0.00062$ \\
		$\frac{1}{(2-u)^2}$ & $(-x)$ & \ref{fig:beta0double1} & $\overline{\mbox{MS}}$ & $0.03476$  & $0.03337$ & $0.00027$ & $-0.00138$ \\
		$\frac{1}{(2-u)^2}$ & $(-x)^2$ & \ref{fig:beta0double2} & $\overline{\mbox{MS}}$ & $-0.00663$  & $-0.00663$ & $0.00111$ & $0$ \\
			$\frac{1}{(2-u)^2}$ & $(-x)^4$ & \ref{fig:beta0double4} & $\overline{\mbox{MS}}$ & $-0.00303$  & $-0.00352$ & $0.00014$ & $-0.00049$ \\
			$B_{\hat D,\lambda=5/3}$ & $W_\tau$ & \ref{fig:beta0Rtau} & $\overline{\mbox{MS}}$ & $0.26290$  & $0.24269$ & $0.00291$ & $-0.02021$ \\
			$B_{\hat D,\lambda=10/3}$ &  $W_\tau$ & \ref{fig:beta0Rtau53} & $\alpha_s^{(5/3)}$ & $0.26290$  & $0.24269$ & $0.00291$ & $-0.02021$ \\
			$B_{\hat D,\lambda=20/3}$ & $W_\tau$ & \ref{fig:beta0Rtau5} & $\alpha_s^{(15/3)}$ & $0.26290$  & $0.24269$ & $0.00291$ & $-0.02021$ \\
			$B_{\hat D,\lambda=10}$ &  $W_\tau$ & \ref{fig:beta0Rtau253} & $\alpha_s^{(25/3)}$ & $0.26290$  & $0.24269$ & $0.00291$ & $-0.02021$ \\
		\hline
	\end{tabular}
	\caption{\label{tab:largebeta0} Numerical values for the FOPT Borel sum $\delta^{(0),{\rm FOPT}}_{\rm Borel}$, the CIPT Borel sum $\delta^{(0),{\rm CIPT}}_{\rm Borel}$, the FOPT Borel sum ambiguity $\delta^{\rm FOPT}$ and the asymptotic separation $\Delta=\delta^{(0),{\rm CIPT}}_{\rm Borel}-\delta^{(0),{\rm FOPT}}_{\rm Borel}$  in the large-$\beta_0$ approximation for the analyses in Sec.~\ref{sec:large-beta}.}
\end{table}

The horizontal red lines represent the 
respective FOPT Borel sum from Eqs.~(\ref{eq:BorelFOPT2}) and (\ref{eq:IRBorelIntFOPTsum}), and the horizontal blue lines  indicate the CIPT Borel sum obtained from adding the appropriate asymptotic separation terms $\Delta$ from Eqs.~(\ref{eq:Sepabeta0}) to the FOPT Borel sum. The red band represents the FOPT Borel sum ambiguity related to adding $\pm\delta^{\rm FOPT}$ given in Eqs.~(\ref{eq:ambitbeta0}) to the FOPT Borel sum. Since for $m=0,1,4$ and all other values $m\neq 2$ the gluon condensate OPE correction vanishes, $\delta^{\rm FOPT}$ vanishes as well. Only for $m=2$ the gluon condensate OPE correction as well as  $\delta^{\rm FOPT}$ are nonzero.
For $m=0,1,4$ the FOPT series have oscillations at lower orders and converge\footnote{We have checked this to be true for any other $m\neq 2$.  Using the usual root test, one finds that the radius of convergence is $\alpha_s(m_\tau^2)=4/9$.} to the FOPT Borel sum value for orders $n\gtrsim 13$. They are ambiguity-free, which is consistent with $\delta^{\rm FOPT}$ being zero as well.
In contrast, the corresponding CIPT series are asymptotic and divergent. We have checked that this is true for any $m\neq 2$ and also for any smaller value of $\alpha_s(m_\tau^2)$. At intermediate orders the CIPT series approach the CIPT Borel sum and eventually diverge. For $m=0$, $m=1$ and $m=4$ the order ranges of closest approach to the CIPT Borel sum are $5\lesssim n \lesssim 17$,  $10\lesssim n \lesssim 17$ and $5\lesssim n \lesssim 8$, respectively. Both, the FOPT and CIPT moment series clearly exhibit a stable convergence/asymptotic regime. The corresponding Borel sum values are distinctly different, and the difference is correctly quantified by the asymptotic separation $\Delta$. 

The fact that the CIPT series are divergent for any value of $\alpha_s(m_\tau^2)$ while there is no associated standard OPE correction to compensate for this behavior, means that the CIPT expansion method cannot provide physically consistent theoretical predictions when the nonperturbative corrections of the Adler function are parametrized with the standard form shown in Eq.~(\ref{eq:DOPE}). It is inconceivable that these observations are a pure artifact of the large-$\beta_0$ approximation, and it is therefore imperative to conclude that the CIPT expansion method is not consistent with the standard OPE approach. This fully corroborates the conclusions on the unusual properties of the CIPT expansion we made in Sec.~\ref{sec:anatomy} based on the analytic propertiers of the CIPT Borel representation.  

For $m=p=2$ (lower left panel of Fig.~\ref{fig:beta0simplepole}) the FOPT and CIPT series are both asymptotic. We see that the FOPT series oscillates around its Borel sum ambiguity band up to order $n\sim 10$ and then diverges. The CIPT series shows an extended linearly increasing behavior for orders $3\lesssim n\lesssim 17$ and diverges beyond. The FOPT Borel sum represents the series value somewhere on the lower side of the linear regime. While the value of the FOPT Borel sum is certainly compatible with the behavior of the FOPT and CIPT series, its value and the FOPT ambiguity could not be easily determined by eye from the behavior of the series. The results show, however, that our practical definition of a vanishing asymptotic separation for $m=p$ is compatible with the quite unstable behavior of the FOPT and CIPT moment series.
Since the gluon condensate OPE correction is nonzero for $m=p=2$, there is in principle no inconsistency between the CIPT and FOPT expansion. However, we observe that the large-order diverging behavior of the FOPT and CIPT series differ substantially. Given that the CIPT expansion is not compatible with the standard OPE corrections for $m\neq 2$, this different high-order behavior can be seen as a confirmation that this is true for $m=p=2$ as well.

\begin{figure} 
	\centering
	\begin{subfigure}[b]{0.48\textwidth}
		\includegraphics[width=\textwidth]{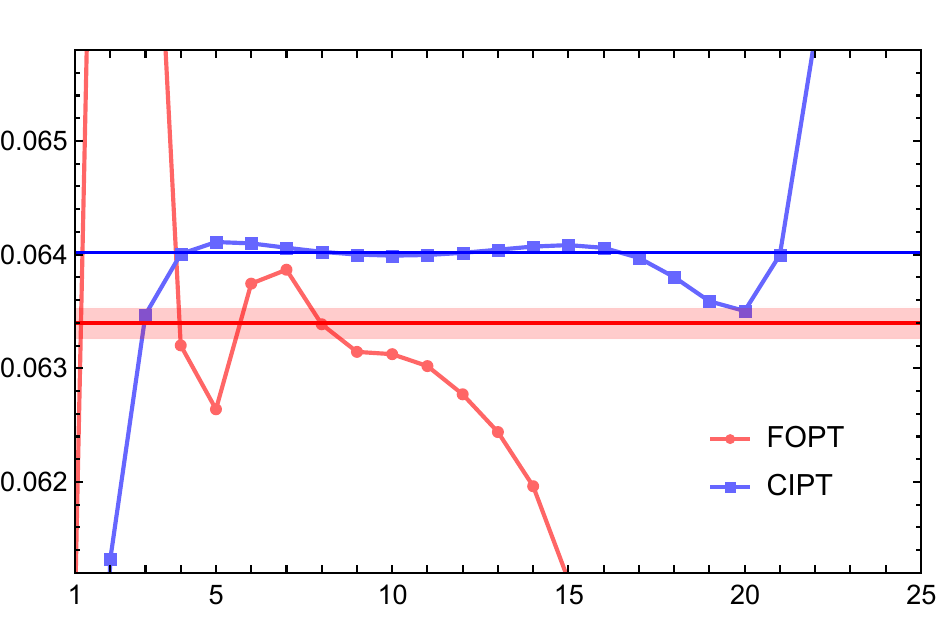}
		\caption{\label{fig:beta0double0} Double pole, p=2, $W(x)=1$, large-$\beta_0$}
	\end{subfigure}
	~
	\begin{subfigure}[b]{0.48\textwidth}
		\includegraphics[width=\textwidth]{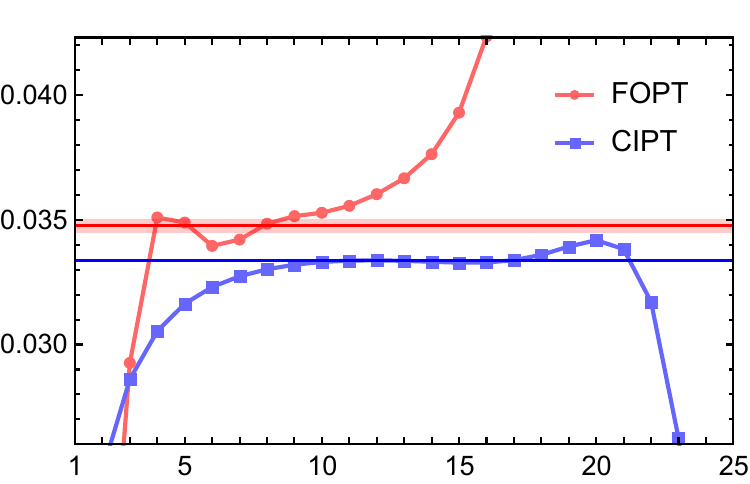}
		\caption{\label{fig:beta0double1} Double pole, p=2, $W(x)=(-x)$, large-$\beta_0$}
	\end{subfigure}
	
	\begin{subfigure}[b]{0.48\textwidth}
		\includegraphics[width=\textwidth]{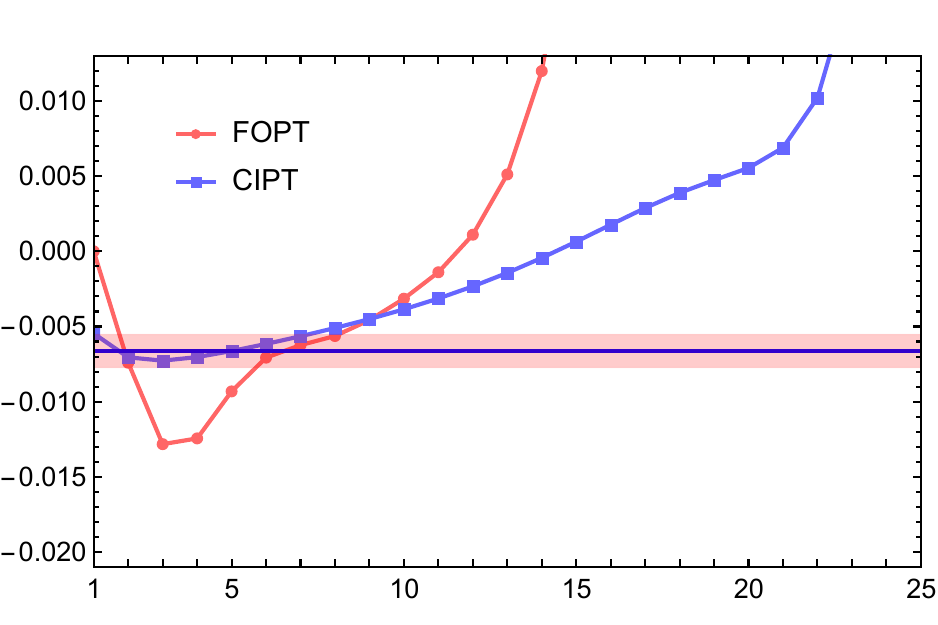}
		\caption{\label{fig:beta0double2} Double pole, p=2, $W(x)=(-x)^2$, large-$\beta_0$}
	\end{subfigure}
	~
	\begin{subfigure}[b]{0.48\textwidth}
		\includegraphics[width=\textwidth]{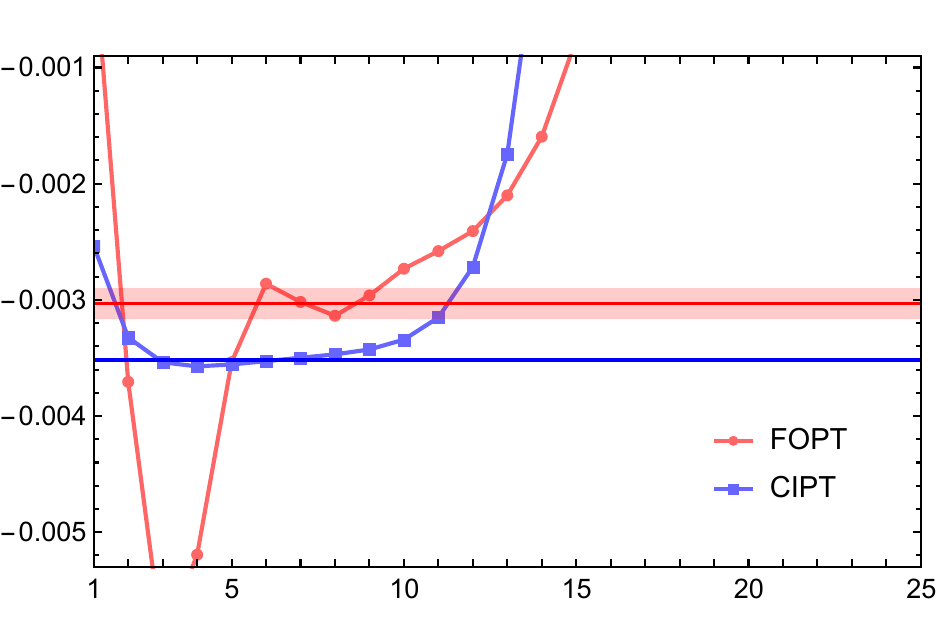}
		\caption{\label{fig:beta0double4} Double pole, p=2, $W(x)=(-x)^4$, large-$\beta_0$}
	\end{subfigure}
\caption{\label{fig:beta0doublepole} 
   Moments $\delta_{\{(-x)^m,2,1\}}^{(0),{\rm FOPT}}(m_\tau^2)$ (red) and $\delta_{\{(-x)^m,2,1\}}^{(0),{\rm CIPT}}(m_\tau^2)$ (blue) in the large-$\beta_0$ approximation for a pure $p=2$ double renormalon pole and weight functions $W(x)=(-x)^m$ with $m=0,1,2,4$ as a function of the order up to which the series are summed. The red and blue horizonal lines represent the Borel sums of the FOPT and CIPT series, respectively, and the red bands indicate the conventional Borel ambiguity of the FOPT series. }
\end{figure}

Next, let us consider a generic double pole IR renormalon Borel term for $p=2$,  $B^{\rm IR}_{\hat D,2,2}(u)=1/(2-u)^2$. Such a term is not contained in the Borel function of the Adler function, but double poles arise for all other IR renormalon singularities. For double poles the large-$\beta_0$ contour integration does not eliminate the renormalon in the FOPT Borel sum and the associated FOPT as well as CIPT series are both asymptotic. The results are shown in Figs.~\ref{fig:beta0doublepole} , again considering the monomial weight function $W(x)=(-x)^m$ for the cases $m=0,1,2,4$ and using the same conventions as in Fig.~\ref{fig:beta0simplepole}. Here, the FOPT Borel sums all exhibit an ambiguity band because $\delta^{\rm FOPT}$ is always nonzero, see Eqs.~(\ref{eq:ambitbeta0}). For $m=0,1,4$ the FOPT series show some oscillatory behavior for orders $n\lesssim 10$ around their respective Borel sum ambiguity band. On the other hand, the CIPT series clearly approach their CIPT Borel sums for orders $7\lesssim n \lesssim 15$,  $10\lesssim n \lesssim 15$ and $5\lesssim n \lesssim 8$, respectively -- a behavior very similar to the single pole case. We again see that the asymptotic separation (added to the FOPT Borel sum) correctly quantifies the values the CIPT series approach at intermediate orders. It is also clearly visible that the asymptotic separation is much larger than the FOPT ambiguity based on the standard definition used for $\delta^{\rm FOPT}$.
For $m=p=2$ (lower left panel) the FOPT and CIPT series again exhibit a quite different character in analogy to the single pole case. Both series show no clear sign of a stable value at intermediate orders, and their overall behavior is compatible with our practical definition of a vanishing asymptotic separation.

We have checked that the behavior we just discussed for $p=2$ IR renormalon poles and weight functions $W(x)=(-x)^m$ with $m=0,1,2,4$
is generic for any (positive integer) values of $p$ and $m$ and any (perturbative) value of $s_0$. 
The asymptotic separation always provides an adequate description of the discrepancy between the FOPT and CIPT moment series at intermediate orders (at least to the extend that a stable value is reached at some intermediate orders for both expansion methods).
We in particular find that for $m\neq p$ 
the asymptotic separation is always larger than the corresponding FOPT Borel sum ambiguity defined by $\delta^{\rm FOPT}$.

\subsection{Hadronic Tau Decay Width}
\label{sec:large-beta-TauWidth}

We now discuss the perturbative series for the hadronic tau decay width $R_\tau$, which is the spectral function moment for the kinematic weight function $W_\tau(x) =(1-x)^3(1+x) = 1-2x+2x^3-x^4$. The Borel function of the Adler function is known in closed form and reads~\cite{Broadhurst:1992si}
\begin{align}
\label{eq:AdlerBorelb0}
B_{\lambda}(u)[\hat D] & \, = \,\frac{128}{3\beta_0}\,\frac{e^{\lambda u}}{2-u}\,\sum_{k=2}^\infty \, \frac{(-1)^k \,k}{[k^2-(1-u)^2]^2}
\,,
\end{align}
where $\lambda=5/3$ in the $\overline{\mbox{MS}}$ scheme for the strong coupling. The Borel function has a single pole at $u=2$ and double poles at integer values for $u$ larger than 2 and all negative integers. In the $\overline{\rm MS}$ scheme the IR renormalon at $p=2$ dominates the behavior of the series for orders $n$ up to around $10$. The UV renormalon poles add a mildly oscillating behavior at these lower orders before they dominate the divergent behavior of the series beyond. The terms of the FOPT and CIPT moment series as well as the FOPT Borel sum, its ambiguity and the asymptotic separation can be obtained in a straightforward way using the method described in Sec.~\ref{sec:large-beta-generic}, identifying the corresponding residues
and summing all terms. The respective numerical values obtained in the following analysis are shown in Tab.~\ref{tab:largebeta0}.

\begin{figure} 
	\centering
	\begin{subfigure}[b]{0.48\textwidth}
		\includegraphics[width=\textwidth]{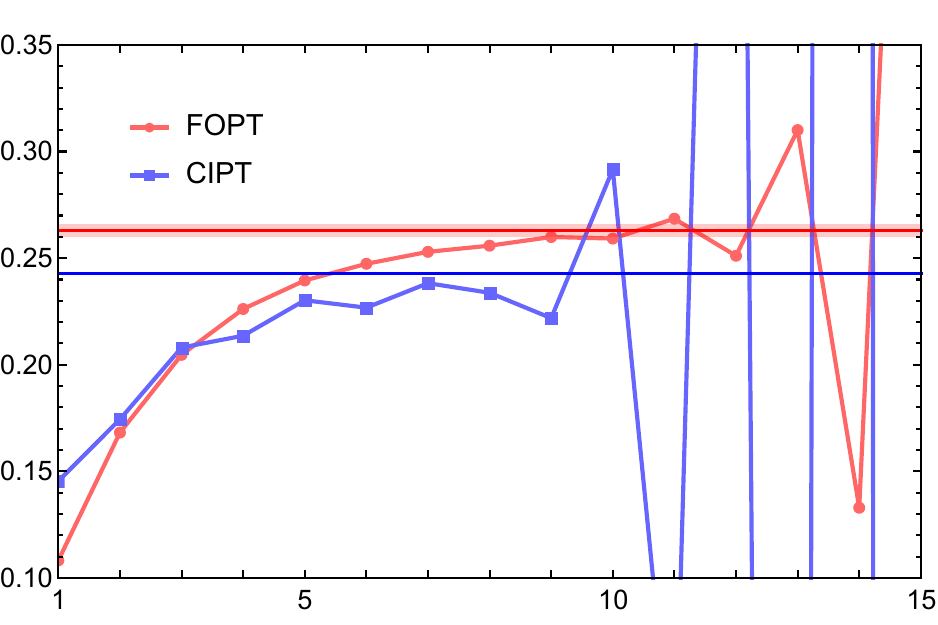}
		\caption{\label{fig:beta0Rtau} $\delta^{(0)}_{W_\tau}(m_\tau^2)$, $\alpha_s^{\overline{\rm MS}}$, large-$\beta_0$}
	\end{subfigure}
    ~
	\begin{subfigure}[b]{0.48\textwidth}
		\includegraphics[width=\textwidth]{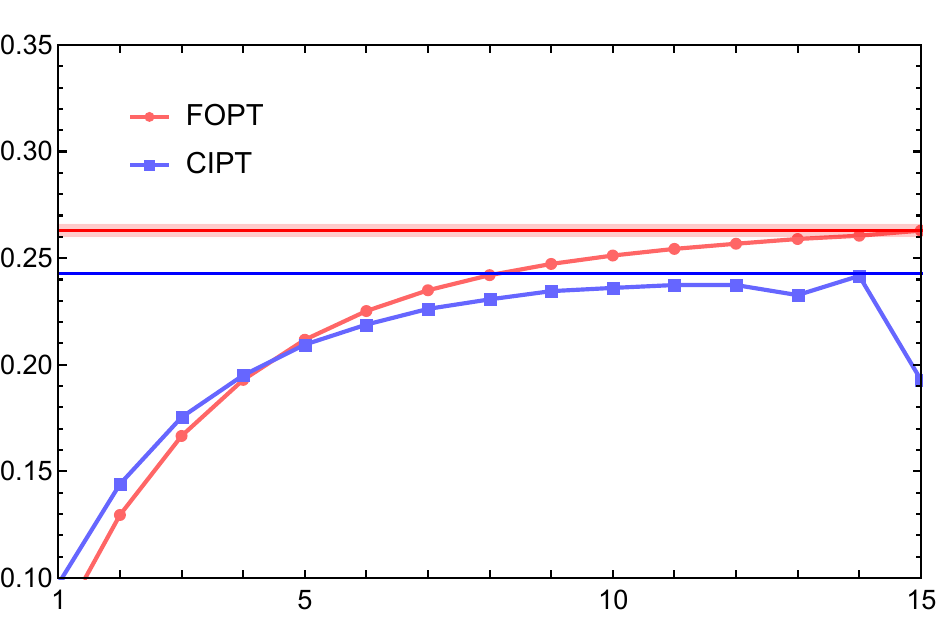}
		\caption{\label{fig:beta0Rtau53} $\delta^{(0)}_{W_\tau}(m_\tau^2)$, $\alpha_s^{(5/3)}$, large-$\beta_0$}
	\end{subfigure}

    \begin{subfigure}[b]{0.48\textwidth}
    	\includegraphics[width=\textwidth]{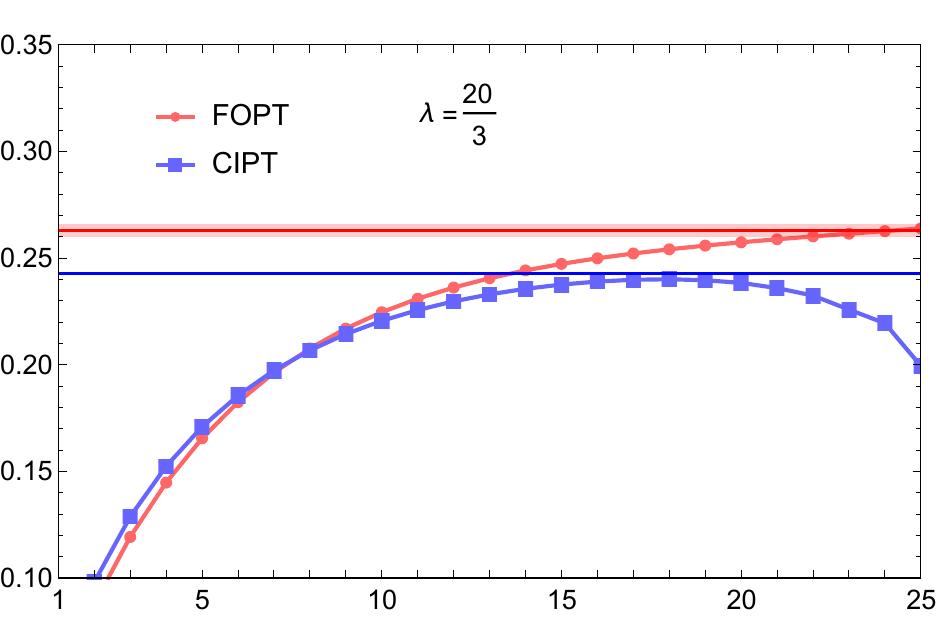}
    	\caption{\label{fig:beta0Rtau5} $\delta^{(0)}_{W_\tau}(m_\tau^2)$, $\alpha_s^{(15/3)}$, large-$\beta_0$}
    \end{subfigure}
    ~
    \begin{subfigure}[b]{0.48\textwidth}
    	\includegraphics[width=\textwidth]{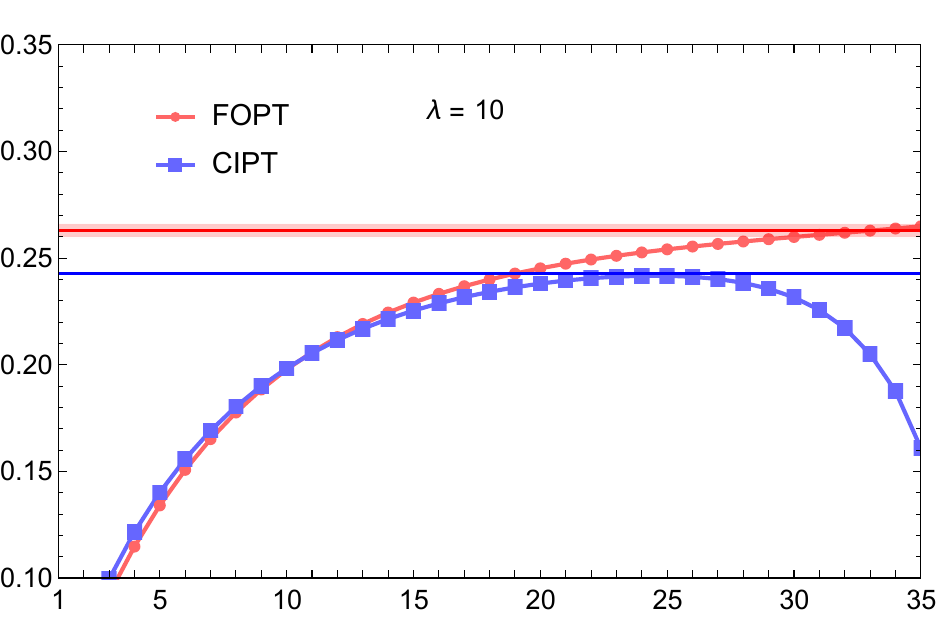}
    	\caption{\label{fig:beta0Rtau253} $\delta^{(0)}_{W_\tau}(m_\tau^2)$, $\alpha_s^{(25/3)}$, large-$\beta_0$}
    \end{subfigure}
	\caption{\label{fig:Rtau_beta0} Moments $\delta^{(0),{\rm FOPT}}_{W_\tau}(m_\tau^2)$ (red) and  $\delta^{(0),{\rm CIPT}}_{W_\tau}(m_\tau^2)$ (blue) in the large-$\beta_0$ approximation as a function of the order up to which the series are summed. Panel (a) shows the result in the $\overline{\mbox{MS}}$ scheme and panel (b) in a scheme where the coupling is defined by $a^{(5/3)}(x) = a(x)/(1+5/3\, a(x))$. Panels (c) and (d) refer to schemes where the coupling it defined by $a^{(15/3)}(x) = a(x)/(1+3\times5/3\, a(x))$ and $a^{(25/3)}(x) = a(x)/(1+5\times5/3\, a(x))$, respectively.
	The red and blue horizonal lines represent the Borel sums of the FOPT and CIPT series, respectively, and the red bands indicate the conventional Borel ambiguity of the FOPT series.}
\end{figure}

The outcome for $\delta^{(0)}_{W_\tau}$ for the FOPT and CIPT series is shown in Fig.~\ref{fig:beta0Rtau} using the same conventions as for the generic examinations in Sec.~\ref{sec:large-beta-generic}. The FOPT series reaches the FOPT Borel sum (red horizontal line) at orders $n=9$ and $10$ where it also approaches a kind of stable asymptotic behavior. At orders beyond it starts to oscillate wildly due to the impact of the UV renormalon contributions. The difference of the values of the FOPT series in the asymptotic region at orders $n=9$ and $10$ to its Borel sum is consistent with the FOPT ambiguity indicated by the narrow red band.
The CIPT series shows a convergent and mildly oscillating behavior for orders $n\lesssim 8$. For orders beyond that, the oscillations quickly become large due to the dominance of UV renormalon contributions. The value of the CIPT series at order $n=7$ is compatible with the CIPT Borel sum determined from the FOPT Borel sum plus the asymptotic separation, but the behavior of the CIPT series appears to indicate that its value in the stable intermediate order regime, before the diverging oscillations related to the UV renormalons set it, is slightly below the blue line. 

From the behavior of the CIPT series originating from a single pole at $p=2$ shown in Fig.~\ref{fig:beta0simplepole} it is actually easy to understand why the CIPT series for $R_\tau$ happens to be systematically below its CIPT Borel sum for orders where the oscillatory behavior from the UV renormalon poles is not yet sizeable. This behavior originates from the different properties of the series associated to the monomial terms with $m=1$ and $m=3,4$: the former closely approaches its Borel sum at orders $n\gtrsim 7$ (see Fig.~\ref{fig:beta0simple1}) while the latter approach their corresponding Borel sum already at lower orders and deflect from them when $n\gtrsim 7$  (see Fig.~\ref{fig:beta0simple4}). The net effect, accounting for the signs of the different monomial terms in the weight function $W_\tau(x)$, is that the CIPT series bounces back from its Borel sum at a finite distance before the start of the diverging oscillatory behavior caused by the UV renormalons. However, as we show below, this feature is less pronounced when we consider different schemes for the strong coupling.

It is straightforward to check that adopting a different scheme for the strong coupling leaves the results for the Borel sums, the FOPT Borel sum ambiguity as well as the asymptotic separation strictly invariant. This is according to the principles of the renormalon calculus mentioned at the beginning of Sec.~\ref{sec:Borelrepresentation}. All qualitative statements just made remain intact as well, apart from the facts that the order region where the series approach their Borel sums and the onset of the oscillatory behavior of the series is shifted toward higher orders if schemes are adopted where the strong coupling value is decreased and 
UV renormalon contributions are suppressed by the change of scheme. For example, if we adopt a scheme where the Adler function's Borel function has the form of Eq.~(\ref{eq:AdlerBorelb0}) with $\lambda=10/3$, which corresponds to using the coupling $a^{(5/3)}(x) = a(x)/(1+5/3\, a(x))$ and $\alpha_s^{(5/3)}(m_\tau^2)=0.24185$, the outcome shown in Fig.~\ref{fig:beta0Rtau53} is obtained.
We see that the region where the series approach their Borel sums is shifted to higher orders and we can also observe more clearly that the CIPT series bounces back from its Borel sum around the twelfth order. In Figs.~\ref{fig:beta0Rtau5} and \ref{fig:beta0Rtau253} the analogue results are shown for $\lambda=20/3$ ($\alpha_s^{(15/3)}(m_\tau^2)=0.15332$) and 
$\lambda=10$ ($\alpha_s^{(25/3)}(m_\tau^2)=0.11224$), respectively. For these $\lambda$ values, the regions where the series approach their Borel sums is shifted to even higher orders. Interestingly, we also observe that the CIPT series approaches more closely the Borel sum for larger values of $\lambda$ and that the bounce back feature diminishes. This shows that the bounce back feature of the CIPT series is a scheme-dependent issue and that our analytic result for the asymptotic separation also applies for the kinematic weight function relevant for $R_\tau$.

We note, that the invariance of the asymptotic separation under changes of $\lambda$ for this geometric kind of scheme modification follows from the identify $f_p(\gamma,a)=e^{p\eta}\sum_{i=0}^\infty  f_p(\gamma-i,a/(1+a\eta))(-\eta)^i/(i!)$ for $f_p(\gamma,a)\equiv e^{-p/a} a^{1-\gamma}/\Gamma(\gamma)$, and also applies beyond the large-$\beta_0$ approximation.
Overall we find that the asymptotic behavior of the FOPT as well as the CIPT moment series each are compatible with their respective Borel sums. The essential point is that even though the $p=2$ renormalon does not contribute to the ambiguity of the FOPT Borel sum, it provides the dominant contribution  to the asymptotic separation. In fact for the case of $R_\tau$ the exact value for the asymptotic separation is $-0.0202094$ from which $-0.0202591$  (which is $99.8\%$) comes from the $p=2$ gluon condensate renormalon. This shows that, in practice, the asymptotic separation is dominated entirely by the gluon condensate renormalon. Conversely, if the gluon condensate renormalon would be absent, the asymptotic separation would still exist, but it would be so small numerically, that it may as well be neglected from the practical perspective. This feature also applies in full QCD. In other words, the asymptotic separation is sizeable only if the normalization of the $p=2$ gluon condensate renormalon is sizeable.
The dominance of the gluon condensate renormalon concerning the FOPT-CIPT discrepancy can be also seen from the dependence of the asymptotic separation, and the gap in the asymptotic intermediate order values of both expansions, on $s_0$. Both scale with $\Lambda_{\rm QCD}^4/s_0^2$ to very good approximation - a fact that has not been noted or appreciated in the literature prior to this work.

\section{Application Accounting for the Full QCD $\beta$-Function}
\label{sec:realistic}

In this section we demonstrate that the expression for the asymptotic separation derived in Sec.~\ref{sec:asymptoticseparation} correctly quantifies the disparity in the observable asymptotic behavior of the FOPT and CIPT spectral function moment series accounting for all known terms in the QCD $\beta$-function up to five loops. Since in full QCD the exact expression for the Borel function of the Adler function is unknown, one has to rely on models. In the context of the FOPT Borel representation and using the generic Borel functions in Eqs.~(\ref{eq:BorelDir}) and (\ref{eq:BorelDuv}), it is straightforward to construct such Borel function models that quantify precisely the higher order behavior associated with the individual terms in the standard OPE. For a dimension-$d$ term in the OPE of Eq.~(\ref{eq:DOPE}) these generic Borel functions are linear combinations of the IR renormalon terms $B^{\rm IR}_{\hat D,p,\gamma}(u) \, = \, \frac{1}{(p-u)^\gamma}$ shown in Eq.~(\ref{eq:BorelDir}) for $p=d/2$ and specific values of $\gamma$. The generic Borel functions for UV renormalons are related to process-dependent matrix elements with insertions of higher-dimensional local operators and constructed in an analogous way. (See e.g. Ref.~\cite{Beneke:1998ui} for a review.)
Given the dimension of the local operators entering the condensate (or process-dependent) matrix elements, their anomalous dimension and their Wilson coefficients, the form of these generic Borel functions can be perturbatively determined in an unambiguous way. However, their precise normalization within the exact Borel function is unknown and represents the most relevant quantitative model-dependent aspect. Furthermore, the information on the Wilson coefficients (or process-dependent matrix elements) and the anomalous dimension of the local operators is in general very limited (i.e.\ known with a precision that is typically much lower than the perturbative orders known for $\delta^{(0)}_{W_i}(s_0)$). Borel models for the Adler function in full QCD thus comprise simplifying assumptions on the properties of the OPE terms and particular (preferential or plausible) choices of the normalization factors. The latter are furthermore fixed such that the model reproduces the coefficients of the perturbative series that have been computed exactly. In the following we consider several Borel models for the Adler function.  Since the construction of these models is not the subject of the article, we only describe their content briefly and refer to Ref.~\cite{Beneke:2008ad,MasterThesisRegner} for details on their construction. The explicit formulas for the Borel models used in the following examinations can be found in App.~\ref{app:models}. We emphasize that we do not intend to enter any considerations on the phenomenological soundness of the Borel models that are being discussed (in the sense that they may or may not approximate the true QCD Borel function of the Adler function). The main focus of this section is to show that the results for the asymptotic separation describe the differences in the asymptotic intermediate-order behavior of the resulting FOPT and CIPT spectral function moment series well for any given model.

\subsection{$R_\tau$ for a Multirenormalon Borel Model}
\label{sec:Rtaumultirenormalon}

We first consider a multirenormalon Borel model for the Adler function containing generic Borel functions for two IR renormalons related to $d=4$ and $d=6$ OPE terms, associated to $p=2$ and $p=3$, respectively, and one UV renormalon related to dimension-$6$ local operators, associated to $p=-1$. The $p=2$ IR renormalon is related to the gluon condensate OPE term and we account for the known ${\cal O}(\alpha_s)$ Wilson coefficient correction. For the UV renormalon we assume an anomalous dimension consistent with the emergence of a double pole in the limit of the large-$\beta_0$ approximation, which is known to exist from Eq.~(\ref{eq:AdlerBorelb0}). For the $p=3$ IR renormalon, higher order corrections to the Wilson coefficient are neglected and the anomalous dimension is assumed to vanish. For the construction of the model we follow Ref.~\cite{Beneke:2008ad}, where the Adler function coefficients $c_{3,1}$, $c_{4,1}$ and $c_{5,1}$ in Eq.~(\ref{eq:AdlerseriesCIPT})\footnote{For $c_{n\le 4,1}$ we use the known exact results~\cite{Gorishnii:1990vf,Surguladze:1990tg,Baikov:2008jh} and for $6$-loop coefficient we adopt the estimate $c_{5,1}=283$ from  Ref.~\cite{Beneke:2008ad}.} are used to fix the coefficients of the three generic renormalon Borel functions and a function linear in $u$ is added to achieve consistency with the known expressions for the coefficients $c_{1,1}$ and $c_{2,1}$. The expression for this Borel model in the $\overline{\rm MS}$ scheme for the strong coupling, referred to as $B_{\hat D,{\rm mr}}(u)$, is shown in Eq.~(\ref{eq:modelmr}), where we included the effects of the $5$-loop correction to the QCD $\beta$-function that were not yet available in Ref.~\cite{Beneke:2008ad}. 

In this Borel model the $p=2$ gluon condensate renormalon cut is implemented with a sizeable normalization, so that the gluon condensate renormalon provides a sizeable contribution to the series coefficients already at low orders.
This is comparable to the situation in the large-$\beta_0$ approximation. In this Borel model the normalization of the $p=-1$ UV renormalon is rather small, because of the absence of any visible oscillatory behavior in the known coefficients $c_{3,1}$, $c_{4,1}$ and $c_{5,1}$.  

\begin{figure} 
	\centering
	\begin{subfigure}[b]{0.48\textwidth}
		\includegraphics[width=\textwidth]{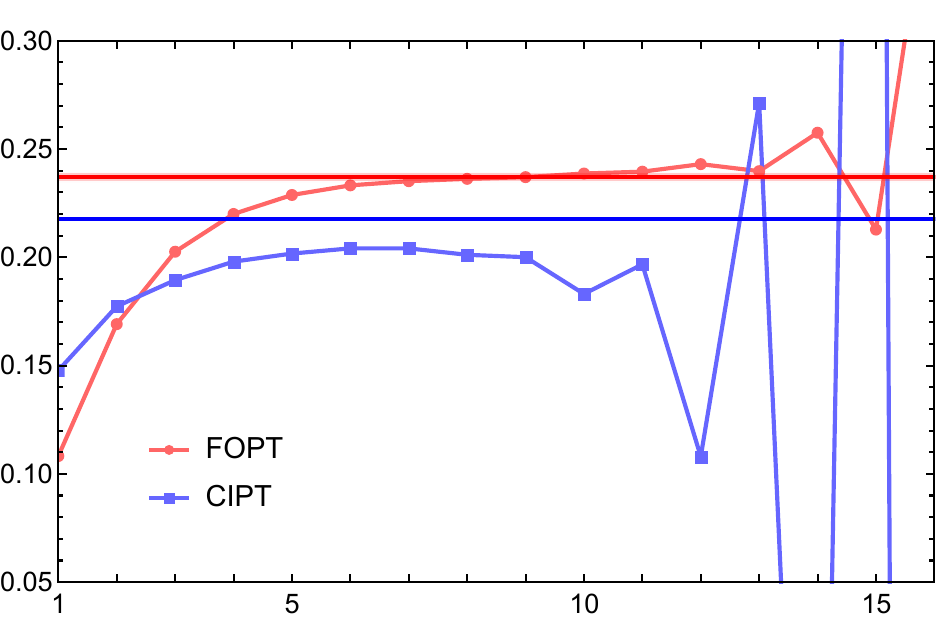}
		\caption{\label{fig:Rtau53} $\delta^{(0)}_{W_\tau}(m_\tau^2)$, $B_{\hat D,{\rm mr}}$, $\alpha_s^{\overline{\rm MS}}$, full $\beta$-function}
	\end{subfigure}
    ~
	\begin{subfigure}[b]{0.48\textwidth}
		\includegraphics[width=\textwidth]{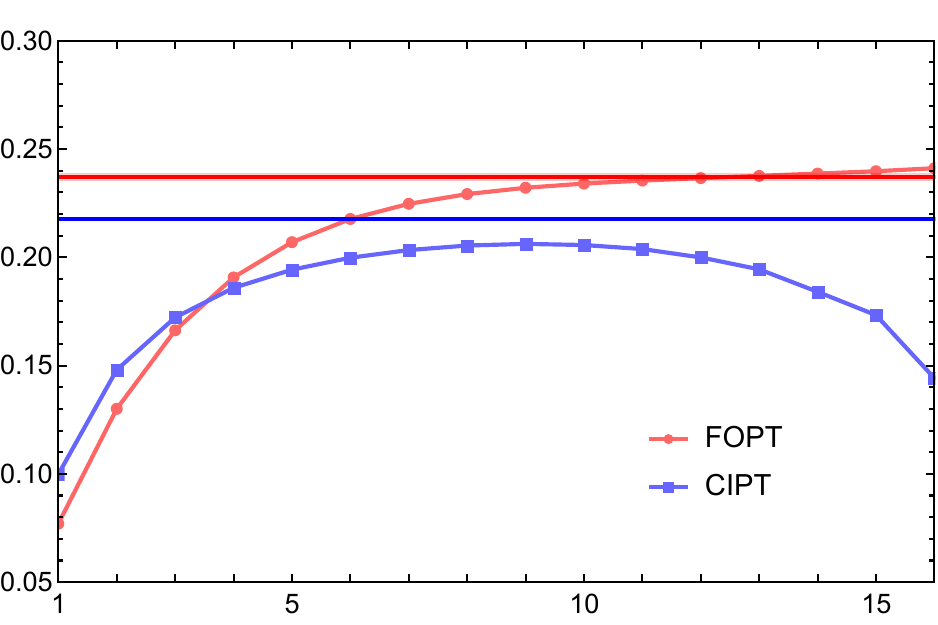}
		\caption{\label{fig:Rtau103} $\delta^{(0)}_{W_\tau}(m_\tau^2)$, $B_{\hat D,{\rm mr}}$, $\alpha_s^{(5/3)}$, full $\beta$-function}
	\end{subfigure}
    \begin{subfigure}[b]{0.48\textwidth}
       \includegraphics[width=\textwidth]{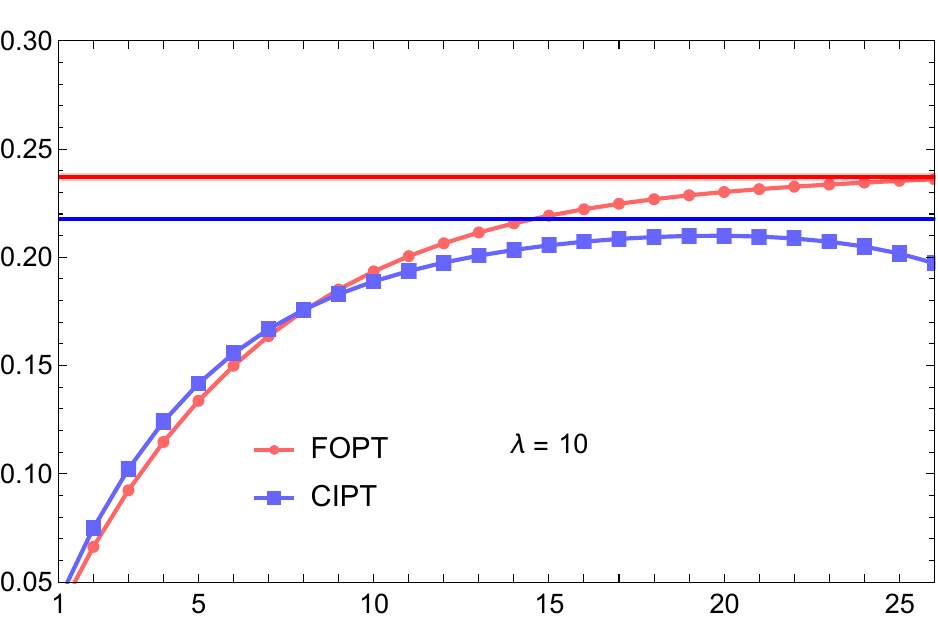}
       \caption{\label{fig:Rtau253} $\delta^{(0)}_{W_\tau}(m_\tau^2)$, $B_{\hat D,{\rm mr}}$, $\alpha_s^{(25/3)}$, full $\beta$-function}
    \end{subfigure}
    ~
    \begin{subfigure}[b]{0.48\textwidth}
       \includegraphics[width=\textwidth]{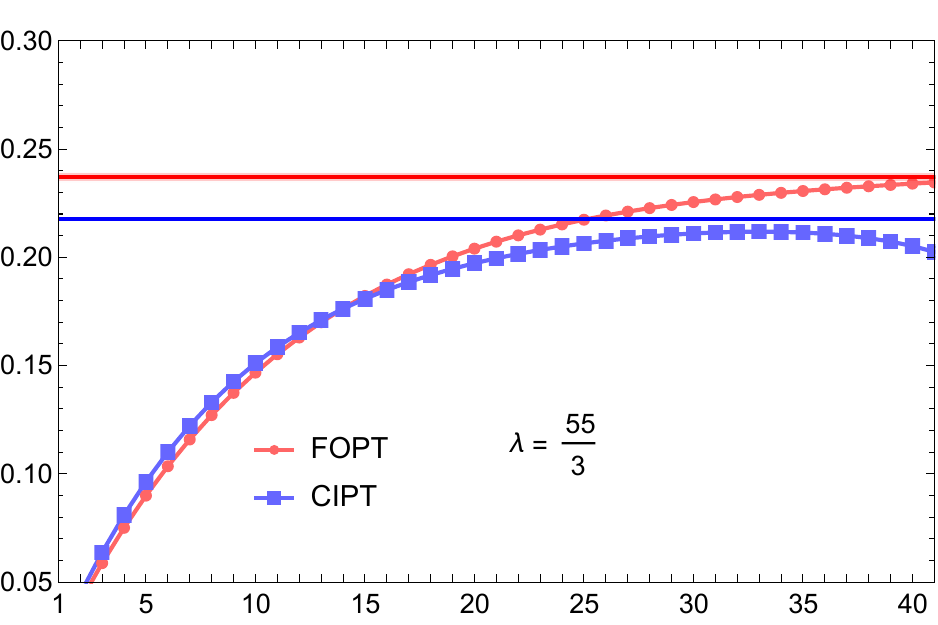}
       \caption{\label{fig:Rtau503} $\delta^{(0)}_{W_\tau}(m_\tau^2)$, $B_{\hat D,{\rm mr}}$, $\alpha_s^{(50/3)}$, full $\beta$-function}
    \end{subfigure}

	\caption{\label{fig:Rtau} Moments $\delta^{(0),{\rm FOPT}}_{W_\tau}(m_\tau^2)$ (red) and  $\delta^{(0),{\rm CIPT}}_{W_\tau}(m_\tau^2)$ (blue)  
	based on the multirenormalon Borel model $B_{\hat D,{\rm mr}}(u)$ accounting for the 5-loop QCD $\beta$-function as a function of the order up to which the series are summed.  Panel (a) shows the result in the $\overline{\mbox{MS}}$ scheme and panel (b) in a scheme where the coupling is defined by $a^{(5/3)}(x) = a(x)/(1+5/3\, a(x))$. Panels (c) and (d) refer to schemes where the coupling is defined by $a^{(25/3)}(x) = a(x)/(1+5\times5/3\, a(x))$ and $a^{(50/3)}(x) = a(x)/(1+10\times5/3\, a(x))$, respectively. The red and blue horizonal lines represent the Borel sums of the FOPT and CIPT series, respectively, and the red bands indicate the conventional Borel ambiguity of the FOPT series.}
\end{figure}

\begin{figure} 
	\centering
	\begin{subfigure}[b]{0.48\textwidth}
		\includegraphics[width=\textwidth]{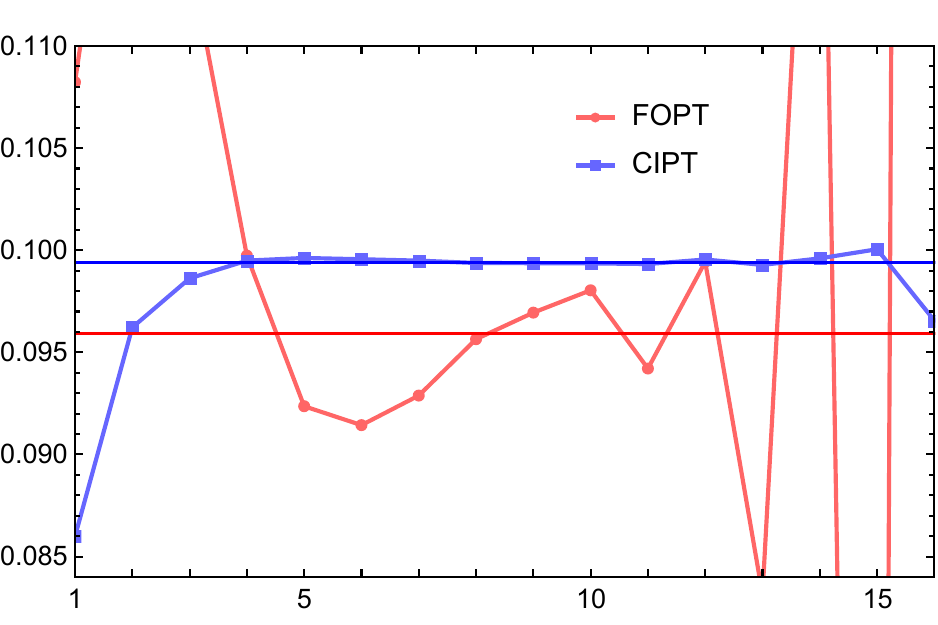}
		\caption{\label{fig:Rtau0} $\delta^{(0)}_{1}(m_\tau^2)$, $B_{\hat D,{\rm mr}}$, $\alpha_s^{\overline{\rm MS}}$, full $\beta$-function}
	\end{subfigure}
	~ 
	\begin{subfigure}[b]{0.48\textwidth}
		\includegraphics[width=\textwidth]{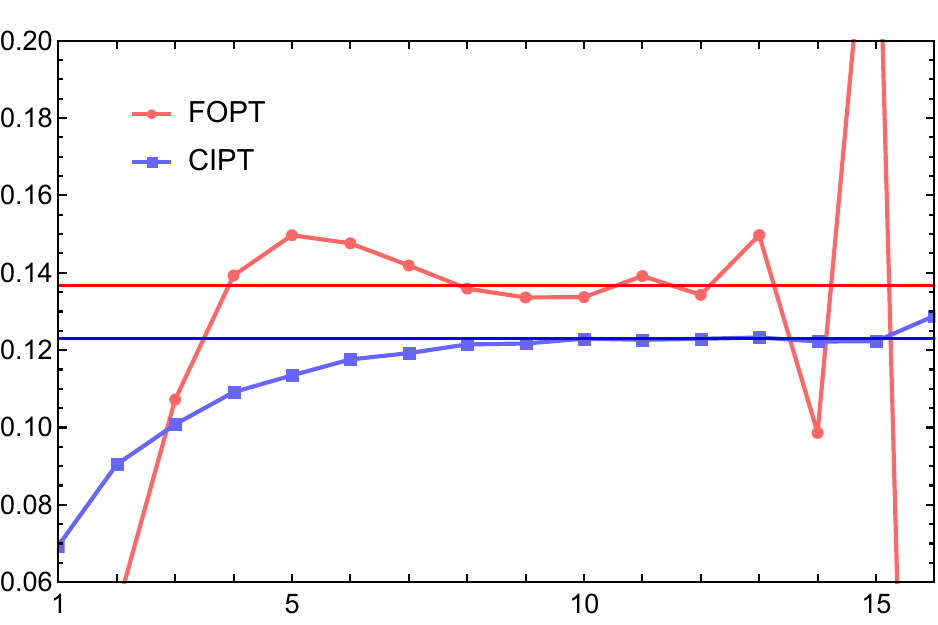}
		\caption{\label{fig:Rtau1} $\delta^{(0)}_{-2x}(m_\tau^2)$, $B_{\hat D,{\rm mr}}$, $\alpha_s^{\overline{\rm MS}}$, full $\beta$-function}
	\end{subfigure}
	
	\begin{subfigure}[b]{0.48\textwidth}
		\includegraphics[width=\textwidth]{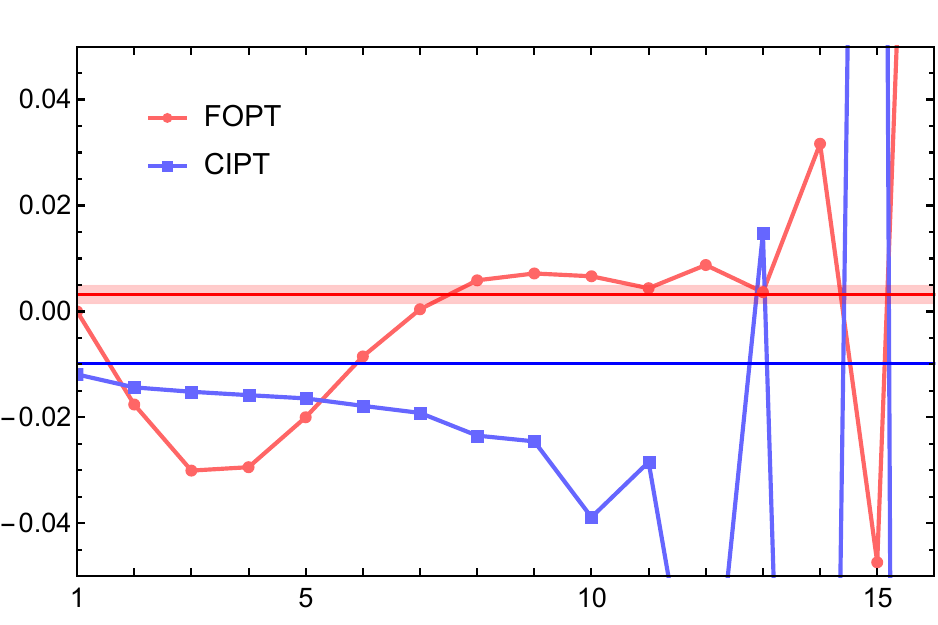}
		\caption{\label{fig:Rtau3}  $\delta^{(0)}_{2x^3}(m_\tau^2)$, $B_{\hat D,{\rm mr}}$, $\alpha_s^{\overline{\rm MS}}$, full $\beta$-function}
	\end{subfigure}
	~
	\begin{subfigure}[b]{0.48\textwidth}
		\includegraphics[width=\textwidth]{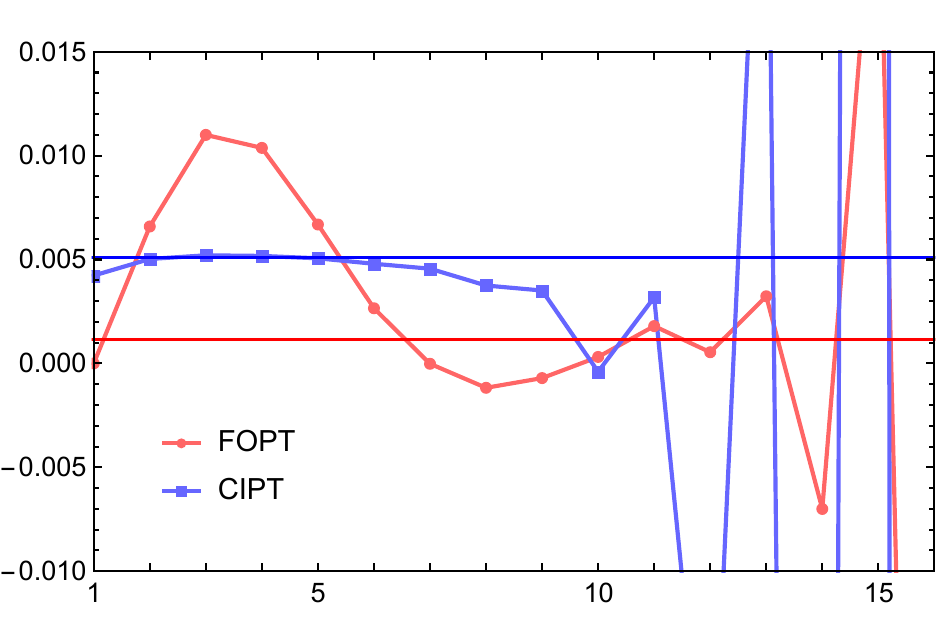}
		\caption{\label{fig:Rtau4}  $\delta^{(0)}_{-x^4}(m_\tau^2)$, $B_{\hat D,{\rm mr}}$, $\alpha_s^{\overline{\rm MS}}$, full $\beta$-function}
	\end{subfigure}
	\caption{\label{fig:Rtaubreakdown} 
		Moments $\delta_{W(x)}^{(0),{\rm FOPT}}(m_\tau^2)$ (red) and $\delta_{W(x)}^{(0),{\rm CIPT}}(m_\tau^2)$ (blue) based on the multirenormalon Borel model $B_{\hat D,{\rm mr}}(u)$ for $W(x)=1$, $-2x$, $2x^3$ and $-x^4$ accounting for the 5-loop QCD $\beta$-function as a function of the order up to which the series are summed. The red and blue horizonal lines represent the Borel sums of the FOPT and CIPT series, respectively, and the red bands indicate the conventional Borel ambiguity of the FOPT series. }
\end{figure}

The outcome for the hadronic tau decay width series  $\delta^{(0)}_{W_\tau}$ in FOPT (red) and CIPT (blue) for $\alpha_s$ in the $\overline{\mbox{MS}}$ scheme and using the multirenormalon model $B_{\hat D,{\rm mr}}(u)$ is shown in Fig.~\ref{fig:Rtau53} as a function of the truncation order using the label conventions from Sec.~\ref{sec:large-beta}. The FOPT and CIPT Borel sums are represented by the red and blue horizontal lines, respectively, and the FOPT Borel sum ambiguity is shown as the narrow (almost invisible) red band. The corresponding numerical values (for Fig.~\ref{fig:Rtau53} and all other figures in this section) are collected in Tab.~\ref{tab:5loopbeta}.
We see that the FOPT series evolves within its narrow ambiguity band for orders $6\lesssim  n \lesssim 11$. The CIPT series approaches its Borel sum, obtained from the FOPT Borel sum plus the asymptotic separation, reaching a minimal distance for orders $n=6$ and $7$, but then bounces back before the oscillating behavior caused by the UV renormalon sets in for $n\gtrsim 10$. The minimal distance is about half of the asymptotic separation. The reason for this behavior can again be traced back to the interplay of contributions in the CIPT series coming from the different polynomial terms in the weight function $W_\tau(x)=1-2x+2x^3-x^4$. The individual contributions of the four polynomial terms is shown in Figs.~\ref{fig:Rtaubreakdown}. We see that for $W(x)=1$ the CIPT series essentially equals its Borel sum for the large range of orders $4 \lesssim m \lesssim 13$. For $W(x)=-2x$ it approaches the Borel sum from below and reaches it for orders $9 \lesssim m \lesssim 13$. For $W(x)=2x^3$ on the other hand, the series is close to its Borel sum only for $n=1$ and then continually drops down until the UV renormalon behavior sets in for $n\gtrsim 10$. Finally, for $W(x)=-x^4$ the series is close to the Borel sum for orders $n\lesssim 7$ and only drops slightly before the  UV renormalon behavior sets in for $n\gtrsim 10$. We see that the asymptotic separation quantifies the different asymptotic behavior of the FOPT and CIPT series very well individually, but the region of order where the CIPT series approach the respective Borel sums closely differs. Since it just so happens that the CIPT series for each polynomial weight function are always below their Borel values
for orders $n\lesssim 10$, the net effect is that the complete CIPT series cannot come very close to its Borel sum before the  UV renormalon behavior sets in. This bounce back behavior is very similar to the corresponding observation we already discussed for the large-$\beta_0$ approximation in Sec.~\ref{sec:large-beta-TauWidth}, but it is more pronounced in full QCD and when the complete QCD $\beta$-function is accounted for. 
Interestingly, like for the large-$\beta_0$ approximation, we find that this feature can be diminished when a different scheme for the strong coupling is adopted. 

In Fig.~\ref{fig:Rtau103} the results for the FOPT and CIPT series are shown in the scheme for the strong coupling defined by $a^{(5/3)}(x) = a(x)/(1+5/3\,a(x))$ so that $\alpha_s^{(5/3)}(m_\tau^2)=0.24185$. The respective Borel sums, the asymptotic separation and the FOPT Borel sum ambiguity are unchanged (as they must by construction), but, in full analogy to the large-$\beta_0$ approximation, the regions of closest approach to the respective Borel sums and the onset of the UV renormalon behavior are pushed toward higher orders. We also observe that the CIPT series approaches its Borel sum more closely. These features become even more pronounced, when we consider the series when the strong couplings are defined by $a^{(25/3)}(x) = a(x)/(1+25/3\,a(x))$ (with $\alpha_s^{(25/3)}(m_\tau^2)=0.11224$) and $a^{(50/3)}(x) = a(x)/(1+50/3\,a(x))$  (with $\alpha_s^{(50/3)}(m_\tau^2)=0.06721$), which are displayed in Figs.~\ref{fig:Rtau253} and \ref{fig:Rtau503}, respectively. Such kind of extreme schemes are of little practical use, since the resulting series converge rather slowly, but we can adopt them within a particular model where all orders can be determined.  
We thus again find that the bounce back feature of the CIPT series is a scheme-dependent issue and that the asymptotic separation also applies for the kinematic weight function relevant for $R_\tau$. The fact that the maximum of the CIPT series, which one may interpret as the best possible approximation to its true value depends on the scheme for the strong coupling adds one additional unusual and remarkable property to the CIPT expansion method that may cast some doubts on its consistency.

Furthermore, the results shown in Figs.~\ref{fig:Rtau} and \ref{fig:Rtaubreakdown} illustrate very clearly that the asymptotic separation is much larger than the FOPT Borel sum ambiguity, showing that our observations made in the large-$\beta_0$ approximation apply also in full QCD. As we demonstrate in the next subsection, this feature turns out to be true for any model with a sizeable $p=2$ gluon condensate renormalon cut. The results show, that only if the Adler function's Borel function indeed has a sizeable $p=2$ gluon condensate renormalon cut, the asymptotic separation can provide an explanation for the disparity between the FOPT-CIPT discrepancy problem for $R_\tau$.

\begin{table}[h]
	\center
	\begin{tabular}{|c|c|c|c|r|r|r|r|}
		\hline
		\multicolumn{8}{|c|}{$5$-loop $\beta_0$-function}\\
		\hline
		$B(u)$ & $W(x)$ & Figure & Scheme &  \multicolumn{1}{|c|}{$\delta^{(0),{\rm FOPT}}_{\rm Borel}$}  &  \multicolumn{1}{|c|}{$\delta^{(0),{\rm CIPT}}_{\rm Borel}$} &
		\multicolumn{1}{|c|}{$\delta^{\rm FOPT}$}  & \multicolumn{1}{|c|}{$\Delta$}  \\
		\hline
		$B_{\hat D,{\rm mr}}$ & $W_\tau$  & \ref{fig:Rtau53} & $\overline{\mbox{MS}}$ & 0.237055 & 0.217754 &  0.001781 & -0.019301\\
		$e^{\frac{5}{3} u}B_{\hat D,{\rm mr}}$  & $W_\tau$ & \ref{fig:Rtau103} & $\alpha_s^{(5/3)}$ & 0.237055 & 0.217754 &  0.001781 & -0.019301\\
		$e^{\frac{25}{3} u}B_{\hat D,{\rm mr}}$ & $W_\tau$  & \ref{fig:Rtau253} & $\alpha_s^{(25/3)}$ & 0.237055 & 0.217754 &  0.001781 & -0.019301\\
		$e^{\frac{50}{3} u}B_{\hat D,{\rm mr}}$  & $W_\tau$ & \ref{fig:Rtau503} & $\alpha_s^{(50/3)}$ & 0.237055 & 0.217754 &  0.001781 & -0.019301\\
		$B_{\hat D,{\rm mr}}$  & $1$ & \ref{fig:Rtau0} & $\overline{\mbox{MS}}$ & 0.095924 & 0.099407 & -0.000029 & 0.003484\\
		$B_{\hat D,{\rm mr}}$  & $-2x$ & \ref{fig:Rtau1} & $\overline{\mbox{MS}}$ & 0.136775 & 0.123065 & 0.000103 & -0.013710\\
		$B_{\hat D,{\rm mr}}$  & $2x^3$ & \ref{fig:Rtau3} & $\overline{\mbox{MS}}$ & 0.003205 & -0.009829  & 0.001738 & -0.013034\\
		$B_{\hat D,{\rm mr}}$  & $-x^4$ & \ref{fig:Rtau4} & $\overline{\mbox{MS}}$ & 0.001152 & 0.005111  & -0.000031 & 0.003959\\
	   $B_{\hat D,{p=2}}$ & $W_\tau$ & \ref{fig:p2wtau} & $\overline{\mbox{MS}}$ & 0.224200 & 0.212073 & 0.000034 & -0.012127\\
		$B_{\hat D,{p=3}}$ & $W_\tau$ & \ref{fig:p3wtau} & $\overline{\mbox{MS}}$ & 0.202604 & 0.201964 & -0.003809 & -0.000640 \\
		$B_{\hat D,{p=2}}$& $(-x)^2$ & \ref{fig:p2m2} & $\overline{\mbox{MS}}$ & -0.002891 & -0.002891 & 0.004035 & 0\\
		$B_{\hat D,{p=3}}$ &  $(-x)^3$ & \ref{fig:p3m3} & $\overline{\mbox{MS}}$ & 0.010218 & 0.010218 & 0.001909 & 0\\
		$B_{\hat D,{\rm mr}}$  & $W_{c=-1}$ & \ref{fig:cm1} & $\overline{\mbox{MS}}$ & 0.277555 & 0.278899 & 0.023019 & 0.001343\\
		$B_{\hat D,{\rm mr}}$  & $W_{c=0}$& \ref{fig:c0} & $\overline{\mbox{MS}}$ & 0.219556 & 0.219821 & 0.011102 &  0.000265\\
		$B_{\hat D,{\rm mr}}$  &  $W_{c=0.75}$ & \ref{fig:c05} & $\overline{\mbox{MS}}$ & 0.176056 & 0.175513 & 0.002164 & -0.000544\\
		$B_{\hat D,{\rm mr}}$  &  $W_{c=1}$ & \ref{fig:c1} & $\overline{\mbox{MS}}$ & 0.161557 & 0.160743 & -0.000816 & -0.000814\\
		\hline
	\end{tabular}
	\caption{\label{tab:5loopbeta} Numerical values for the FOPT Borel sum $\delta^{(0),{\rm FOPT}}_{\rm Borel}$, the CIPT Borel sum $\delta^{(0),{\rm CIPT}}_{\rm Borel}$, the FOPT Borel sum ambiguity $\delta^{\rm FOPT}$ and the asymptotic separation $\Delta=\delta^{(0),{\rm CIPT}}_{\rm Borel}-\delta^{(0),{\rm FOPT}}_{\rm Borel}$  accounting for the full 5-loop $\beta$-function for the analyses in Sec.~\ref{sec:realistic}.}
\end{table}

\subsection{Single-Renormalon Borel Models}
\label{sec:singlerenormalon}

\begin{figure} 
	\centering
	\begin{subfigure}[b]{0.48\textwidth}
		\includegraphics[width=\textwidth]{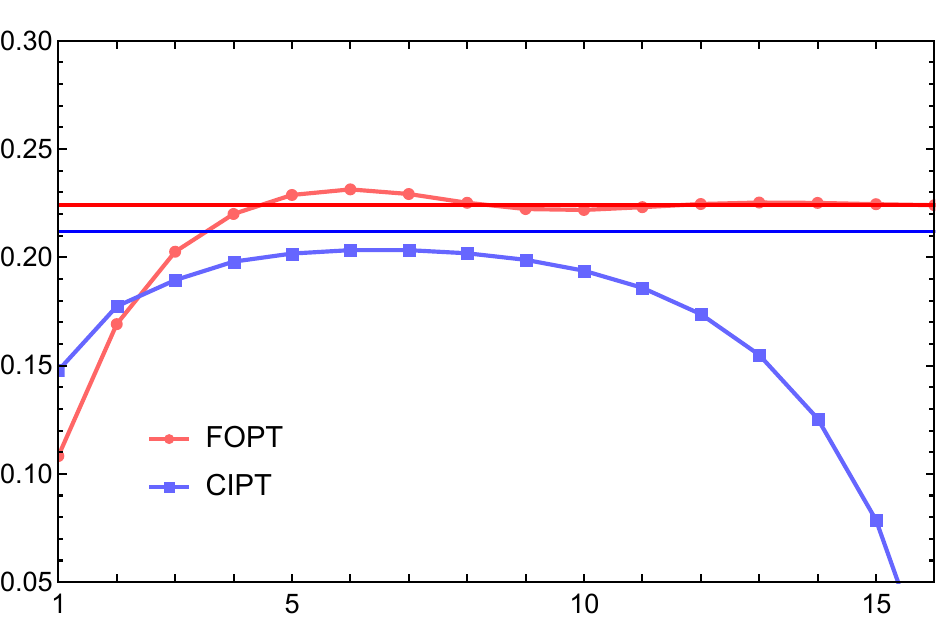}
		\caption{\label{fig:p2wtau} $\delta^{(0)}_{W_\tau}(m_\tau^2)$, $B_{\hat D,p=2}$, $\alpha_s^{\overline{\rm MS}}$, full $\beta$-function}
	\end{subfigure}
	\begin{subfigure}[b]{0.48\textwidth}
		\includegraphics[width=\textwidth]{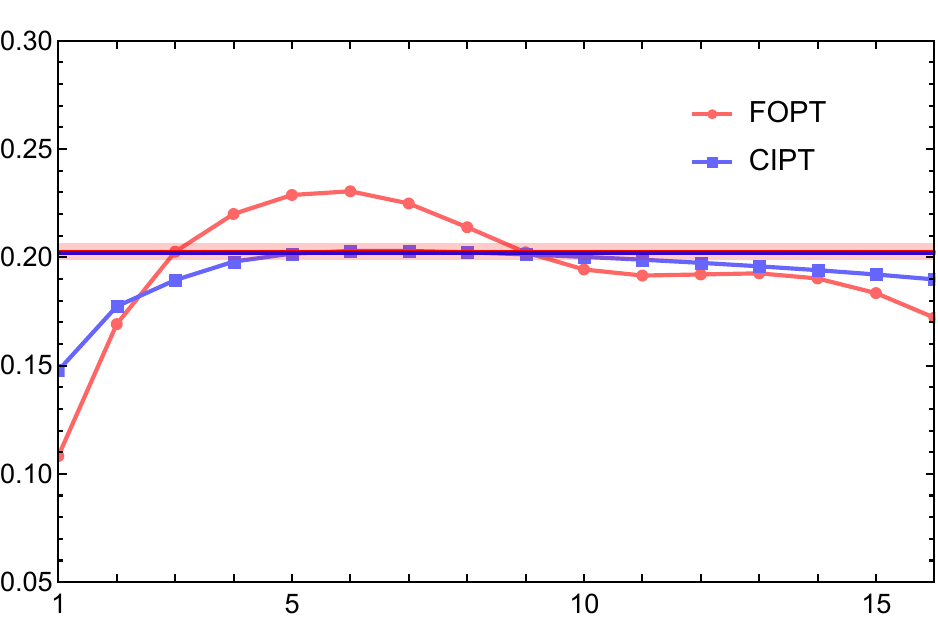}
		\caption{\label{fig:p3wtau} $\delta^{(0)}_{W_\tau}(m_\tau^2)$, $B_{\hat D,p=3}$, $\alpha_s^{\overline{\rm MS}}$, full $\beta$-function}
	\end{subfigure}
	
	\begin{subfigure}[b]{0.48\textwidth}
		\includegraphics[width=\textwidth]{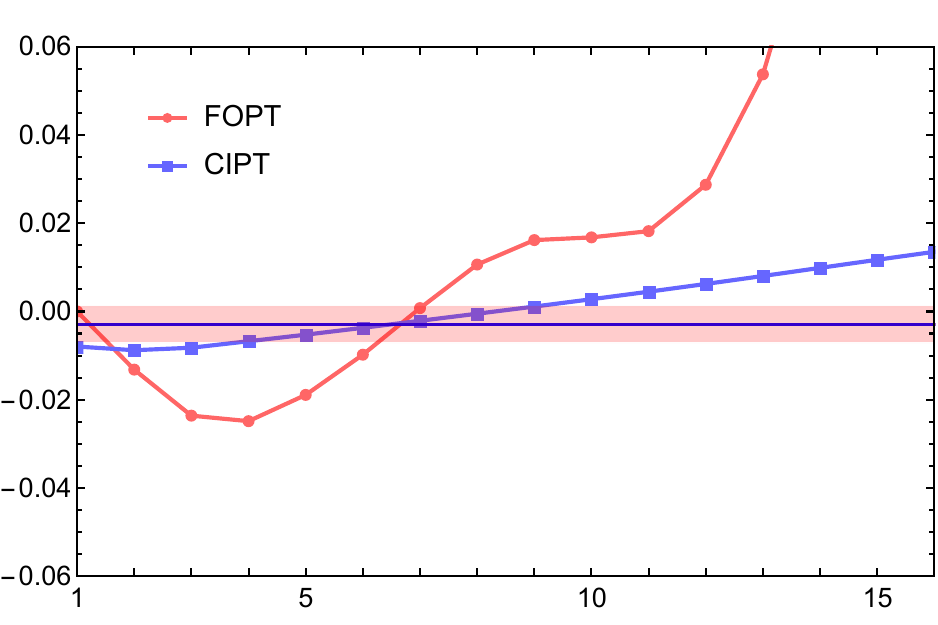}
		\caption{\label{fig:p2m2} $\delta^{(0)}_{(-x)^2}(m_\tau^2)$, $B_{\hat D,p=2}$, $\alpha_s^{\overline{\rm MS}}$, full $\beta$-function}
	\end{subfigure}
	~
	\begin{subfigure}[b]{0.48\textwidth}
		\includegraphics[width=\textwidth]{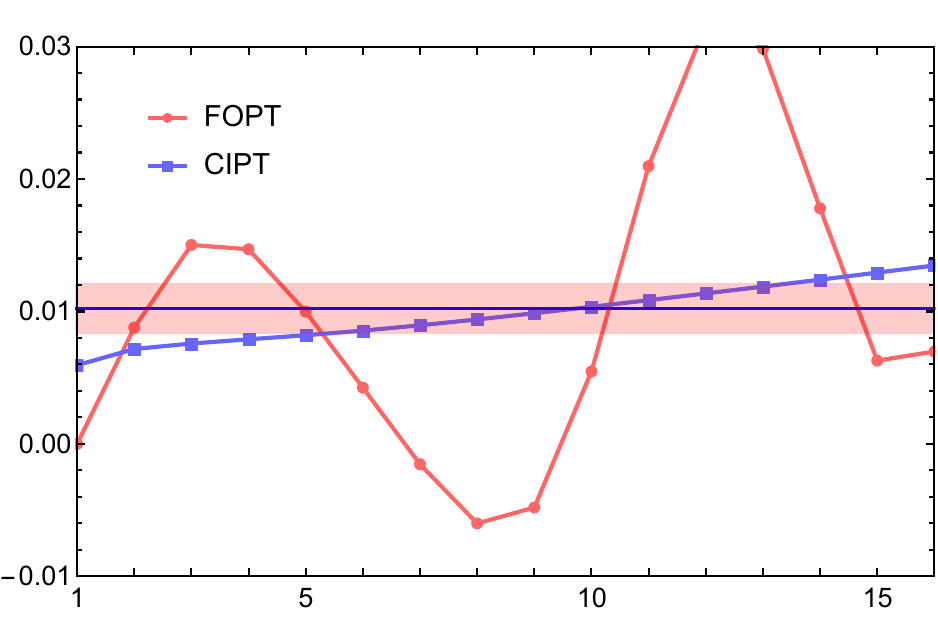}
		\caption{\label{fig:p3m3} $\delta^{(0)}_{(-x)^3}(m_\tau^2)$, $B_{\hat D,p=3}$, $\alpha_s^{\overline{\rm MS}}$, full $\beta$-function}
	\end{subfigure}
	\caption{\label{fig:p23models} 
		Moments $\delta_{W_\tau}(m_\tau^2)^{(0),{\rm FOPT}}(m_\tau^2)$ (red) and $\delta_{W_\tau}^{(0),{\rm CIPT}}(m_\tau^2)$ (blue) based on the single renormalon Borel models $B_{\hat D,p=2}(u)$ (panel (a)) and $B_{\hat D,p=3}(u)$ (panel (b)) accounting for the 5-loop QCD $\beta$-function as a function of the order up to which the series are summed. Panel (c) shows the two series based on $B_{\hat D,p=2}(u)$ with $W(x)=(-x)^2$ and panel (d) shows the two series based on $B_{\hat D,p=3}(u)$ with $W(x)=(-x)^3$. The red and blue horizonal lines represent the Borel sums of the FOPT and CIPT series, respectively, and the red bands indicate the conventional Borel ambiguity of the FOPT series. }
\end{figure}

Let us now discuss (in the $\overline{\rm MS}$ scheme) two Borel models involving only single IR renormalon contributions, one based on a generic Borel function for a $p=2$ (dimension-4 gluon condensate) renormalon and one for a $p=3$ (dimension-6 condensate) renormalon. In both cases we assume that the Wilson coefficients of the associated OPE terms are unity and have vanishing anomalous dimensions. The models are constructed by first fixing the coefficient of the generic Borel functions such that the $6$-loop reduced Adler function coefficient is reproduced exactly and then by adding a third order polynmial in $u$ such that the lower order coefficients are reproduced as well. We refer to these Borel models as $B_{\hat D,p=2}(u)$ and $B_{\hat D,p=3}(u)$, and their explicit expressions are shown in Eqs.~(\ref{eq:modelp2}) and (\ref{eq:modelp3}), respectively. Both account for the QCD $\beta$-function at the $5$-loop level. 
These models are interesting because the numerical impact of the asymptotic separation for a $p=3$ renormalon substantially differs from that of a $p=2$ renormalon, and because we can once more study the case $m=p$ where we adopt the practical definition that the asymptotic separation vanishes. The study visualizes that the asymptotic separation is numerically sizeable and thus practically relevant only if the Borel function of the Adler function indeed has a sizeable gluon condensate cut, as we have already pointed out before.
In the following we refer to these models also as the $p=2$ and $p=3$ models, respectively.

In Figs.~\ref{fig:p2wtau} and \ref{fig:p3wtau} the outcome for the FOPT and CIPT series for the hadronic decay rate weight function $W_\tau(x)$ are shown for the $p=2$ and the $p=3$ models, respectively. The results for the $p=2$ model are quite similar to the multirenormalon model discussed in Sec.~\ref{sec:Rtaumultirenormalon}, apart from the fact that the oscillating high-order behavior due to a UV renormalon exhibited in the multirenormalon model is absent. This illustrates the dominance of the $p=2$ renormalon in the multirenormalon model for low orders. In particular, the asymptotic separation is again much larger than the FOPT ambiguity band, and the CIPT series again remains below its Borel sum. The result shows that these features are inherent properties of the FOPT and CIPT series if the renormalon associated to the gluon condensate term in the Adler function Borel function is sizeable. 

For the $p=3$ model the FOPT Borel sum is much smaller than for the $p=2$ model and compatible with the CIPT series  for the low orders $5\lesssim n \lesssim 9$. This behavior is caused by the small size of the coefficients $c_{n>5,1}$ predicted by the $p=3$ model in contrast to the $p=2$ (or the multirenormalon) model. It was shown in Ref.~\cite{Beneke:2008ad} that all models with a strongly suppressed (or vanishing) gluon condensate renormalon norm have this property.
This situation is nicely described by the asymptotic separation as well. Here, due to the absence of the gluon condensate renormalon, the asymptotic separation is tiny and an order of magnitude smaller than the width of the FOPT ambiguity band for a $p=3$ renormalon. So in this case the FOPT and CIPT Borel sums both are well within the FOPT Borel sum ambiguity and essentially agree.
The strong suppression of the asymptotic separation in comparison to the $p=2$ model arises from its power behavior  $\sim\Lambda_{\rm QCD}^6/s_0^3$ and due to cancellations among the contributions from the different $m$ values. Interestingly, for the $p=3$ IR renormalon model the FOPT series does not behave very well, in the sense that it is far away from its Borel sum at the 5-loop level and that very large higher order corrections arise for orders beyond. Eventually, these higher order corrections bring the FOPT series in agreement with its Borel sum and the CIPT series. The results demonstrate that -- if the Adler function's true Borel function indeed has a strongly suppressed gluon condensate renormalon -- 
the asymptotic separation still exists, but is very tiny and not relevant from the practical perspective. In this situation, the discrepancy between the FOPT-CIPT discrepancy problem is unrelated to the IR properties of both expansion methods and may be interpreted as an accidental feature of the perturbative coefficients and the truncation order. However, it was argued in Ref.~\cite{Beneke:2012vb} that the scenario of a strongly suppressed gluon condensate renormalon normalization in full QCD is not plausible as the order-behavior of the known coefficients $c_{n<5,1}$ are fully compatible with a sizeable norm value. 

In Figs.~\ref{fig:p2m2}  and \ref{fig:p3m3}  the outcome for the FOPT and CIPT series for the $p=2$ and $p=3$ renormalon models are shown for the weight functions $W(x)=(-x)^p$, where we have defined the asymptotic separation to vanish and the FOPT and CIPT Borel sums agree identically. Here, the FOPT series are quite unstable and do not show any stability regime. The CIPT series in both cases shows a slow and steady linear rise and the Borel sum is well within this linear regime. We have confirmed that this behavior is generic for any series generated by a generic $p$ renormalon for the weight function $W(x)=(-x)^p$. This confirms once more that our practical definition of a vanishing asymptotic separation for $m=p$ is also adequate in the context of full QCD due to the instability of the FOPT and CIPT moment series.

\subsection{Moments with Small Asymptotic Separation}
\label{sec:smallseparation}

Having predictive control over the difference of the Borel sums of the FOPT and CIPT series, it is obvious that this may provide a powerful theoretical tool to design moments where the asymptotic separation is going to be small, even in the case that the Adler function's Borel function contains a sizeable gluon condensate cut. Such a tool was absent prior to this article, and using it in connection with the already known constraints related to suppressing the effects of the OPE and duality violation corrections in Eq.~(\ref{eq:momdef}) may uncover new classes of spectral function moments useful for phenomenological analyses and high-precision strong coupling determinations. 
As we have said earlier, it is a priori impossible to construct moments where the asymptotic separation is guaranteed to vanish exactly since the Borel function of the Adler function is unkown (beyond the large-$\beta_0$ approximation). However, we have also seen from the examinations above that the gluon condensate renormalon provides the dominant numerical contribution to the asymptotic separation. So an adequate strategy is to construct moments based on weight functions which ensure that the contribution of the $p=2$ renormalon to the asymptotic separation is suppressed. In the following examination we use the large-$\beta_0$ expressions for the asymptotic separation given in Eq.~(\ref{eq:Sepabeta0}) as a handy analytic identifier for suitable moments. We stress that the main purpose of the following analysis is to demonstrate the practicability of the expressions for the analytic separation and not to provide a thorough phenomenological analysis. 

\begin{figure} 
	\centering
	\begin{subfigure}[b]{0.48\textwidth}
		\includegraphics[width=\textwidth]{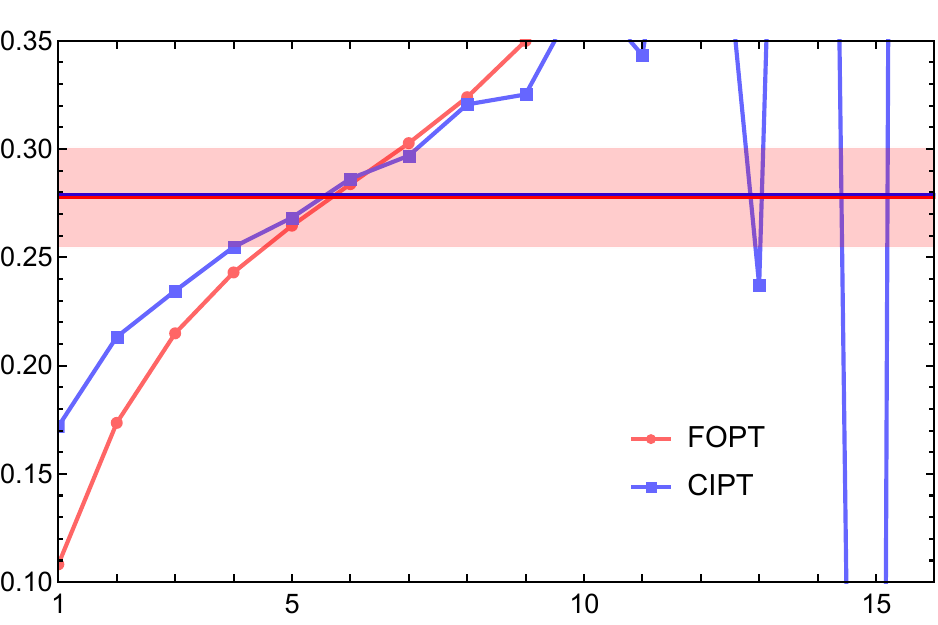}
		\caption{\label{fig:cm1} $\delta^{(0)}_{W_{c=-1}}(m_\tau^2)$, $B_{\hat D,{\rm mr}}$, $\alpha_s^{\overline{\rm MS}}$, full $\beta$-function}
	\end{subfigure}
	~ 
	\begin{subfigure}[b]{0.48\textwidth}
		\includegraphics[width=\textwidth]{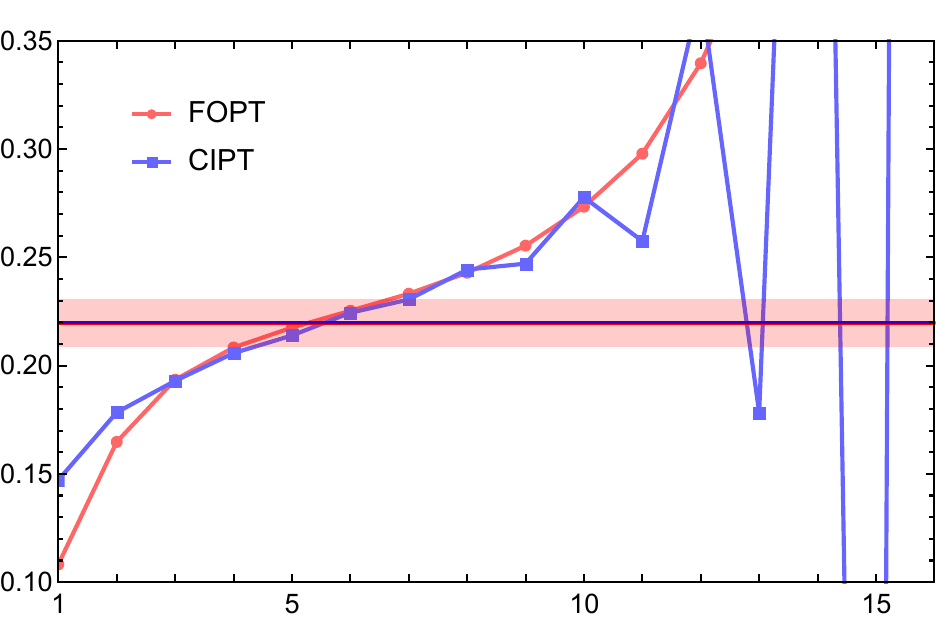}
		\caption{\label{fig:c0}  $\delta^{(0)}_{W_{c=0}}(m_\tau^2)$, $B_{\hat D,{\rm mr}}$, $\alpha_s^{\overline{\rm MS}}$, full $\beta$-function}
	\end{subfigure}
	
	\begin{subfigure}[b]{0.48\textwidth}
		\includegraphics[width=\textwidth]{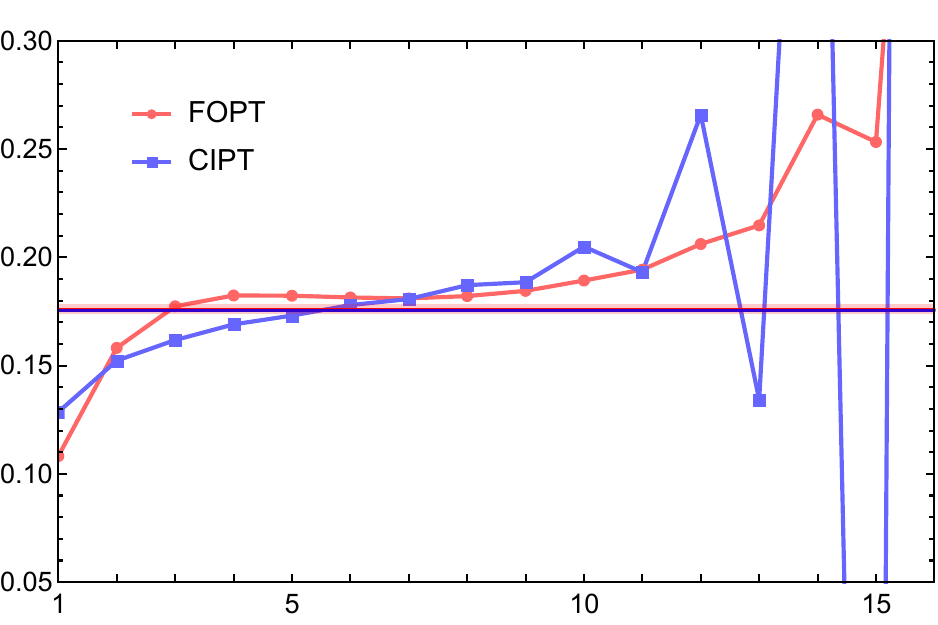}
		\caption{\label{fig:c05}  $\delta^{(0)}_{W_{c=0.75}}(m_\tau^2)$, $B_{\hat D,{\rm mr}}$, $\alpha_s^{\overline{\rm MS}}$, full $\beta$-function}
	\end{subfigure}
	~
	\begin{subfigure}[b]{0.48\textwidth}
		\includegraphics[width=\textwidth]{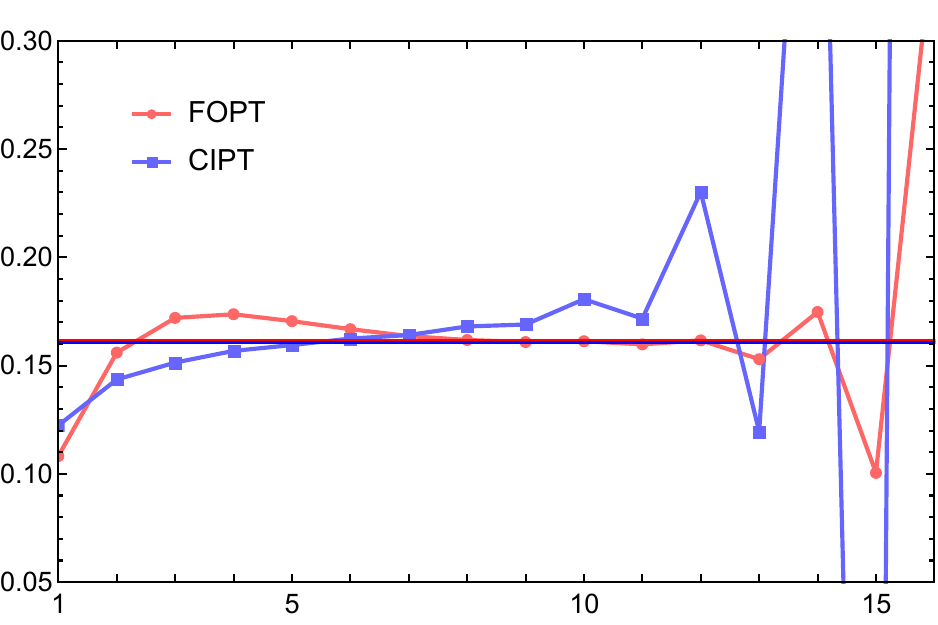}
		\caption{\label{fig:c1}  $\delta^{(0)}_{W_{c=1}}(m_\tau^2)$, $B_{\hat D,{\rm mr}}$, $\alpha_s^{\overline{\rm MS}}$, full $\beta$-function}
	\end{subfigure}
	\caption{\label{fig:magicmoments} 
		Moments $\delta_{W_c}^{(0),{\rm FOPT}}(m_\tau^2)$ (red) and $\delta_{W_c}^{(0),{\rm CIPT}}(m_\tau^2)$ (blue) ) based on the multirenormalon Borel model $B_{\hat D,{\rm mr}}(u)$ for $W_c=(1-x)^2(1+c x+x^2)$ for $c=-1,0,0.75,1$ accounting for the 5-loop QCD $\beta$-function as a function of the order up to which the series are summed. The red and blue horizonal lines represent the Borel sums of the FOPT and CIPT series, respectively, and the red bands indicate the conventional Borel ambiguity of the FOPT series. }
\end{figure}

Let us consider exemplarily (in the $\overline{\rm MS}$ scheme) the polynomial weight function $W_c(x)=(1-x)^2(1+c x+x^2)$, which is pinched (i.e.\ has a double zero at $x=1$) and leads to spectral function moments with vanishing net asymptotic separation from the $p=2$ gluon condensate renormalon in the large-$\beta_0$ approximation for any value of $c$. The value of $c$ modulates the contribution of the $x^2$ term in $W_c(x)$ and thus the size of the FOPT Borel sum ambiguity, which is minimal for $c$ values close to unity. In Fig.~\ref{fig:magicmoments}  the results for the FOPT (red) and CIPT (blue) spectral function moment series arising from the multirenormalon Borel model 
of Sec.~\ref{sec:Rtaumultirenormalon} are shown for $c=-1$, $0$, $0.75$ and $1$. The colored horizonal lines represent the FOPT and CIPT Borel sums (which are almost on top of each other) and the red band indicates the FOPT Borel sum ambiguity. Again the corresponding numerical values for the Borel sums, the FOPT Borel ambiguity and the asymptotic separation are collected in Tab.~\ref{tab:5loopbeta}. 
Comparing to Fig.~\ref{fig:Rtau} where the series for the hadronic tau decay rate with the weight function $W_\tau(x)$ are displayed\footnote{For easier visual comparison we have displayed the same range of values on the $y$-axes in all panels of Figs.~\ref{fig:Rtau} and \ref{fig:magicmoments}.}, the FOPT and CIPT series for the weight function $W_c(x)$ are substantially closer as long as $c<1$. Furthermore, for $c\lesssim 0$ and orders $4\lesssim n \lesssim 7$ both series provide essentially equivalent descriptions. For $c=1$ the FOPT and CIPT series show some moderate discrepancy for orders $2<n<5$ and are compatible with their Borel sum only at order $n=7$. For all values of $c$ the asymptotic separation is strongly suppressed and ranges in size between $0.0002$ and $0.001$. On the other hand, the FOPT Borel sum ambiguity amounts to sizeable $0.023$ for $c=-1$ and continually decreases down to $-0.0008$ for $c=1$ where it agrees in size with the asymptotic separation. From the numbers shown in Tab.~\ref{tab:5loopbeta} we see that for $c=0.75$ the FOPT Borel sum ambiguity and the asymptotic separation amount to $1.2\%$  and $0.3\%$, respectively, of the Borel sum value. 
Overall, the results confirm the effectiveness of the asymptotic separation for the targeted design of moments where the FOPT and CIPT series show a more consistent intermediate and high order behavior than the kinematic moment $R_\tau$.

\section{Implications and the Operator Product Expansion}
\label{sec:implications}

In Sec.~\ref{sec:borelspace} we noted that the analytic expression in Eq.~(\ref{eq:IRBorelIntCIPT}), which arises generically in the CIPT Borel representation of Eq.~(\ref{eq:BorelCIPT}) when carrying out the $\bar u$ integration first, 
may be interpreted such that the Borel sum of the (perturbative) Adler function $\hat D(s)$ in powers of $\alpha_s(-s)$ may have an unphysical cut along the negative real $s$-axis. However, this interpretation 
is not compulsory because the contour integration is an integral part of the CIPT spectral function moments and thus also of the CIPT Borel representation, see Sec~\ref{sec:Borelrepresentation}. So the form of the CIPT Borel representation does a priori not make any statement of the Borel sum of the Adler function in the complex plane.
Nevertheless, it is an interesting question to which extent the structure of the CIPT Borel representation, which differs from that of the FOPT Borel representation in Eq.(\ref{eq:BorelFOPT}), may be relevant for the Borel sum of the Adler function $\hat D(s)$ for which one may either adopt an expansion in powers of $\alpha_s(-s)$ or in powers of $\alpha_s(s_0)=\alpha_s(|s|)$. This is what we intend to explore in this section.
We emphasize that the following discussion does not affect the results we have discussed in the previous sections on the spectral function moments.  

At this point, let us step back from the spectral function moments and have a closer look at the perturbative Adler function $\hat D(s)$ in the complex $s$-plane, which is the underlying quantity from which the spectral function moment series are derived. 
The structures of the FOPT and CIPT spectral function moment Borel representations imply the following two definitions for Borel sums for the perturbative Adler function:	
\begin{align}
\label{eq:AdlerBorelFOPT}
\hat D_{\rm Borel}^{\rm FOPT}(s) = &\,{\rm PV}\int_{0}^\infty \!\! {\rm d} u \,\,  B[\hat D](u)\,e^{-\frac{4\pi u}{\beta_0\alpha_s(-s)}} \,,\\
\label{eq:AdlerBorelCIPT}
\hat D_{\rm Borel}^{\rm CIPT}(s) = &\,\int_0^\infty \!\! {\rm d} \bar u \,\,
\big({\textstyle \frac{\alpha_s(-s)}{\alpha_s(|s|)}}\big)\,
B[\hat D]\Big({\textstyle \frac{\alpha_s(-s)}{\alpha_s(|s|)}} \bar u\Big)\,e^{-\frac{4\pi \bar u}{\beta_0\alpha_s(|s|)}}\,, 
\end{align}
We label these two Borel sums as ``FOPT" and ``CIPT" only due to their correspondence to Eqs.~(\ref{eq:BorelFOPT}) and 
(\ref{eq:BorelCIPT}), respectively, but of course no contour integration is carried out anywhere. 
The expression $\hat D_{\rm Borel}^{\rm FOPT}(s)$ is the traditional formula for the Borel sum of the Adler function that has been used in many previous studies. The expression for $\hat D_{\rm Borel}^{\rm CIPT}(s)$ is new.

The expression for $\hat D_{\rm Borel}^{\rm CIPT}(s)$ is already well-defined everywhere in the complex $s$ plane except along the negative real $s$-axis.
It is equivalent to the Borel integral $\int_0^\infty \!\! {\rm d}u  B[\hat D](u)\,e^{-\frac{4\pi u}{\beta_0\alpha_s(-s)}}$ using the regularization prescription, where the integration path is deformed above the real $u$-axis for ${\rm Im}[\alpha_s(-s)]={\rm Im}[s]>0$ (path~1b in Fig.~\ref{fig:u-contourIR}) and below the real $u$-axis for ${\rm Im}[\alpha_s(-s)]={\rm Im}[s]<0$ (path~1a in Fig.~\ref{fig:u-contourIR}). The traditional Borel sum $\hat D_{\rm Borel}^{\rm FOPT}(s)$ imposes the PV prescription arising from the average of paths~1a and 1b, to be well-defined. As was already shown in Sec.~\ref{sec:borelspace},
for a generic IR renormalon Borel function contribution with the form $B^{\rm IR}_{\hat D,p,\gamma}(u) = 1/(p-u)^\gamma$, this leads to the following generic contributions for the Borel sum of the Adler function:  
\begin{align}
\label{eq:IRBorelDFOPTsum}
\hat D_{p,\gamma}^{\rm FOPT}(s) \, = \, & 
-\Big({\textstyle -\frac{\alpha_s(-s)\beta_0}{4\pi}}\Big)^{1-\gamma}\,\Gamma\Big(1-\gamma,{\textstyle - \frac{4\pi\,p}{\alpha_s(-s)\beta_0}}\Big)\,e^{-\frac{4\pi\,p}{\alpha_s(-s)\beta_0}} \\ \nonumber  &
\quad - {\rm sig}[{\rm Im}[s]]\,\frac{i\pi}{\Gamma(\gamma)}\,\Big({\textstyle \frac{\alpha_s(-s)\beta_0}{4\pi}}\Big)^{1-\gamma}\,
e^{-\frac{4\pi\,p}{\alpha_s(-s)\beta_0}}\,,\\
\label{eq:IRBorelDCIPTsum}
\hat D_{p,\gamma}^{\rm CIPT}(s) \, = \, &
-\Big({\textstyle -\frac{\alpha_s(-s)\beta_0}{4\pi}}\Big)^{1-\gamma}\,\Gamma\Big(1-\gamma,{\textstyle - \frac{4\pi\,p}{\alpha_s(-s)\beta_0}}\Big)\,e^{-\frac{4\pi\,p}{\alpha_s(-s)\beta_0}}\,.
\end{align}
The expression for $\hat D_{p,\gamma}^{\rm FOPT}(s)$ satisfies the Schwartz reflection principle (i.e.\ $(\hat D_{p,\gamma}^{\rm FOPT}(s))^*=\hat D_{p,\gamma}^{\rm FOPT}(-s)$) and is analytic except along the positive real $s$-axis, where it has a cut, and at the Landau pole located on the negative real $s$-axis. It thus has the analytic properties one also expects from the physical Adler function for perturbative values of $s$.
The expression for $\hat D_{p,\gamma}^{\rm CIPT}(s)$ satisfies the Schwartz reflection principle as well and has the same nonanalytic structures as $\hat D_{p,\gamma}^{\rm FOPT}(s)$. But it exhibits in addition a cut along the negative real $s$-axis, as we have already pointed in in Sec..~\ref{sec:borelspace}.
	
Since the physical Adler function is known be analytic along the Euclidean negative real $s$-axis, this may be taken as a reason to dismiss $\hat D_{p,\gamma}^{\rm CIPT}(s)$ as a viable generic term in the Adler function's Borel sum. However, this unphysical cut is $\Lambda_{\rm QCD}^{2p}/s_0^p$ power suppressed, so that it could in principle be compensated by OPE corrections that do not have the standard form shown in Eq.~(\ref{eq:DOPE}), but exhibit an analogous cut. We therefore continue with our considerations.

The question we would like to explore is, how well the two definitions for the Borel sum describe the actual behavior of the Adler function series in the complex plane when either expanded in powers of $\alpha_s(-s)$ or in powers of $\alpha_s(s_0)=\alpha_s(|s|)$:
\begin{eqnarray}
\label{eq:AdlerseriesFOPT2}
\hat D^{\rm FOPT}(s) & \, = \, &
\, \sum\limits_{n=1}^\infty\,
\big({\textstyle \frac{\alpha_s(s_0)}{\pi}}\big)^n \, \sum\limits_{k=1}^{n} k\, c_{n,k}\,\ln^{k-1}({\textstyle \frac{-s}{s_0}}) \,.\\
\label{eq:AdlerseriesCIPT2}
\hat D^{\rm CIPT}(s) & \, =  \,& \, \sum\limits_{n=1}^\infty
c_{n,1} \,\big({\textstyle \frac{\alpha_s(-s)}{\pi}}\big)^n\,, 
\end{eqnarray}
We label these two expansions as ``FOPT" and ``CIPT", but we do it merely due to their correspondence to the expansions for the FOPT and CIPT spectral function moments. We also remind the reader that both series are identical along the Euclidean axis $s=-|s|$. 
Since we can apply the manipulations in Sec.~\ref{sec:Borelrepresentation} on the FOPT moment series also for the most part without accounting for contour integration, we can show that $\hat D_{\rm Borel}^{\rm FOPT}(s)$ is the proper Borel representation for the `FOPT' series in powers of $\alpha_s(s_0)$.
But in the absence of the contour integration (which provided unique criteria which contributions should be regarded as part of the series coefficients and which variable should be regarded as the expansion parameter for the CIPT moment series) we do not have compelling arguments which of the Borel representations is appropriate for the `CIPT' expansion in powers of $\alpha_s(-s)$. At the perturbative level, the $u$ and $\bar u$ Taylor expansions of Eqs.~(\ref{eq:AdlerBorelFOPT}) and (\ref{eq:AdlerBorelCIPT}) both lead to the `CIPT' series in Eq.~(\ref{eq:AdlerseriesCIPT2}).
So let us compare the intermediate order behavior of the `FOPT' and `CIPT' expansions numerically with the Borel sums based on the prescriptions shown in Eqs.~(\ref{eq:AdlerBorelFOPT}) and (\ref{eq:AdlerBorelCIPT}) and using $\hat D_{p,\gamma}^{\rm FOPT}(s)$ and $\hat D_{p,\gamma}^{\rm CIPT}(s)$, respectively, for the Borel sums arising from the IR renormalons. (We recall that the Borel sums for UV renormalons are identical, see Sec.~\ref{sec:borelspace}).
We use the multirenormalon Borel function model already employed in Sec.~\ref{sec:Rtaumultirenormalon} (and given in Eq.~(\ref{eq:modelmr})), following the arguments of Ref.~\cite{Beneke:2008ad,Beneke:2012vb} that it provides a realistic estimate on the qualiative high-order behavior for the perturbative Adler function.

\begin{figure} 
	\centering
	\begin{subfigure}[b]{0.48\textwidth}
		\includegraphics[width=\textwidth]{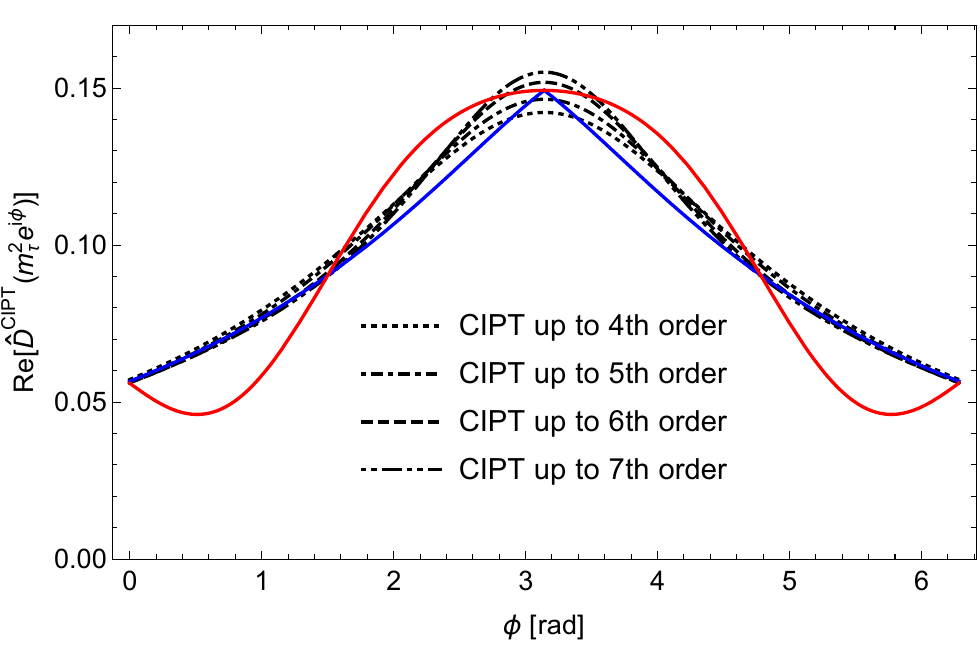}
		\caption{\label{fig:AdlerCIPTRe} }
	\end{subfigure}
	~ 
	\begin{subfigure}[b]{0.48\textwidth}
		\includegraphics[width=\textwidth]{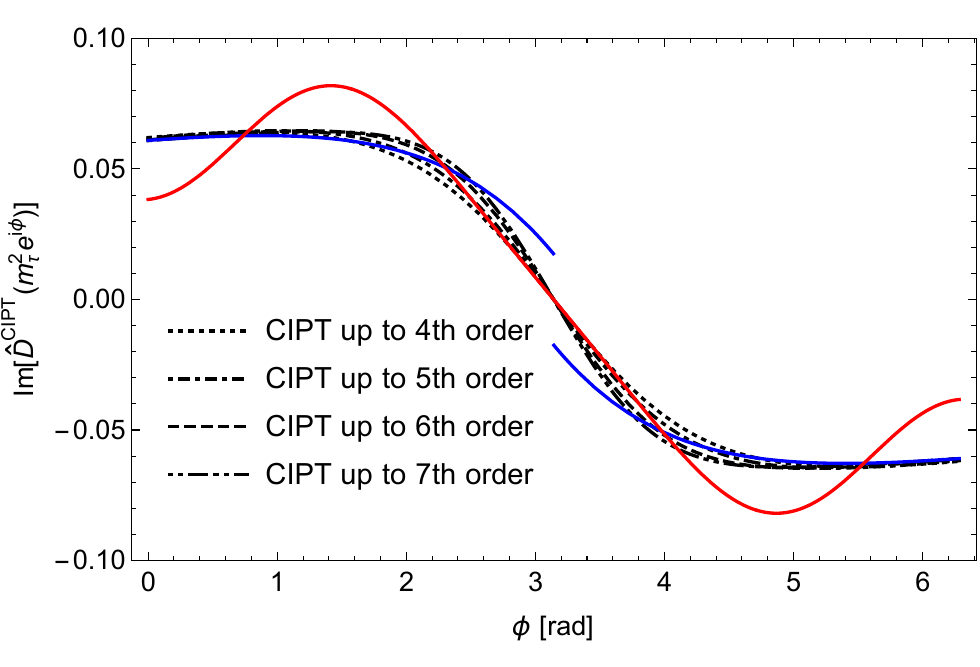}
		\caption{\label{fig:AdlerCIPTIm} }
	\end{subfigure}
	
	\begin{subfigure}[b]{0.48\textwidth}
		\includegraphics[width=\textwidth]{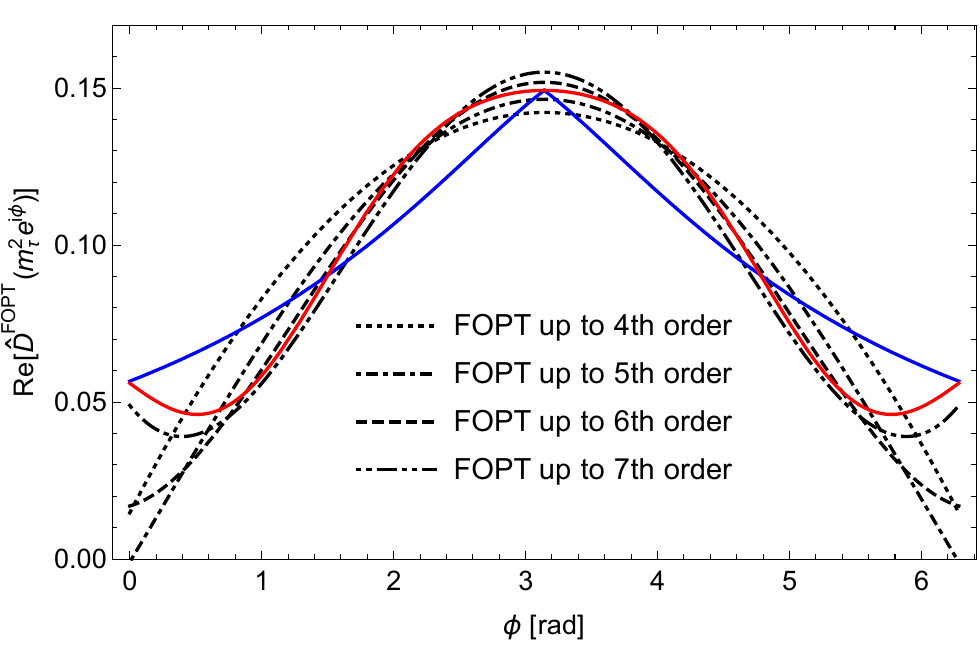}
		\caption{\label{fig:AdlerFOPTRe} }
	\end{subfigure}
	~
	\begin{subfigure}[b]{0.48\textwidth}
		\includegraphics[width=\textwidth]{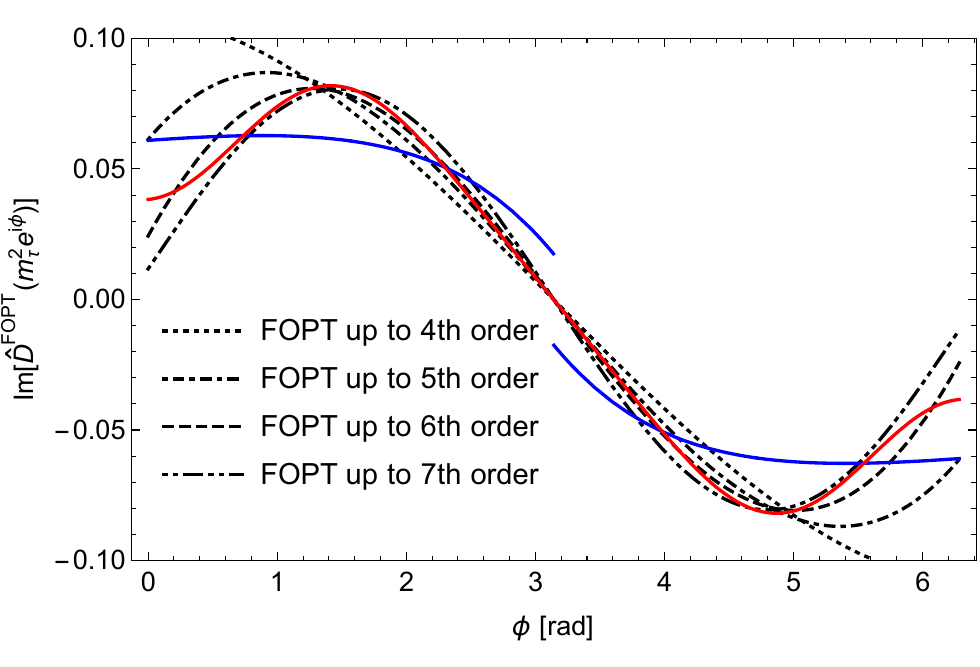}
		\caption{\label{fig:AdlerFOPTIm} }
	\end{subfigure}
	\caption{\label{fig:AdlerfctBorel} 
		Real and imaginary parts of the Adler function series $\hat D^{\rm CIPT}(m_\tau^2 e^{i\phi})$ and $\hat D^{\rm FOPT}(m_\tau^2 e^{i\phi})$ for a summation of the perturbative series up to ${\cal O}(\alpha_s^4)$ (black dotted), ${\cal O}(\alpha_s^5)$ (black dot-dashed ), ${\cal O}(\alpha_s^6)$ (black dashed) and ${\cal O}(\alpha_s^7)$ (wide-dot-dashed) as a function of $\phi$ for $\alpha_s(m^2_\tau)=0.34$. The results beyond the 4th order are obtained from the Borel model given in 
		Eq.~(\ref{eq:modelmr}). Also displayed are the corresponding CIPT (blue) and FOPT (red) Borel sums.		
		}
\end{figure}

In Fig.~\ref{fig:AdlerfctBorel} the real (left panels) and imaginary parts (right panels) of the Adler function $\hat D(s)$ for $s=m_{\tau}^2 e^{i \phi}$ are shown as a function of $\phi$ at the 4th (5-loop) to the 7th (8-loop) order\footnote{This range of order covers the exactly known 5-loop result, and the `CIPT' and `FOPT' series both
show a converging behavior.} (black lines) in the `CIPT' (upper panels) and `FOPT' (lower panels) expansions compared to the CIPT (blue) and FOPT (red) Borel sums based on the generic expressions from Eqs.~(\ref{eq:IRBorelDCIPTsum}) and (\ref{eq:IRBorelDFOPTsum}) for the renormalon contributions in the Borel model function. 

The `CIPT' series shows a very well converging behavior particularly close to the positive real $s$-axis ($\phi=0,2\pi$). But it is of course still an asymptotic series that diverges eventually. The `FOPT' series, on the other hand, is comparatively unstable close to the positive real $s$-axis. Along the negative real Euclidean $s$-axis  ($\phi=\pi$), where the Adler function is real-valued, both series are identical, but everywhere else the `CIPT' and the `FOPT' series appear to disagree and to approach different functions. This behavior for the same kind of model has been subject to a dedicate analysis in Refs.~\cite{Beneke:2008ad} (see also Ref.~\cite{Davier:2008sk}), but in their analysis only the traditional FOPT Borel sum was available for comparison.
We see that the `CIPT' series agrees very well with the CIPT Borel sum 
for the real as well as the imaginary part everywhere in the complex plane except for a narrow region close to the Euclidean negative real $s$-axis ($0.8\pi\lesssim \phi\lesssim 1.2\pi$). Close to the Euclidean axis the real parts of the CIPT Borel sum and the `CIPT' series agree quite well, but the CIPT Borel sum's imaginary part exhibits the unphysical cut we have mentioned before, while the imaginary part of the `CIPT' series is continuous and approaches zero at $\phi=\pi$ smoothly. Even though the imaginary part of the `CIPT' series has a zero at any order on the Euclidean axis, it should be noted that its slope at the zero increases strongly at higher orders. It is therefore not excluded that the cut in the Borel sum is compatible with the `CIPT' series expansion. This may deserve further investigations, which is, however, beyond the scope of this work.

Overall, apart from the disagreement concerning the imaginary part close to the negative real $s$-axis, it is clearly visible that the `CIPT' series behavior is in a much better agreement with the CIPT Borel sum than with the FOPT Borel sum. It is an interesting fact that in the computation of the CIPT Borel sum for spectral function moments the region close to the negative real $s$-axis is avoided due to the contour deformation. So the region where the disagreement arises does not contribute in the CIPT moment Borel representation. 
As far as the `FOPT' series is concerned, even though it is less stable compared to the `CIPT' series close to the positive real $s$-axis, it clearly shows -- as anticipated -- much better overall agreement with the FOPT Borel sum than with the CIPT Borel sum.
The observations are the same for larger values of $|s|$ where the `CIPT' and the `FOPT' Adler function series are more stable due to the smaller value of the strong coupling: The differences between `FOPT' and `CIPT' is still visible, but the overall difference between both expansions and the two Borel sums decreases with the factor $e^{-\frac{8\pi}{\alpha_s(-s)\beta_0}}\sim \Lambda_{\rm QCD}^4/s^2$. We remind the reader, that this scaling behavior arises from the existence of the sizeable gluon condensate renormalon cut contained in the Borel model, which dominates the discrepancy between both types of expansions. These observations are the same for any Borel model that contains a sizeable gluon condensate cut.

The argumentations for the existence of the asymptotic separation for the FOPT and CIPT spectral function moment expansions rely strictly on the contour integration over the polynomial weight functions. The inconsistency of the CIPT moment series expansion with the standard form of the OPE corrections for the Adler function shown in Eq.~(\ref{eq:DOPE}) is imperative and can be clearly demonstrated at the level of the spectral function moments, which include the contour integration as an essential ingredient. Numerically, the asymptotic separation and the CIPT inconsistency are dominated by the gluon condensate IR renormalon yielding an overall scaling of these effects $\sim\Lambda_{\rm QCD}^4/s_0^2$. 
From the observations in this section, however, we find that it is not excluded that the FOPT-CIPT
discrepancy for the moments series may already arise at the level of the perturbative expansions of the Adler function $\hat D(s)$ in the complex plane
either in powers of $\alpha_s(s_0)$ or $\alpha_s(-s)$.
If true, this would imply that it is the expansion either in powers of $\alpha_s(s_0)$ or $\alpha_s(-s)$ that may already generate a different IR sensitivity. The fact that the difference in the two types of expansions shown in Figs.~\ref{fig:AdlerfctBorel} exhibits the analogous scaling  $\sim\Lambda_{\rm QCD}^4/s^2$ provides a strong evidence in favor of this view. This would entail, that the OPE corrections of the standard form in Eq.~(\ref{eq:DOPE}) may not apply to the perturbative expansion of the Adler function in powers of $\alpha_s(-s)$. At this point, the situation is, however, not conclusive without further conceptual insights.

\section{Conclusions}
\label{sec:conclusions}

The observation that the FOPT and CIPT perturbation series for some hadronic $\tau$ decay spectral function moments such as the total decay width each seem to approach systematically different values has been a long-standing problem in the literature. So far neither a fully satisfactory resolution of this matter nor an analytic quantification of the observed disparity has been identified.
In this article we have shown that the Borel representations of the FOPT and CIPT series expansions for the spectral function moments in general differ. The difference is related to an intrinsic differing analytic regularization concerning the IR renormalon cuts contained in the Borel function of the underlying Adler function that arises from the contour integration over the moments weight function. In other words, the FOPT and CIPT series expansions inherently correspond to a different arrangement of IR sensitive contributions associated to the divergent structures related to IR renormalons. We have shown that this difference, which we call the ``asymptotic separation", can be analytically calculated for any concrete Borel function model.
The asymptotic separation is proportional to the exponential of the inverse strong coupling and thus power-suppressed in accordance to the OPE operator dimension the IR renormalon is associated to. We have shown for concrete models of the Adler function's Borel function that the
differing asymptotic behavior of the FOPT and CIPT series at intermediate orders, where a definite series value is approached, is described very well by the asymptotic separation. 
We also found that the asymptotic separation is only numerically sizeable, if the Adler function contains a gluon condensate renormalon with a sizeable normalization. For other higher-dimensional IR renormalons the asymptotic separation exists as well, but it is very small numerically. We demonstrated that one can systematically construct spectral function moments where the asymptotic separation is strongly suppressed, even in the presence of a sizeable gluon condensate cut.

Our findings imply that the OPE power corrections in the context of using either the FOPT and CIPT expansion differ. This difference does not refer to an infrared regularization scheme dependence of the nonperturbative vacuum matrix element in the OPE, but to a non-trivial difference concerning the analytic structure of the OPE corrections. The FOPT expansion is consistent in context of using the standard form of the Adler function's  OPE power corrections, where each OPE term has the generic form  $\frac{1}{(-s)^{d/2}} \sum_i  C(\alpha_s(-s)) \langle  {\cal O}\rangle$, with $\langle  {\cal O}\rangle$ being a dimension-$d$ nonperturbative matrix element, $C$ the Wilson coefficient and $s$ the Adler function's invariant mass. The CIPT expansion is inconsistent with this standard form of the power corrections. One of the most compelling facts demonstrating this peculiar feature of the CIPT expansion method is that 
spectral function moment series can be devised where the gluon condensate OPE corrections and the associated divergent renormalon behavior of the perturbative series are eliminated within the FOPT approach, while the corresponding series in the CIPT expansion is still divergent. This peculiar property of the CIPT expansion method can also be demonstrated from the analytic properties of the CIPT moment's Borel representation, which we have analysed in detail in this work.

The extent to which these findings are phenomenologically relevant for the FOPT and CIPT series for hadronic $\tau$ decay spectral function moments at the 5-loop level depends on whether the Adler function's Borel function has a gluon condensate renormalon with a sizeable normalization. If this normalization is sizeable, the results of this work can provide an explanation for the long-standing FOPT-CIPT moment series discrepancy problem. In any case,
the findings of this article contribute toward a more refined understanding of the conceptual aspects of the FOPT and CIPT expansion approaches for the perturbative $\tau$ hadronic spectral function moments and the practical implementation of OPE power corrections
for strong coupling determinations from hadronic $\tau$ decays.

\section*{Acknowledgments}
We thank Matthias Jamin and Diogo Boito for helpful discussions and comments to the manuscript.
We acknowledge partial support by the FWF Austrian Science Fund under the Doctoral Program ``Particles and Interactions'' No.\ W1252-N27 and under the Project No. P32383-N27. We also thank the Erwin-Schr\"odinger International Institute for Mathematics and Physics for partial support. 
\vspace*{0.3cm}

\appendix

\section{Coefficients}
\label{app:exponential}

\begin{align}
f_1(u,L_x) = &\, - u \, \bar{\beta}_1\,L_x\\
f_2(u,L_x) = &\,  u \, \left[ - \bar{\beta}_2\, L_x \, - \, \frac{1}{2} \, \bar{\beta}_1 \, L_x^2 \right] + \, \frac{1}{2} \, u^2 \, \bar{\beta}_1^2 \, L_x^2  \\
f_3(u,L_x) = &\,  u \, \left[- \bar{\beta}_3\, L_x \, + \, \left(  \frac{1}{2} \, \bar{\beta}_1^2 \, - \, \bar{\beta}_2  \right) L_x^2 \, - \, \frac{1}{3} \, \bar{\beta}_1 \, L_x^3  \right] \\
&\,+ u^2  \left[ \bar{\beta}_1\, \bar{\beta}_2 \, L_x^2 \, + \,  \frac{1}{2} \, \bar{\beta}_1^2 \,  L_x^3  \right]  -  \frac{1}{6} \, u^3 \, \bar{\beta}_1^3 \, L_x^3 \nonumber \\
f_4(u,L_x) = &\,  u \, \left[- \bar{\beta}_4\, L_x \, + \, \frac{3}{2} \, \left(   \bar{\beta}_1 \, \bar{\beta}_2 \, - \, \bar{\beta}_3  \right) L_x^2 \, + \,  \left(  \frac{2}{3} \,  \bar{\beta}_1^2 \, - \, \bar{\beta}_2  \right) L_x^3 \, - \, \frac{1}{4} \, \bar{\beta}_1 \, L_x^4  \right] \nonumber \\
&\,+ u^2 \, \left[ \,  \left( \frac{1}{2} \,  \bar{\beta}_2^2  \, + \, \bar{\beta}_1 \, \bar{\beta}_3  \right) L_x^2  \, - \, \frac{1}{2} \, \bar{\beta}_1 \left(  \bar{\beta}_1^2 \, - \, 3 \, \bar{\beta}_2  \right) L_x^3 \, + \, \frac{11}{24} \, \bar{\beta}_1^2 \, L_x^4  \right] \nonumber \\
&\,+ u^3  \left[ - \, \frac{1}{2} \, \bar{\beta}_1^2 \, \bar{\beta}_2 \, L_x^3 \, - \,  \frac{1}{4} \, \bar{\beta}_1^3 \,  L_x^4  \right]  +  \frac{1}{24} \, u^4 \, \bar{\beta}_1^4 \, L_x^4 \\
\bar{\beta}_n \equiv &\, \frac{\beta_n}{\beta_0^{n+1}} 
\end{align}

\section{Borel Models}
\label{app:models}

\begin{align}
\label{eq:modelmr}
B_{\hat D,{\rm mr}}(u)  \, = \, &\, b^{(0)}_{\rm mr} + b^{(1)}_{\rm mr}\,u \, + \,
N^{(2)}_{\rm mr}\sum_{i=0}^{3} \,a^{(2,i)}_{\rm mr}\,B^{\rm IR}_{\hat D,2,\gamma^{(2,i)}}(u) 
\, + \,
N^{(3)}_{\rm mr}\sum_{i=0}^{3} \,a^{(3,i)}_{3}\,B^{\rm IR}_{\hat D,3,\gamma^{(3,i)}}(u)
\nonumber \\ &
\, + \,
N^{(-1)}_{\rm mr}\sum_{i=0}^{3} \,a^{(-1,i)}_{\rm mr}\,B^{\rm UV}_{\hat D,-1,\tilde\gamma^{(-1,i)}}(u)
\\
\label{eq:modelp2}
B_{\hat D,p=2}(u)  \, = \, &\,\sum_{j=0}^3 b^{(j)}_{2}\,u^j \, + \,
N^{(2)}_2\sum_{i=0}^{3} \,a^{(2,i)}_2\,B^{\rm IR}_{\hat D,2,\gamma^{(2,i)}}(u) 
\\
\label{eq:modelp3}
B_{\hat D,p=3}(u)  \, = \, &\,\sum_{j=0}^3 b^{(j)}_{3}\,u^j \, + \,
N^{(2)}_3\sum_{i=0}^{3} \,a^{(3,i)}_{3}\,B^{\rm IR}_{\hat D,3,\gamma^{(3,i)}}(u) 
\end{align}

\begin{align}
N^{(2)}_{\rm mr} &= 4.4669405   & N^{(3)}_{\rm mr} &= -20.9245713  \quad & N^{(-1)}_{\rm mr} &=-0.023602
\\ \nonumber
N^{(2)}_2 &= 2.6819313  & N^{(2)}_3 &= 56.5347352
\end{align}
\begin{align}
\gamma^{(2,0)} &= \frac{209}{81}  & \gamma^{(2,1)} &=  \frac{128}{81} \quad & \gamma^{(2,2)} &=  \frac{47}{81} \quad & \gamma^{(2,3)} &=  -\frac{34}{81}
\\
\gamma^{(3,0)} &= \frac{91}{27}  & \gamma^{(3,1)} &=  \frac{64}{27} \quad & \gamma^{(3,2)} &=  \frac{37}{27} \quad & \gamma^{(3,3)} &=  \frac{10}{27} 
\nonumber\\ 
\tilde \gamma^{(-1,0)} &= \frac{98}{81}  & \tilde \gamma^{(-1,1)} &= \frac{17}{81} \quad & \tilde \gamma^{(-1,2)} &= -\frac{64}{81} \quad & \tilde \gamma^{(-1,3)} &= -\frac{145}{81} 
\nonumber
\end{align}
\begin{align}
a^{(2,0)}_{\rm mr} &=1 & a^{(2,1)}_{\rm mr} &=-0.499518 & a^{(2,2)}_{\rm mr} &=-0.726053 & a^{(2,3)}_{\rm mr} &=-2.154561 
\\
a^{(3,0)}_{3} &=1 & a^{(3,1)}_{3} &=-0.327643 & a^{(3,2)}_{3} &=-0.341418 & a^{(3,3)}_{3} &=1.028346  
\nonumber \\
a^{(-1,0)}_{\rm mr} &=1 & a^{(-1,1)}_{\rm mr} &=-1.233479 & a^{(-1,2)}_{\rm mr} &=-3.037540 & a^{(-1,3)}_{\rm mr} &=-0.173072 
\nonumber\\
a^{(2,0)}_2 &=1 & a^{(2,1)}_2 &=-0.327643 & a^{(2,2)}_2 &=-0.879417 & a^{(2,3)}_2 &=-1.585526 
\nonumber
\end{align}
\begin{align}
b^{(0)}_{\rm mr} &=28.154241 & b^{(1)}_{\rm mr} &=-0.985951
\\
b^{(0)}_2 &=7.555621  & b^{(1)}_2 &=-0.758615 & b^{(2)}_2 =&-0.0807798  &  b^{(3)}_2 &=0.0453495 
\nonumber\\
b^{(0)}_3 &=-33.99899  & b^{(1)}_3 &=-2.981131 & b^{(2)}_3 =&-0.571259  &  b^{(3)}_3 &=-0.0641748 
\nonumber
\end{align}

\bibliography{sources}

\begin{thebibliography}{35}%
\makeatletter
\providecommand \@ifxundefined [1]{%
 \@ifx{#1\undefined}
}%
\providecommand \@ifnum [1]{%
 \ifnum #1\expandafter \@firstoftwo
 \else \expandafter \@secondoftwo
 \fi
}%
\providecommand \@ifx [1]{%
 \ifx #1\expandafter \@firstoftwo
 \else \expandafter \@secondoftwo
 \fi
}%
\providecommand \natexlab [1]{#1}%
\providecommand \enquote  [1]{``#1''}%
\providecommand \bibnamefont  [1]{#1}%
\providecommand \bibfnamefont [1]{#1}%
\providecommand \citenamefont [1]{#1}%
\providecommand \href@noop [0]{\@secondoftwo}%
\providecommand \href [0]{\begingroup \@sanitize@url \@href}%
\providecommand \@href[1]{\@@startlink{#1}\@@href}%
\providecommand \@@href[1]{\endgroup#1\@@endlink}%
\providecommand \@sanitize@url [0]{\catcode `\\12\catcode `\$12\catcode
  `\&12\catcode `\#12\catcode `\^12\catcode `\_12\catcode `\%12\relax}%
\providecommand \@@startlink[1]{}%
\providecommand \@@endlink[0]{}%
\providecommand \url  [0]{\begingroup\@sanitize@url \@url }%
\providecommand \@url [1]{\endgroup\@href {#1}{\urlprefix }}%
\providecommand \urlprefix  [0]{URL }%
\providecommand \Eprint [0]{\href }%
\providecommand \doibase [0]{http://dx.doi.org/}%
\providecommand \selectlanguage [0]{\@gobble}%
\providecommand \bibinfo  [0]{\@secondoftwo}%
\providecommand \bibfield  [0]{\@secondoftwo}%
\providecommand \translation [1]{[#1]}%
\providecommand \BibitemOpen [0]{}%
\providecommand \bibitemStop [0]{}%
\providecommand \bibitemNoStop [0]{.\EOS\space}%
\providecommand \EOS [0]{\spacefactor3000\relax}%
\providecommand \BibitemShut  [1]{\csname bibitem#1\endcsname}%
\let\auto@bib@innerbib\@empty
\bibitem [{\citenamefont {Barate}\ \emph {et~al.}(1998)\citenamefont {Barate}
  \emph {et~al.}}]{Barate:1998uf}%
  \BibitemOpen
  \bibfield  {author} {\bibinfo {author} {\bibfnamefont {R.}~\bibnamefont
  {Barate}} \emph {et~al.} (\bibinfo {collaboration} {ALEPH}),\ }\href
  {\doibase 10.1007/s100520050217} {\bibfield  {journal} {\bibinfo  {journal}
  {Eur. Phys. J. C}\ }\textbf {\bibinfo {volume} {4}},\ \bibinfo {pages} {409}
  (\bibinfo {year} {1998})}\BibitemShut {NoStop}%
\bibitem [{\citenamefont {Schael}\ \emph {et~al.}(2005)\citenamefont {Schael}
  \emph {et~al.}}]{Schael:2005am}%
  \BibitemOpen
  \bibfield  {author} {\bibinfo {author} {\bibfnamefont {S.}~\bibnamefont
  {Schael}} \emph {et~al.} (\bibinfo {collaboration} {ALEPH}),\ }\href
  {\doibase 10.1016/j.physrep.2005.06.007} {\bibfield  {journal} {\bibinfo
  {journal} {Phys. Rept.}\ }\textbf {\bibinfo {volume} {421}},\ \bibinfo
  {pages} {191} (\bibinfo {year} {2005})},\ \Eprint
  {http://arxiv.org/abs/hep-ex/0506072} {arXiv:hep-ex/0506072} \BibitemShut
  {NoStop}%
\bibitem [{\citenamefont {Davier}\ \emph {et~al.}(2014)\citenamefont {Davier},
  \citenamefont {H{\"o}cker}, \citenamefont {Malaescu}, \citenamefont {Yuan},\
  and\ \citenamefont {Zhang}}]{Davier:2013sfa}%
  \BibitemOpen
  \bibfield  {author} {\bibinfo {author} {\bibfnamefont {M.}~\bibnamefont
  {Davier}}, \bibinfo {author} {\bibfnamefont {A.}~\bibnamefont {H{\"o}cker}},
  \bibinfo {author} {\bibfnamefont {B.}~\bibnamefont {Malaescu}}, \bibinfo
  {author} {\bibfnamefont {C.-Z.}\ \bibnamefont {Yuan}}, \ and\ \bibinfo
  {author} {\bibfnamefont {Z.}~\bibnamefont {Zhang}},\ }\href {\doibase
  10.1140/epjc/s10052-014-2803-9} {\bibfield  {journal} {\bibinfo  {journal}
  {Eur. Phys. J. C}\ }\textbf {\bibinfo {volume} {74}},\ \bibinfo {pages}
  {2803} (\bibinfo {year} {2014})},\ \Eprint {http://arxiv.org/abs/1312.1501}
  {arXiv:1312.1501 [hep-ex]} \BibitemShut {NoStop}%
\bibitem [{\citenamefont {Ackerstaff}\ \emph {et~al.}(1999)\citenamefont
  {Ackerstaff} \emph {et~al.}}]{Ackerstaff:1998yj}%
  \BibitemOpen
  \bibfield  {author} {\bibinfo {author} {\bibfnamefont {K.}~\bibnamefont
  {Ackerstaff}} \emph {et~al.} (\bibinfo {collaboration} {OPAL}),\ }\href
  {\doibase 10.1007/s100529901061} {\bibfield  {journal} {\bibinfo  {journal}
  {Eur. Phys. J. C}\ }\textbf {\bibinfo {volume} {7}},\ \bibinfo {pages} {571}
  (\bibinfo {year} {1999})},\ \Eprint {http://arxiv.org/abs/hep-ex/9808019}
  {arXiv:hep-ex/9808019} \BibitemShut {NoStop}%
\bibitem [{\citenamefont {Braaten}\ \emph {et~al.}(1992)\citenamefont
  {Braaten}, \citenamefont {Narison},\ and\ \citenamefont
  {Pich}}]{Braaten:1991qm}%
  \BibitemOpen
  \bibfield  {author} {\bibinfo {author} {\bibfnamefont {E.}~\bibnamefont
  {Braaten}}, \bibinfo {author} {\bibfnamefont {S.}~\bibnamefont {Narison}}, \
  and\ \bibinfo {author} {\bibfnamefont {A.}~\bibnamefont {Pich}},\ }\href
  {\doibase 10.1016/0550-3213(92)90267-F} {\bibfield  {journal} {\bibinfo
  {journal} {Nucl. Phys. B}\ }\textbf {\bibinfo {volume} {373}},\ \bibinfo
  {pages} {581} (\bibinfo {year} {1992})}\BibitemShut {NoStop}%
\bibitem [{\citenamefont {Le~Diberder}\ and\ \citenamefont
  {Pich}(1992)}]{LeDiberder:1992jjr}%
  \BibitemOpen
  \bibfield  {author} {\bibinfo {author} {\bibfnamefont {F.}~\bibnamefont
  {Le~Diberder}}\ and\ \bibinfo {author} {\bibfnamefont {A.}~\bibnamefont
  {Pich}},\ }\href {\doibase 10.1016/0370-2693(92)90172-Z} {\bibfield
  {journal} {\bibinfo  {journal} {Phys. Lett. B}\ }\textbf {\bibinfo {volume}
  {286}},\ \bibinfo {pages} {147} (\bibinfo {year} {1992})}\BibitemShut
  {NoStop}%
\bibitem [{\citenamefont {Boito}\ \emph {et~al.}(2015)\citenamefont {Boito},
  \citenamefont {Golterman}, \citenamefont {Maltman}, \citenamefont {Osborne},\
  and\ \citenamefont {Peris}}]{Boito:2014sta}%
  \BibitemOpen
  \bibfield  {author} {\bibinfo {author} {\bibfnamefont {D.}~\bibnamefont
  {Boito}}, \bibinfo {author} {\bibfnamefont {M.}~\bibnamefont {Golterman}},
  \bibinfo {author} {\bibfnamefont {K.}~\bibnamefont {Maltman}}, \bibinfo
  {author} {\bibfnamefont {J.}~\bibnamefont {Osborne}}, \ and\ \bibinfo
  {author} {\bibfnamefont {S.}~\bibnamefont {Peris}},\ }\href {\doibase
  10.1103/PhysRevD.91.034003} {\bibfield  {journal} {\bibinfo  {journal} {Phys.
  Rev. D}\ }\textbf {\bibinfo {volume} {91}},\ \bibinfo {pages} {034003}
  (\bibinfo {year} {2015})},\ \Eprint {http://arxiv.org/abs/1410.3528}
  {arXiv:1410.3528 [hep-ph]} \BibitemShut {NoStop}%
\bibitem [{\citenamefont {Pich}\ and\ \citenamefont
  {Rodriguez-Sanchez}(2016)}]{Pich:2016bdg}%
  \BibitemOpen
  \bibfield  {author} {\bibinfo {author} {\bibfnamefont {A.}~\bibnamefont
  {Pich}}\ and\ \bibinfo {author} {\bibfnamefont {A.}~\bibnamefont
  {Rodriguez-Sanchez}},\ }\href {\doibase 10.1103/PhysRevD.94.034027}
  {\bibfield  {journal} {\bibinfo  {journal} {Phys. Rev. D}\ }\textbf {\bibinfo
  {volume} {94}},\ \bibinfo {pages} {034027} (\bibinfo {year} {2016})},\
  \Eprint {http://arxiv.org/abs/1605.06830} {arXiv:1605.06830 [hep-ph]}
  \BibitemShut {NoStop}%
\bibitem [{\citenamefont {Shifman}\ \emph {et~al.}(1979)\citenamefont
  {Shifman}, \citenamefont {Vainshtein},\ and\ \citenamefont
  {Zakharov}}]{Shifman:1978bx}%
  \BibitemOpen
  \bibfield  {author} {\bibinfo {author} {\bibfnamefont {M.~A.}\ \bibnamefont
  {Shifman}}, \bibinfo {author} {\bibfnamefont {A.}~\bibnamefont {Vainshtein}},
  \ and\ \bibinfo {author} {\bibfnamefont {V.~I.}\ \bibnamefont {Zakharov}},\
  }\href {\doibase 10.1016/0550-3213(79)90022-1} {\bibfield  {journal}
  {\bibinfo  {journal} {Nucl. Phys. B}\ }\textbf {\bibinfo {volume} {147}},\
  \bibinfo {pages} {385} (\bibinfo {year} {1979})}\BibitemShut {NoStop}%
\bibitem [{\citenamefont {Gross}\ and\ \citenamefont
  {Neveu}(1974)}]{Gross:1974jv}%
  \BibitemOpen
  \bibfield  {author} {\bibinfo {author} {\bibfnamefont {D.~J.}\ \bibnamefont
  {Gross}}\ and\ \bibinfo {author} {\bibfnamefont {A.}~\bibnamefont {Neveu}},\
  }\href {\doibase 10.1103/PhysRevD.10.3235} {\bibfield  {journal} {\bibinfo
  {journal} {Phys. Rev. D}\ }\textbf {\bibinfo {volume} {10}},\ \bibinfo
  {pages} {3235} (\bibinfo {year} {1974})}\BibitemShut {NoStop}%
\bibitem [{\citenamefont {'t~Hooft}(1979)}]{tHooft:1977xjm}%
  \BibitemOpen
  \bibfield  {author} {\bibinfo {author} {\bibfnamefont {G.}~\bibnamefont
  {'t~Hooft}},\ }\href@noop {} {\bibfield  {journal} {\bibinfo  {journal}
  {Subnucl. Ser.}\ }\textbf {\bibinfo {volume} {15}},\ \bibinfo {pages} {943}
  (\bibinfo {year} {1979})}\BibitemShut {NoStop}%
\bibitem [{\citenamefont {David}(1984)}]{David:1983gz}%
  \BibitemOpen
  \bibfield  {author} {\bibinfo {author} {\bibfnamefont {F.}~\bibnamefont
  {David}},\ }\href {\doibase 10.1016/0550-3213(84)90235-9} {\bibfield
  {journal} {\bibinfo  {journal} {Nucl. Phys. B}\ }\textbf {\bibinfo {volume}
  {234}},\ \bibinfo {pages} {237} (\bibinfo {year} {1984})}\BibitemShut
  {NoStop}%
\bibitem [{\citenamefont {Mueller}(1985)}]{Mueller:1984vh}%
  \BibitemOpen
  \bibfield  {author} {\bibinfo {author} {\bibfnamefont {A.~H.}\ \bibnamefont
  {Mueller}},\ }\href {\doibase 10.1016/0550-3213(85)90485-7} {\bibfield
  {journal} {\bibinfo  {journal} {Nucl. Phys. B}\ }\textbf {\bibinfo {volume}
  {250}},\ \bibinfo {pages} {327} (\bibinfo {year} {1985})}\BibitemShut
  {NoStop}%
\bibitem [{\citenamefont {Beneke}(1999)}]{Beneke:1998ui}%
  \BibitemOpen
  \bibfield  {author} {\bibinfo {author} {\bibfnamefont {M.}~\bibnamefont
  {Beneke}},\ }\href {\doibase 10.1016/S0370-1573(98)00130-6} {\bibfield
  {journal} {\bibinfo  {journal} {Phys. Rept.}\ }\textbf {\bibinfo {volume}
  {317}},\ \bibinfo {pages} {1} (\bibinfo {year} {1999})},\ \Eprint
  {http://arxiv.org/abs/hep-ph/9807443} {arXiv:hep-ph/9807443} \BibitemShut
  {NoStop}%
\bibitem [{\citenamefont {Ball}\ \emph {et~al.}(1995)\citenamefont {Ball},
  \citenamefont {Beneke},\ and\ \citenamefont {Braun}}]{Ball:1995ni}%
  \BibitemOpen
  \bibfield  {author} {\bibinfo {author} {\bibfnamefont {P.}~\bibnamefont
  {Ball}}, \bibinfo {author} {\bibfnamefont {M.}~\bibnamefont {Beneke}}, \ and\
  \bibinfo {author} {\bibfnamefont {V.~M.}\ \bibnamefont {Braun}},\ }\href
  {\doibase 10.1016/0550-3213(95)00392-6} {\bibfield  {journal} {\bibinfo
  {journal} {Nucl. Phys. B}\ }\textbf {\bibinfo {volume} {452}},\ \bibinfo
  {pages} {563} (\bibinfo {year} {1995})},\ \Eprint
  {http://arxiv.org/abs/hep-ph/9502300} {arXiv:hep-ph/9502300} \BibitemShut
  {NoStop}%
\bibitem [{\citenamefont {Neubert}(1996)}]{Neubert:1995gd}%
  \BibitemOpen
  \bibfield  {author} {\bibinfo {author} {\bibfnamefont {M.}~\bibnamefont
  {Neubert}},\ }\href {\doibase 10.1016/0550-3213(96)00002-8} {\bibfield
  {journal} {\bibinfo  {journal} {Nucl. Phys. B}\ }\textbf {\bibinfo {volume}
  {463}},\ \bibinfo {pages} {511} (\bibinfo {year} {1996})},\ \Eprint
  {http://arxiv.org/abs/hep-ph/9509432} {arXiv:hep-ph/9509432} \BibitemShut
  {NoStop}%
\bibitem [{\citenamefont {Beneke}\ and\ \citenamefont
  {Jamin}(2008)}]{Beneke:2008ad}%
  \BibitemOpen
  \bibfield  {author} {\bibinfo {author} {\bibfnamefont {M.}~\bibnamefont
  {Beneke}}\ and\ \bibinfo {author} {\bibfnamefont {M.}~\bibnamefont {Jamin}},\
  }\href {\doibase 10.1088/1126-6708/2008/09/044} {\bibfield  {journal}
  {\bibinfo  {journal} {JHEP}\ }\textbf {\bibinfo {volume} {09}},\ \bibinfo
  {pages} {044} (\bibinfo {year} {2008})},\ \Eprint
  {http://arxiv.org/abs/0806.3156} {arXiv:0806.3156 [hep-ph]} \BibitemShut
  {NoStop}%
\bibitem [{\citenamefont {Gorishnii}\ \emph {et~al.}(1991)\citenamefont
  {Gorishnii}, \citenamefont {Kataev},\ and\ \citenamefont
  {Larin}}]{Gorishnii:1990vf}%
  \BibitemOpen
  \bibfield  {author} {\bibinfo {author} {\bibfnamefont {S.}~\bibnamefont
  {Gorishnii}}, \bibinfo {author} {\bibfnamefont {A.}~\bibnamefont {Kataev}}, \
  and\ \bibinfo {author} {\bibfnamefont {S.}~\bibnamefont {Larin}},\ }\href
  {\doibase 10.1016/0370-2693(91)90149-K} {\bibfield  {journal} {\bibinfo
  {journal} {Phys. Lett. B}\ }\textbf {\bibinfo {volume} {259}},\ \bibinfo
  {pages} {144} (\bibinfo {year} {1991})}\BibitemShut {NoStop}%
\bibitem [{\citenamefont {Surguladze}\ and\ \citenamefont
  {Samuel}(1991)}]{Surguladze:1990tg}%
  \BibitemOpen
  \bibfield  {author} {\bibinfo {author} {\bibfnamefont {L.~R.}\ \bibnamefont
  {Surguladze}}\ and\ \bibinfo {author} {\bibfnamefont {M.~A.}\ \bibnamefont
  {Samuel}},\ }\href {\doibase 10.1103/PhysRevLett.66.560} {\bibfield
  {journal} {\bibinfo  {journal} {Phys. Rev. Lett.}\ }\textbf {\bibinfo
  {volume} {66}},\ \bibinfo {pages} {560} (\bibinfo {year} {1991})},\ \bibinfo
  {note} {[Erratum: Phys.Rev.Lett. 66, 2416 (1991)]}\BibitemShut {NoStop}%
\bibitem [{\citenamefont {Baikov}\ \emph {et~al.}(2008)\citenamefont {Baikov},
  \citenamefont {Chetyrkin},\ and\ \citenamefont {Kuhn}}]{Baikov:2008jh}%
  \BibitemOpen
  \bibfield  {author} {\bibinfo {author} {\bibfnamefont {P.}~\bibnamefont
  {Baikov}}, \bibinfo {author} {\bibfnamefont {K.}~\bibnamefont {Chetyrkin}}, \
  and\ \bibinfo {author} {\bibfnamefont {J.~H.}\ \bibnamefont {Kuhn}},\ }\href
  {\doibase 10.1103/PhysRevLett.101.012002} {\bibfield  {journal} {\bibinfo
  {journal} {Phys. Rev. Lett.}\ }\textbf {\bibinfo {volume} {101}},\ \bibinfo
  {pages} {012002} (\bibinfo {year} {2008})},\ \Eprint
  {http://arxiv.org/abs/0801.1821} {arXiv:0801.1821 [hep-ph]} \BibitemShut
  {NoStop}%
\bibitem [{\citenamefont {Boito}\ \emph {et~al.}(2021)\citenamefont {Boito},
  \citenamefont {Golterman}, \citenamefont {Maltman}, \citenamefont {Peris},
  \citenamefont {Rodrigues},\ and\ \citenamefont {Schaaf}}]{Boito:2020xli}%
  \BibitemOpen
  \bibfield  {author} {\bibinfo {author} {\bibfnamefont {D.}~\bibnamefont
  {Boito}}, \bibinfo {author} {\bibfnamefont {M.}~\bibnamefont {Golterman}},
  \bibinfo {author} {\bibfnamefont {K.}~\bibnamefont {Maltman}}, \bibinfo
  {author} {\bibfnamefont {S.}~\bibnamefont {Peris}}, \bibinfo {author}
  {\bibfnamefont {M.~V.}\ \bibnamefont {Rodrigues}}, \ and\ \bibinfo {author}
  {\bibfnamefont {W.}~\bibnamefont {Schaaf}},\ }\href {\doibase
  10.1103/PhysRevD.103.034028} {\bibfield  {journal} {\bibinfo  {journal}
  {Phys. Rev. D}\ }\textbf {\bibinfo {volume} {103}},\ \bibinfo {pages}
  {034028} (\bibinfo {year} {2021})},\ \Eprint
  {http://arxiv.org/abs/2012.10440} {arXiv:2012.10440 [hep-ph]} \BibitemShut
  {NoStop}%
\bibitem [{\citenamefont {Zyla}\ \emph {et~al.}(2020)\citenamefont {Zyla} \emph
  {et~al.}}]{Pdg2020}%
  \BibitemOpen
  \bibfield  {author} {\bibinfo {author} {\bibfnamefont {P.}~\bibnamefont
  {Zyla}} \emph {et~al.} (\bibinfo {collaboration} {Particle Data Group}),\
  }\href@noop {} {\bibfield  {journal} {\bibinfo  {journal} {to be published in
  Prog. Theor. Exp. Phys.}\ }\textbf {\bibinfo {volume} {2020}} (\bibinfo
  {year} {2020})}\BibitemShut {NoStop}%
\bibitem [{\citenamefont {Pivovarov}(1991)}]{Pivovarov:1991rh}%
  \BibitemOpen
  \bibfield  {author} {\bibinfo {author} {\bibfnamefont {A.}~\bibnamefont
  {Pivovarov}},\ }\href {\doibase 10.1007/BF01625906} {\bibfield  {journal}
  {\bibinfo  {journal} {Sov. J. Nucl. Phys.}\ }\textbf {\bibinfo {volume}
  {54}},\ \bibinfo {pages} {676} (\bibinfo {year} {1991})},\ \Eprint
  {http://arxiv.org/abs/hep-ph/0302003} {arXiv:hep-ph/0302003} \BibitemShut
  {NoStop}%
\bibitem [{\citenamefont {Caprini}\ and\ \citenamefont
  {Fischer}(2009)}]{Caprini:2009vf}%
  \BibitemOpen
  \bibfield  {author} {\bibinfo {author} {\bibfnamefont {I.}~\bibnamefont
  {Caprini}}\ and\ \bibinfo {author} {\bibfnamefont {J.}~\bibnamefont
  {Fischer}},\ }\href {\doibase 10.1140/epjc/s10052-009-1142-8} {\bibfield
  {journal} {\bibinfo  {journal} {Eur. Phys. J. C}\ }\textbf {\bibinfo {volume}
  {64}},\ \bibinfo {pages} {35} (\bibinfo {year} {2009})},\ \Eprint
  {http://arxiv.org/abs/0906.5211} {arXiv:0906.5211 [hep-ph]} \BibitemShut
  {NoStop}%
\bibitem [{\citenamefont {Caprini}\ and\ \citenamefont
  {Fischer}(2011)}]{Caprini:2011ya}%
  \BibitemOpen
  \bibfield  {author} {\bibinfo {author} {\bibfnamefont {I.}~\bibnamefont
  {Caprini}}\ and\ \bibinfo {author} {\bibfnamefont {J.}~\bibnamefont
  {Fischer}},\ }\href {\doibase 10.1103/PhysRevD.84.054019} {\bibfield
  {journal} {\bibinfo  {journal} {Phys. Rev. D}\ }\textbf {\bibinfo {volume}
  {84}},\ \bibinfo {pages} {054019} (\bibinfo {year} {2011})},\ \Eprint
  {http://arxiv.org/abs/1106.5336} {arXiv:1106.5336 [hep-ph]} \BibitemShut
  {NoStop}%
\bibitem [{\citenamefont {Jamin}(2005)}]{Jamin:2005ip}%
  \BibitemOpen
  \bibfield  {author} {\bibinfo {author} {\bibfnamefont {M.}~\bibnamefont
  {Jamin}},\ }\href {\doibase 10.1088/1126-6708/2005/09/058} {\bibfield
  {journal} {\bibinfo  {journal} {JHEP}\ }\textbf {\bibinfo {volume} {09}},\
  \bibinfo {pages} {058} (\bibinfo {year} {2005})},\ \Eprint
  {http://arxiv.org/abs/hep-ph/0509001} {arXiv:hep-ph/0509001} \BibitemShut
  {NoStop}%
\bibitem [{\citenamefont {Descotes-Genon}\ and\ \citenamefont
  {Malaescu}(2010)}]{DescotesGenon:2010cr}%
  \BibitemOpen
  \bibfield  {author} {\bibinfo {author} {\bibfnamefont {S.}~\bibnamefont
  {Descotes-Genon}}\ and\ \bibinfo {author} {\bibfnamefont {B.}~\bibnamefont
  {Malaescu}},\ }\href@noop {} {\  (\bibinfo {year} {2010})},\ \Eprint
  {http://arxiv.org/abs/1002.2968} {arXiv:1002.2968 [hep-ph]} \BibitemShut
  {NoStop}%
\bibitem [{\citenamefont {Beneke}\ \emph {et~al.}(2013)\citenamefont {Beneke},
  \citenamefont {Boito},\ and\ \citenamefont {Jamin}}]{Beneke:2012vb}%
  \BibitemOpen
  \bibfield  {author} {\bibinfo {author} {\bibfnamefont {M.}~\bibnamefont
  {Beneke}}, \bibinfo {author} {\bibfnamefont {D.}~\bibnamefont {Boito}}, \
  and\ \bibinfo {author} {\bibfnamefont {M.}~\bibnamefont {Jamin}},\ }\href
  {\doibase 10.1007/JHEP01(2013)125} {\bibfield  {journal} {\bibinfo  {journal}
  {JHEP}\ }\textbf {\bibinfo {volume} {01}},\ \bibinfo {pages} {125} (\bibinfo
  {year} {2013})},\ \Eprint {http://arxiv.org/abs/1210.8038} {arXiv:1210.8038
  [hep-ph]} \BibitemShut {NoStop}%
\bibitem [{\citenamefont {Benitez-Rathgeb}\ \emph {et~al.}(2022)\citenamefont
  {Benitez-Rathgeb}, \citenamefont {Boito}, \citenamefont {Hoang},\ and\
  \citenamefont {Jamin}}]{Benitez-Rathgeb:2022yqb}%
  \BibitemOpen
  \bibfield  {author} {\bibinfo {author} {\bibfnamefont {M.~A.}\ \bibnamefont
  {Benitez-Rathgeb}}, \bibinfo {author} {\bibfnamefont {D.}~\bibnamefont
  {Boito}}, \bibinfo {author} {\bibfnamefont {A.~H.}\ \bibnamefont {Hoang}}, \
  and\ \bibinfo {author} {\bibfnamefont {M.}~\bibnamefont {Jamin}},\
  }\href@noop {} {\  (\bibinfo {year} {2022})},\ \Eprint
  {http://arxiv.org/abs/2202.10957} {arXiv:2202.10957 [hep-ph]} \BibitemShut
  {NoStop}%
\bibitem [{\citenamefont {Hoang}\ \emph {et~al.}(2018)\citenamefont {Hoang},
  \citenamefont {Jain}, \citenamefont {Lepenik}, \citenamefont {Mateu},
  \citenamefont {Preisser}, \citenamefont {Scimemi},\ and\ \citenamefont
  {Stewart}}]{Hoang:2017suc}%
  \BibitemOpen
  \bibfield  {author} {\bibinfo {author} {\bibfnamefont {A.~H.}\ \bibnamefont
  {Hoang}}, \bibinfo {author} {\bibfnamefont {A.}~\bibnamefont {Jain}},
  \bibinfo {author} {\bibfnamefont {C.}~\bibnamefont {Lepenik}}, \bibinfo
  {author} {\bibfnamefont {V.}~\bibnamefont {Mateu}}, \bibinfo {author}
  {\bibfnamefont {M.}~\bibnamefont {Preisser}}, \bibinfo {author}
  {\bibfnamefont {I.}~\bibnamefont {Scimemi}}, \ and\ \bibinfo {author}
  {\bibfnamefont {I.~W.}\ \bibnamefont {Stewart}},\ }\href {\doibase
  10.1007/JHEP04(2018)003} {\bibfield  {journal} {\bibinfo  {journal} {JHEP}\
  }\textbf {\bibinfo {volume} {04}},\ \bibinfo {pages} {003} (\bibinfo {year}
  {2018})},\ \Eprint {http://arxiv.org/abs/1704.01580} {arXiv:1704.01580
  [hep-ph]} \BibitemShut {NoStop}%
\bibitem [{\citenamefont {Brown}\ \emph {et~al.}(1992)\citenamefont {Brown},
  \citenamefont {Yaffe},\ and\ \citenamefont {Zhai}}]{Brown:1992pk}%
  \BibitemOpen
  \bibfield  {author} {\bibinfo {author} {\bibfnamefont {L.~S.}\ \bibnamefont
  {Brown}}, \bibinfo {author} {\bibfnamefont {L.~G.}\ \bibnamefont {Yaffe}}, \
  and\ \bibinfo {author} {\bibfnamefont {C.-X.}\ \bibnamefont {Zhai}},\ }\href
  {\doibase 10.1103/PhysRevD.46.4712} {\bibfield  {journal} {\bibinfo
  {journal} {Phys. Rev. D}\ }\textbf {\bibinfo {volume} {46}},\ \bibinfo
  {pages} {4712} (\bibinfo {year} {1992})},\ \Eprint
  {http://arxiv.org/abs/hep-ph/9205213} {arXiv:hep-ph/9205213} \BibitemShut
  {NoStop}%
\bibitem [{\citenamefont {Boito}\ \emph {et~al.}(2016)\citenamefont {Boito},
  \citenamefont {Jamin},\ and\ \citenamefont {Miravitllas}}]{Boito:2016pwf}%
  \BibitemOpen
  \bibfield  {author} {\bibinfo {author} {\bibfnamefont {D.}~\bibnamefont
  {Boito}}, \bibinfo {author} {\bibfnamefont {M.}~\bibnamefont {Jamin}}, \ and\
  \bibinfo {author} {\bibfnamefont {R.}~\bibnamefont {Miravitllas}},\ }\href
  {\doibase 10.1103/PhysRevLett.117.152001} {\bibfield  {journal} {\bibinfo
  {journal} {Phys. Rev. Lett.}\ }\textbf {\bibinfo {volume} {117}},\ \bibinfo
  {pages} {152001} (\bibinfo {year} {2016})},\ \Eprint
  {http://arxiv.org/abs/1606.06175} {arXiv:1606.06175 [hep-ph]} \BibitemShut
  {NoStop}%
\bibitem [{\citenamefont {Broadhurst}(1993)}]{Broadhurst:1992si}%
  \BibitemOpen
  \bibfield  {author} {\bibinfo {author} {\bibfnamefont {D.~J.}\ \bibnamefont
  {Broadhurst}},\ }\href {\doibase 10.1007/BF01560355} {\bibfield  {journal}
  {\bibinfo  {journal} {Z. Phys. C}\ }\textbf {\bibinfo {volume} {58}},\
  \bibinfo {pages} {339} (\bibinfo {year} {1993})}\BibitemShut {NoStop}%
\bibitem [{\citenamefont {Regner}(2020)}]{MasterThesisRegner}%
  \BibitemOpen
  \bibfield  {author} {\bibinfo {author} {\bibfnamefont {C.}~\bibnamefont
  {Regner}},\ }\href@noop {} {\bibfield  {journal} {\bibinfo  {journal}
  {University of Vienna}\ } (\bibinfo {year} {2020})}\BibitemShut {NoStop}%
\bibitem [{\citenamefont {Davier}\ \emph {et~al.}(2008)\citenamefont {Davier},
  \citenamefont {Descotes-Genon}, \citenamefont {Hocker}, \citenamefont
  {Malaescu},\ and\ \citenamefont {Zhang}}]{Davier:2008sk}%
  \BibitemOpen
  \bibfield  {author} {\bibinfo {author} {\bibfnamefont {M.}~\bibnamefont
  {Davier}}, \bibinfo {author} {\bibfnamefont {S.}~\bibnamefont
  {Descotes-Genon}}, \bibinfo {author} {\bibfnamefont {A.}~\bibnamefont
  {Hocker}}, \bibinfo {author} {\bibfnamefont {B.}~\bibnamefont {Malaescu}}, \
  and\ \bibinfo {author} {\bibfnamefont {Z.}~\bibnamefont {Zhang}},\ }\href
  {\doibase 10.1140/epjc/s10052-008-0666-7} {\bibfield  {journal} {\bibinfo
  {journal} {Eur. Phys. J. C}\ }\textbf {\bibinfo {volume} {56}},\ \bibinfo
  {pages} {305} (\bibinfo {year} {2008})},\ \Eprint
  {http://arxiv.org/abs/0803.0979} {arXiv:0803.0979 [hep-ph]} \BibitemShut
  {NoStop}%
\end{thebibliography}%

\end{document}